\documentclass[a4paper,fleqn]{cas-sc}

\usepackage{amsmath}
\usepackage{cases}
\usepackage[numbers]{natbib}
\usepackage{mathtools}
\usepackage{interval}
\usepackage{csquotes}
\usepackage{mathrsfs}
\usepackage{bm}
\usepackage{enumerate}
\usepackage[center]{subfigure}
\usepackage{epstopdf}
\usepackage{float}
\usepackage{algorithm}
\usepackage{algorithmicx} 
\usepackage{algpseudocode}

\usepackage{indentfirst}
\usepackage{hypernat}
\usepackage{microtype} 
\usepackage{multicol} 
\usepackage{lettrine} 
\usepackage{xcolor}
\usepackage{hyperref} 

\begin{document}
	\let\WriteBookmarks\relax
	\def\floatpagepagefraction{1}
	\def\textpagefraction{.001}
	
	\shorttitle{Energy-stable finite-volume scheme for wall distance}
	\shortauthors{X. Xu and Q. Wang}

\title[mode=title]{An energy-stable and accuracy-preserving finite volume scheme for wall-distance computations}

\author[1]{Xiaorui Xu}[orcid=0009-0001-1776-8687]
\ead{xxrcfd@csrc.ac.cn}
\author[1]{Qian Wang}[orcid=0000-0001-5409-1663]
\cormark[1]
\ead{qian.wang@csrc.ac.cn}
\cortext[1]{Corresponding author.}

\affiliation[1]{organization={Mechanics Division, Beijing Computational Science Research Center},
	city={Beijing},
	postcode={100193},
	country={China}}

\begin{abstract}
The distance to the nearest wall provides a length scale in models of wall-bounded turbulence. Computing this distance accurately on highly stretched meshes is challenging because small near-wall cells restrict explicit pseudo-time steps and geometric singularities can induce spurious oscillations. To address these challenges, this paper presents a semi-implicit finite-volume scheme based on a scalar auxiliary variable (SAV) for solving the pseudo-time Eikonal equation on unstructured meshes. An upwind weighted least-squares reconstruction supplies the gradients used to evaluate the cell-averaged Eikonal residuals. The artificial-viscosity term is discretized using a two-point face flux whose coefficient vanishes as the regularized residuals in both adjacent cells tend to zero. The SAV update advances the semi-discrete system in pseudo-time by evaluating the residuals and viscosity coefficients explicitly and coupling the wall-distance averages and auxiliary variables implicitly. The analysis establishes unconditional energy stability under homogeneous Dirichlet conditions, permitting a grid-independent global pseudo-time step. The system's M-matrix structure guarantees nonnegative cell averages of the physical wall distance under a separate step restriction, while residual regularization and artificial viscosity preserve the formal order of spatial accuracy. 
Numerical tests confirm second-order spatial convergence for the quarter-cylinder case and demonstrate robust wall-distance computation on highly stretched meshes typical of Reynolds-averaged Navier--Stokes simulations. 
The SAV formulation is independent of the reconstruction scheme, providing a basis for extensions to higher-order finite-volume methods and other nonconservative scalar equations.

\end{abstract}




\maketitle


\section{Introduction}\label{section:Introduction}

The distance to the nearest wall is an important geometric quantity in simulations of wall-bounded turbulent flows, serving as a length scale and entering wall treatments in Reynolds-averaged Navier--Stokes (RANS) closures such as the Spalart--Allmaras model \cite{spalart1992one,allmaras2012modifications} and in hybrid RANS/large-eddy simulation (LES) methods such as the detached-eddy simulation (DES) \cite{tucker2003differential,tucker2005computations}. Distance and signed-distance fields are also used in grid generation \cite{quadros2004laytracks}, overset-grid assembly \cite{kang2013fast}, moving-boundary computations \cite{kakumani2022use}, level-set redistancing \cite{gibou2018review}, medial-axis extraction \cite{xia2010finite}, surface reconstruction in computer graphics \cite{calakli2011ssd}, and automatic feature removal in computer-aided design (CAD) geometries \cite{lee2010surface}. A direct approach to computing wall distance is to search for the nearest surface element for each volume point. Search-based methods can provide accurate reference values, yet a brute-force nearest-surface search requires $O(N_vN_s)$ operations, where $N_v$ and $N_s$ denote the numbers of volume and wall grid points, respectively \cite{tucker2003differential,xia2010finite}. Search procedures also require geometry-dependent data structures, can introduce load-balancing and memory-replication issues in distributed-memory implementations, and can incur repeated search costs for moving or adaptive meshes \cite{tucker2003transport,wang2026parallel}. These limitations have motivated the development of iterative and differential-equation-based approaches, whose mathematical foundation lies in the theory of first-order Hamilton--Jacobi equations \cite{crandall1984two}.

Iterative approaches are often used to solve the steady Eikonal equation. Many of these approaches rely on the causality principle, whereby the solution at any point depends only on upstream information. This principle has motivated the development of three main families of fast Eikonal solvers, represented by the fast marching method (FMM), the fast sweeping method (FSM), and the fast iterative method (FIM). FMM accepts grid points in order of increasing arrival time using a heap-based data structure, but this sequential ordering limits parallel efficiency \cite{sethian1996fast,sethian1999fastmarching}. FSM propagates information through alternating Gauss--Seidel sweeps, whose efficiency depends on the sweeping order and the geometry of the characteristics \cite{tsai2003fast,zhao2005fast,gremaud2006computational,kleb2025automatic}. FIM permits concurrent updates of grid points in an unordered active set, although individual points may require repeated updates before convergence \cite{jeong2008fast}. A lattice Boltzmann formulation has also been developed for steady Eikonal problems and uses explicit local updates suited to parallel computation \cite{douich2026lattice}. Although these methods are effective, integrating them into production CFD codes can require changes to data structures and parallel algorithms, particularly for distributed-memory computations on highly stretched, unstructured boundary-layer meshes. Mesh motion or adaptation may also require updates to the wall-distance field and its supporting data structures. These considerations motivate solving the Eikonal equation directly within the numerical framework used by the flow solver.

Differential-equation-based approaches compute the wall distance by solving a governing equation within the numerical framework of a flow solver. Poisson-type approaches \cite{tucker2003differential,tucker2003transport,tucker2005computations,tucker2011hybrid,xu2011computations} are simple and robust, although they can be less accurate in regions far from the wall \cite{tucker2011hybrid,mor2025central}. Eikonal models represent the wall distance more faithfully and admit both steady and pseudo-time formulations. In the steady formulation, the discrete Eikonal equation can be solved implicitly using Newton-type iterations. These methods can converge rapidly near the solution, but constructing the discrete Jacobian can be complicated, and convergence can be sensitive to the initial guess \cite{mor2025central,de2026finite}. Pseudo-time formulations introduce a pseudo-time derivative and use a pseudo-velocity related to the distance gradient to express the equation in a transport-like form that can be advanced to steady state \cite{tucker2005computations,xia2010finite}. 
Numerical approaches to the pseudo-time Eikonal equation include finite-difference (FD) \cite{sussman1994level,sussman1999efficient}, discontinuous Galerkin (DG) \cite{cheng2007discontinuous,lebouteiller2018accurate}, and finite-volume (FV) schemes \cite{xia2010finite,xu2011computations,huang2021high}.
Pseudo-time marching avoids the need for Newton-type iterations and fits naturally into the time-stepping framework of CFD solvers, which simplifies implementation. A pseudo-time Eikonal formulation is therefore adopted in this work.
However, existing pseudo-time Eikonal schemes can encounter stability difficulties. The Courant--Friedrichs--Lewy (CFL) condition can severely restrict the pseudo-time step in explicit schemes, particularly on meshes with highly distorted cells or large cell aspect ratios. Even when this condition is satisfied, geometric singularities may induce spurious oscillations that destabilize the pseudo-time iteration and cause it to diverge.

To address these stability challenges, we propose a finite-volume scheme based on the scalar auxiliary variable (SAV) approach for the pseudo-time Eikonal equation with vanishing artificial viscosity.
An upwind weighted least-squares reconstruction yields a piecewise linear approximation of the wall-distance field. In each control volume, the cell-averaged Eikonal residual is evaluated from the gradient of the reconstructed polynomial and then regularized to obtain a bounded residual. At each interior face, the artificial-viscosity flux is approximated using a two-point difference of the adjacent cell averages. The viscosity coefficient depends on the magnitudes of the regularized residuals in the two adjacent cells and tends to zero as both residuals vanish. The resulting semi-discrete system is advanced in pseudo-time using a cellwise scalar auxiliary variable. The nonlinear Eikonal residual and viscosity coefficients are evaluated explicitly, while the wall-distance averages and auxiliary variables are updated implicitly through a coupled linear system.

Under homogeneous Dirichlet boundary conditions, the SAV-FV update is unconditionally energy-stable, permitting a global pseudo-time step independent of the grid spacing. Under a separate pseudo-time step restriction, the discrete comparison principle associated with the system's M-matrix structure guarantees nonnegative cell averages of the physical wall distance. For the larger steps used in practice, a lower bound is enforced directly after each update. The analysis also shows that the residual regularization and the vanishing artificial viscosity preserve the formal order of spatial accuracy. Numerical tests confirm second-order convergence for the quarter-cylinder case and demonstrate robust wall-distance computation on highly stretched meshes representative of practical RANS simulations.

To the authors' knowledge, the present work is the first to combine the SAV approach with a cell-centered finite-volume discretization on unstructured meshes. The SAV approach was introduced by Shen et al.~\cite{shen2018sav,shen2018convergence,shen2019efficient} for gradient-flow problems. By representing the nonlinear part of the free energy through a scalar auxiliary variable, their formulation yields linear time-discrete schemes that are unconditionally energy-stable with respect to a modified energy. The approach has since been combined with spectral methods \cite{shen2018convergence,shen2019efficient,li2020fourier}, finite element methods \cite{li2022optimal,yao2024sav,wagner2026gsav}, block-centered finite difference methods \cite{li2019energy}, and finite volume element methods \cite{liu2026savfve}. Huang et al.~\cite{huang2023structure} developed a finite-volume SAV scheme for the degenerate Cahn--Hilliard equation on structured grids, using upwind fluxes based on cell averages and dimensional splitting. Extensions to higher-order reconstructions and unstructured meshes remain to be established.
A key challenge in applying the SAV approach to finite-volume methods is that its energy-stability analysis typically relies on a discrete integration-by-parts identity, which finite-volume reconstructions on general unstructured meshes do not automatically satisfy. The present work addresses this difficulty through a two-point artificial-viscosity flux that provides the required identity, with each interior face contributing a nonpositive multiple of the squared difference between adjacent cell averages to the discrete energy balance. The viscosity coefficient depends on the Eikonal residuals in the adjacent cells, allowing the stabilization to preserve the formal order of spatial accuracy. Since the SAV formulation is independent of the reconstruction scheme, the framework supports extensions to higher-order finite-volume methods on both structured and unstructured meshes and provides a basis for applications to other nonconservative scalar equations.

The remainder of this paper is organized as follows. Section~\ref{section:numerical method} presents the SAV-based finite-volume scheme for the pseudo-time Eikonal equation. Section~\ref{section:Analysis} analyzes its energy stability, bound preservation, and spatial accuracy. Section~\ref{section:numerical_results} presents the numerical results, and Section~\ref{section:conclusions} summarizes the conclusions.


\section{Finite volume scheme based on the SAV approach} \label{section:numerical method}


\subsection{Semi-discrete finite-volume scheme} \label{section:semi-discrete-fv}

The pseudo-time Eikonal equation is given by
\begin{equation}
	\left\{
	\begin{aligned}
		\frac{\partial \phi}{\partial \tau}
		+ \lVert \nabla \phi \rVert &= 1, \\[4pt]
		\left.\phi\right|_{\Gamma} &= 0,
	\end{aligned}
	\right.
	\label{eq:eikonal eqn}
\end{equation}
where $\tau$ denotes the pseudo-time variable, $\phi = \phi(\vec{x},\tau)$ is the distance field defined on the domain $\Omega \subset \mathbb{R}^n$, $\Gamma \subset \partial\Omega$ denotes the wall boundary, and $\lVert \cdot \rVert$ is the Euclidean norm.
Equation \eqref{eq:eikonal eqn} is regularized by introducing a vanishing artificial viscosity term:
\begin{equation}
	\frac{\partial \phi}{\partial \tau}
	+ \lVert \nabla \phi \rVert
	= 1 + \nabla \cdot \left(\epsilon \nabla \phi\right),
	\label{eq:vis-eikonal}
\end{equation}
where $\epsilon$ denotes the diffusion coefficient. As $\epsilon \rightarrow 0$, the solution of \eqref{eq:vis-eikonal} approaches that of \eqref{eq:eikonal eqn}. The artificial viscosity term serves two essential purposes: it smooths the solution \cite{huang2021high,mor2025central} and, more critically, suppresses unbounded growth of the computed distance field at local maxima. Without artificial viscosity, at a smooth local maximum where $\nabla \phi = 0$, equation \eqref{eq:eikonal eqn} reduces to $\partial \phi/\partial \tau = 1$, causing the computed wall distance to increase at the next time step. If the point remains a local maximum with a vanishing gradient, this increase persists over successive time steps, potentially leading to unbounded growth. Artificial viscosity is therefore essential for stabilizing the present numerical formulation.

Let $\Omega_i$ denote a control volume, $|\Omega_i|$ its volume, and $\bar{\phi}_i$ the corresponding cell average of $\phi$, defined by
\begin{equation}
	\bar{\phi}_i = \frac{1}{|\Omega_i|}\int_{\Omega_i}\phi\,dV.
\end{equation}
Averaging \eqref{eq:vis-eikonal} over $\Omega_i$ yields the semi-discrete finite-volume formulation
\begin{equation}
	\frac{d\bar{\phi}_i}{d\tau} + \bar{G}_i(\phi)
	= \frac{1}{|\Omega_i|}
	\int_{\Omega_i}\nabla \cdot (\epsilon \nabla \phi)\,dV,
	\label{eq:cell-average-eikonal}
\end{equation}
with the cell residual defined as
\begin{equation}
	\bar{G}_i(\phi)
	= \frac{1}{|\Omega_i|}
	\int_{\Omega_i}\left(\lVert \nabla \phi \rVert - 1\right)\,dV.
	\label{eq:ori-residual}
\end{equation}
In the implementation, the cell residual is regularized to obtain the bounded form
\begin{equation}
	G_i(\phi)
	= \frac{\bar{G}_i(\phi)}
	{\beta \tilde{d}_i^{-1} + |\bar{G}_i(\phi)|},
	\label{eq:regularized-G}
\end{equation}
where $\tilde{d}_i = \min_{f \in \partial\Omega_i}\lVert \vec{x}_i - \vec{x}_f \rVert$, with $\vec{x}_i$ denoting the centroid of cell $i$ and $\vec{x}_f$ the center of face $f$. The parameter $\beta$ is introduced to improve numerical stability for complex geometries and is set to $10$ in this work. The term $\beta\tilde{d}_i^{-1}$ controls the residual scaling on highly stretched or highly skewed cells. Its role in maintaining stability and preserving the scheme's formal order of accuracy is analyzed in Section \ref{section:Analysis}.
Since $G_i(\phi)=0$ if and only if $\bar{G}_i(\phi)=0$, the regularization preserves the steady-state solutions in the absence of artificial viscosity. The modified semi-discrete formulation considered in this work is
\begin{equation}
	\frac{d\bar{\phi}_i}{d\tau} + G_i(\phi)
	= \frac{1}{|\Omega_i|}
	\int_{\Omega_i}\nabla \cdot (\epsilon \nabla \phi)\,dV.
	\label{eq:eikonal-G_i}
\end{equation}

The steady-state wall-distance field is obtained by advancing the semi-discrete finite-volume scheme \eqref{eq:eikonal-G_i} in pseudo-time until the pseudo-time derivative vanishes. At each pseudo-time step, given the current cell averages, the spatial discretization proceeds in three stages:
\begin{enumerate}[(1)]
	\setlength{\itemsep}{0pt}
	\setlength{\parsep}{0pt}
	\setlength{\parskip}{0pt}
	\item  A polynomial $\phi_i$ is reconstructed in each control volume $\Omega_i$ using the procedure described in Section \ref{section:wls-reconstruction}, yielding a piecewise polynomial approximation of the wall-distance field. 
	\item The cell-averaged residual $\bar{G}_i$ is evaluated by numerically integrating $\lVert\nabla\phi_i\rVert-1$ over $\Omega_i$ and normalizing the result by cell volume $|\Omega_i|$, with gradients at the quadrature points obtained by differentiating the reconstruction polynomial $\phi_i$. The regularized residual $G_i$ is then computed using \eqref{eq:regularized-G}. 
	\item The vanishing artificial-viscosity term is discretized as described in Section \ref{section:artificial-viscosity}.
\end{enumerate}


Once the spatial discretization is complete, the resulting semi-discrete system is advanced by one pseudo-time step using the scalar auxiliary variable (SAV) approach presented in Section \ref{section:sav-discretization}.

\subsection{Upwind weighted least-squares reconstruction} \label{section:wls-reconstruction}

A piecewise polynomial approximation of $\phi$ is constructed over the computational domain using a linear upwind weighted least-squares reconstruction. In each control volume, the reconstruction polynomial is differentiated to obtain $\nabla\phi$, which is then used to evaluate the Eikonal residual. 

In each control volume $\Omega_i$, the solution is approximated by a linear polynomial
\begin{equation}
	\phi_i (\vec{x})
	= \bar{\phi}_i
	+ \sum_{m=1}^{M_i}a_{i,m}\varphi_{i,m}(\vec{x}),
	\label{eq:poly-reconstruction}
\end{equation}
where $M_i=3$ in three dimensions, $\boldsymbol{a}_i=(a_{i,1},a_{i,2},a_{i,3})^T$ is the vector of unknown reconstruction coefficients, and $\varphi_{i,m}$ are linear basis functions. In Cartesian coordinates, these basis functions can be chosen as
\begin{equation}
	\varphi_{i,1}=x-x_i,
	\qquad
	\varphi_{i,2}=y-y_i,
	\qquad
	\varphi_{i,3}=z-z_i.
	\label{eq:linear-basis}
\end{equation}
Since the basis functions are centered at the cell centroid, each has zero cell average. Consequently, the polynomial in \eqref{eq:poly-reconstruction} preserves the prescribed cell average $\bar{\phi}_i$ exactly, ensuring a conservative reconstruction.

\begin{figure} 
	\centering
	\includegraphics[width=0.4\textwidth]{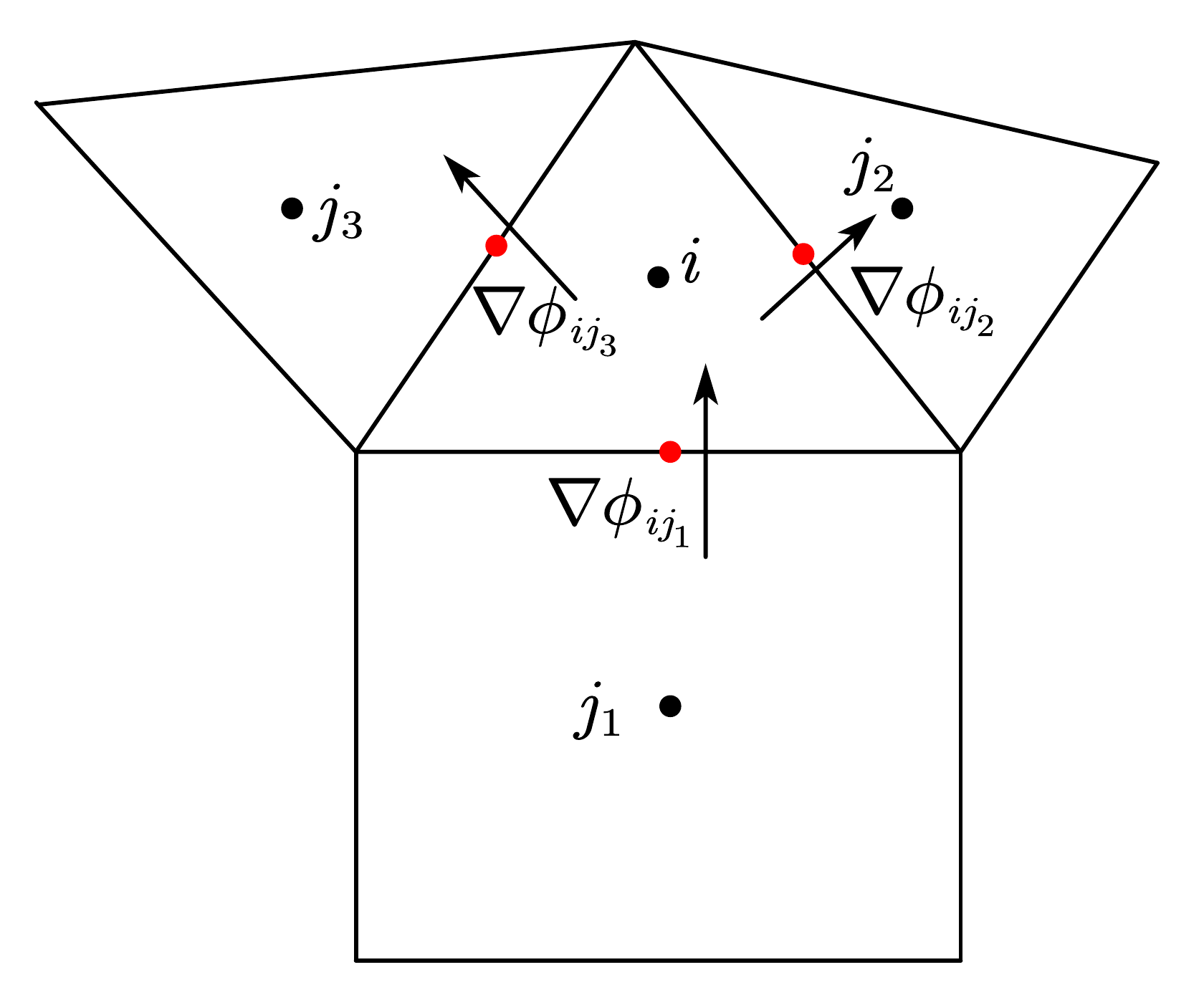}
	\caption{Reconstruction stencil.}
	\label{fig:stencil_wls}
\end{figure}

The reconstruction coefficients $\boldsymbol{a}_i$ are determined from the known cell averages on the stencil $S_i=\{i\}\cup\mathcal{N}(i)$, which comprises the target cell $i$ and its face neighbors, as illustrated in Figure \ref{fig:stencil_wls}. Here, $\mathcal{N}(i)$ denotes the set of cells sharing a face with cell $i$, and $N_i=|\mathcal{N}(i)|$ is the number of such cells. The polynomial $\phi_i$ is extended to each neighboring cell $\Omega_j$, and its average over $\Omega_j$ is fitted to the prescribed cell average $\bar{\phi}_j$.
For each $j\in\mathcal{N}(i)$, the corresponding mean-matching condition is
\begin{equation}
	\frac{1}{|\Omega_j|}
	\int_{\Omega_j}\phi_i(\vec{x})\,dV
	= \bar{\phi}_j.
	\label{eq:mean-matching}
\end{equation}
Substituting \eqref{eq:poly-reconstruction} into \eqref{eq:mean-matching} yields
\begin{equation}
	\sum_{m=1}^{M_i}a_{i,m}\mu_{ij,m}
	= \bar{\phi}_j-\bar{\phi}_i,
	\label{eq:ls-row}
\end{equation}
where
\begin{equation}
	\mu_{ij,m}
	= \frac{1}{|\Omega_j|}
	\int_{\Omega_j}\varphi_{i,m}(\vec{x})\,dV.
\end{equation}
For the Cartesian linear basis in \eqref{eq:linear-basis}, these moments reduce to the differences between the cell-centroid coordinates:
\begin{equation}
	\mu_{ij,1}=x_j-x_i,
	\qquad
	\mu_{ij,2}=y_j-y_i,
	\qquad
	\mu_{ij,3}=z_j-z_i.
\end{equation}
Assembling the conditions for all face neighbors yields a linear system with $N_i$ equations and $M_i$ unknown reconstruction coefficients. These conditions are enforced in a least-squares sense to determine $\boldsymbol{a}_i$.

To improve reconstruction accuracy on highly anisotropic grids \cite{mor2025central} and introduce an upwind bias, a weight $\omega_{ij}$ is assigned to each neighboring cell $j\in\mathcal{N}(i)$. The weighted mean-matching condition corresponding to \eqref{eq:ls-row} is
\begin{equation}
	\omega_{ij}\sum_{m=1}^{M_i}a_{i,m}\mu_{ij,m}
	= \omega_{ij}\left(\bar{\phi}_j-\bar{\phi}_i\right).
	\label{eq:wls-weighted-row}
\end{equation}
Collecting these conditions for all neighboring cells yields the compact matrix form
\begin{equation}
	\mathsf{B}_i\boldsymbol{a}_i=\boldsymbol{c}_i,
	\label{eq:wls-local-system}
\end{equation}
where the coefficient matrix $\mathsf{B}_i\in\mathbb{R}^{N_i\times M_i}$, the unknown vector $\boldsymbol{a}_i\in\mathbb{R}^{M_i}$, and the right-hand-side vector $\boldsymbol{c}_i\in\mathbb{R}^{N_i}$ are given by
\begin{equation}
	\begin{aligned}
		\mathsf{B}_i &=
		\begin{bmatrix}
			\omega_{ij_1}\mu_{ij_1,1}
			& \omega_{ij_1}\mu_{ij_1,2}
			& \cdots
			& \omega_{ij_1}\mu_{ij_1,M_i} \\
			\omega_{ij_2}\mu_{ij_2,1}
			& \omega_{ij_2}\mu_{ij_2,2}
			& \cdots
			& \omega_{ij_2}\mu_{ij_2,M_i} \\
			\vdots & \vdots & \ddots & \vdots \\
			\omega_{ij_{N_i}}\mu_{ij_{N_i},1}
			& \omega_{ij_{N_i}}\mu_{ij_{N_i},2}
			& \cdots
			& \omega_{ij_{N_i}}\mu_{ij_{N_i},M_i}
		\end{bmatrix},
		\\[6pt]
		\boldsymbol{a}_i &=
		\begin{bmatrix}
			a_{i,1} \\
			a_{i,2} \\
			\vdots \\
			a_{i,M_i}
		\end{bmatrix},
		\qquad
		\boldsymbol{c}_i=
		\begin{bmatrix}
			\omega_{ij_1}\left(\bar{\phi}_{j_1}-\bar{\phi}_i\right) \\
			\omega_{ij_2}\left(\bar{\phi}_{j_2}-\bar{\phi}_i\right) \\
			\vdots \\
			\omega_{ij_{N_i}}\left(\bar{\phi}_{j_{N_i}}-\bar{\phi}_i\right)
		\end{bmatrix},
	\end{aligned}
	\label{eq:wls-matrix-vector-form}
\end{equation}
with $j_k$ denoting the $k$th neighboring cell in $\mathcal{N}(i)$. Since the system is generally overdetermined, it is solved in a least-squares sense. The minimum-norm least-squares solution is
\begin{equation}
	\boldsymbol{a}_i=\mathsf{B}_i^{+}\boldsymbol{c}_i,
	\label{eq:wls-pseudo-inverse}
\end{equation}
where $\mathsf{B}_i^{+}$ denotes the Moore--Penrose pseudoinverse \cite{penrose1955generalized}.

The reconstruction weight $\omega_{ij}$ combines a geometric factor with an upwind factor. The geometric weight is defined as the inverse centroid distance,
\begin{equation}
	\omega_{ij}^{0}=\frac{1}{d_{ij}}, \qquad d_{ij}= \lVert \vec{x}_i - \vec{x}_j \rVert,
	\label{eq:distance-weight}
\end{equation}
which balances the geometric scales of the mean-matching equations on meshes with nonuniform centroid spacing, particularly highly stretched meshes.
The upwind factor $\chi_{ij}$ introduces a directional bias based on the local propagation direction of the nonconservative Eikonal operator. Let $\nabla\phi_{ij}$ denote the face-centered gradient obtained by averaging the reconstructed gradients from the two adjacent cells. A neighboring cell $j$ is classified as upstream of cell $i$ when
\begin{equation}
	\nabla\phi_{ij}\cdot(\vec{x}_i-\vec{x}_j)\ge0.
	\label{eq:upwind-criterion}
\end{equation}
Neighbors satisfying this criterion retain their full geometric weight, whereas downstream neighbors are downweighted by a factor $\alpha_{\phi}\ge1$:
\begin{equation}
	\chi_{ij}=
	\begin{cases}
		1,
		& \nabla\phi_{ij}\cdot(\vec{x}_i-\vec{x}_j)\ge0,\\[4pt]
		1/\alpha_{\phi},
		& \nabla\phi_{ij}\cdot(\vec{x}_i-\vec{x}_j)<0.
	\end{cases}
	\label{eq:upwind-weight}
\end{equation}
The resulting reconstruction weight is
\begin{equation}
	\omega_{ij}=\omega_{ij}^{0}\chi_{ij}
	=\frac{\chi_{ij}}{d_{ij}}.
	\label{eq:final-weight}
\end{equation}
This weighting favors information from the locally upstream side and reduces the influence of downstream neighbors without reformulating the Eikonal equation in conservative-flux form.
On meshes whose spacing increases away from the wall, inverse-distance weighting can itself favor neighbors on the smaller-distance side, introducing a geometric bias even when $\alpha_{\phi}=1$. On uniformly spaced stencils, this geometric bias is absent, and $\alpha_{\phi}>1$ is needed to introduce an explicit upwind preference. However, excessively large values of $\alpha_{\phi}$ may amplify errors originating at wall singularities in complex geometries and lead to numerical oscillations. In this work, $\alpha_{\phi}=10$ is recommended to balance convergence speed and robustness. The influence of $\alpha_{\phi}$ on numerical performance is examined in Section \ref{section:convergence-factors}.

Boundary faces are incorporated into the same algebraic formulation by replacing the missing neighboring cell average with a boundary or ghost value determined according to the treatment described in Section \ref{section:boundary-condition}.

%

\subsection{Vanishing artificial viscosity} \label{section:artificial-viscosity}

In this work, a residual-based vanishing artificial viscosity is employed. Its coefficient depends on the local Eikonal residual, providing stabilization where the residual is large and tending to zero as the numerical solution satisfies the original Eikonal equation $\|\nabla\phi\|=1$. This approach stabilizes the solution near geometric singularities while limiting artificial diffusion in smooth, well-resolved regions, thereby preserving the accuracy of the finite-volume scheme.

For an interior cell $i$, the cell-averaged artificial-viscosity term is discretized as
\begin{equation}
	\begin{aligned}
		\frac{1}{|\Omega_i|}
		\int_{\Omega_i}\nabla\cdot(\epsilon\nabla\phi)\,dV
		&= \frac{1}{|\Omega_i|}
		\oint_{\partial\Omega_i}\epsilon\nabla\phi\cdot\vec{n}\,dS
		\\[4pt]
		&\approx \frac{1}{|\Omega_i|}
		\sum_{j\in\mathcal{N}(i)}
		\frac{\epsilon_{ij}|S_{ij}|}{d_{ij}}
		\left(\bar{\phi}_j-\bar{\phi}_i\right)
		\\[4pt]
		&= \sum_{j\in\mathcal{N}(i)}
		w_{ij}\left(\bar{\phi}_j-\bar{\phi}_i\right),
	\end{aligned}
	\label{eq:vanishing-artifical-viscosity}
\end{equation}
where $\vec{n}$ denotes the outward unit normal to the cell boundary, $|S_{ij}|$ is the area of the face shared by cells $i$ and $j$. The face viscosity coefficient is defined as
\begin{equation}
	\epsilon_{ij}
	= \frac{d_{ij}}{2}
	\left(|\hat{G}_i|+|\hat{G}_j|\right),
	\qquad
	\hat{G}_k=\frac{G_k}{1+|G_k|},
	\quad k\in\{i,j\}.
\end{equation}
The corresponding diffusion weights are therefore
\begin{equation}
	w_{ij}
	= \frac{\epsilon_{ij}|S_{ij}|}{|\Omega_i|d_{ij}}
	= \frac{\left(|\hat{G}_i|+|\hat{G}_j|\right)|S_{ij}|}
	{2|\Omega_i|}.
	\label{eq:vis-weight}
\end{equation}

For a boundary cell $i$, at each physical boundary face, the same flux approximation is used with $\hat{G}_j=\hat{G}_i$. The missing neighboring cell average is replaced by either a boundary value or a ghost-cell average, as specified for each boundary type in Section~\ref{section:boundary-condition}.



\subsection{Pseudo-time stepping based on the SAV formulation} \label{section:sav-discretization}

The SAV formulation introduces the positive auxiliary function
\begin{equation}
	f(\tau)=\exp\left(-\frac{\tau}{\tau+c}\right),
	\qquad c>0,
	\label{eq:sav-function}
\end{equation}
where $\tau$ denotes the accumulated pseudo time and $c$ is set to $1$ in this work. This function satisfies 
\begin{equation}
	f'(\tau)=-\alpha(\tau)f(\tau), \quad 
	\alpha(\tau)=\frac{c}{(\tau+c)^2},
\end{equation}
allowing the auxiliary-variable equation to be expressed in a linear relaxation form.
For a global pseudo-time step $\Delta\tau$, let $\tau^{n+1}=\tau^n+\Delta\tau$ and define $f^{n+1}=f(\tau^{n+1})$, $\alpha^{n+1}=\alpha(\tau^{n+1})$, and $G_i^n=G_i(\phi^n)$. The artificial viscosity term is treated implicitly as $\nabla\cdot(\epsilon^n\nabla\phi^{n+1})$, with the diffusion coefficient evaluated at pseudo-time level $n$. The proposed semi-implicit SAV finite-volume (SAV-FV) scheme then reads
\begin{numcases}{}
	\frac{\bar{\phi}_i^{n+1}-\bar{\phi}_i^n}{\Delta\tau}
	+ \frac{r_i^{n+1}}{f^{n+1}}G_i^n
	= \sum_{j\in\mathcal{N}(i)}w_{ij}^n
	\left(\bar{\phi}_j^{n+1}-\bar{\phi}_i^{n+1}\right),
	\label{eq:sav-phi-update}
	\\[6pt]
	\frac{r_i^{n+1}-r_i^n}{\Delta\tau}
	= -\alpha^{n+1}r_i^{n+1}
	+ \frac{\bar{\phi}_i^{n+1}}{f^{n+1}}G_i^n,
	\label{eq:sav-r-update}
\end{numcases}
where $r_i$ denotes the cellwise scalar auxiliary variable. 
The nonlinear Eikonal residual $G_i^n$ is treated explicitly, while the cell averages and auxiliary variables at pseudo-time level $n+1$ are coupled implicitly.
Eliminating $r_i^{n+1}$ from \eqref{eq:sav-phi-update}--\eqref{eq:sav-r-update} yields a linear system for the updated cell averages, with the equation for cell $i$ given by
\begin{equation}
\left[\frac{1}{\Delta \tau}
+\frac{\Delta \tau (G_i^n)^2}{(1+\alpha^{n+1}\Delta \tau)(f^{n+1})^2} + \sum_{j\in\mathcal{N}(i)}w_{ij}^n \right]
\bar{\phi}_i^{n+1}
- \sum_{j\in\mathcal{N}(i)}w_{ij}^n \bar{\phi}_j^{n+1}
=\frac{\bar{\phi}_i^n}{\Delta \tau}
-\frac{r_i^nG_i^n}{(1+\alpha^{n+1}\Delta \tau)f^{n+1}} .
\label{eq:linear-sav-cell}
\end{equation}
Assembling \eqref{eq:linear-sav-cell} over all cells yields the global sparse linear system
\begin{equation}
	A\bar{\boldsymbol{\phi}}^{n+1}=\boldsymbol{b}.
	\label{eq:linear-system}
\end{equation}
For an interior cell $i$, the matrix entries and right-hand-side component are
\begin{align}
	A_{ii}&=\frac{1}{\Delta\tau}
	+\frac{\Delta\tau(G_i^n)^2}
	{(1+\alpha^{n+1}\Delta\tau)(f^{n+1})^2}
	+\sum_{j\in\mathcal{N}(i)}w_{ij}^n,
	\label{eq:matrix-diagonal}
	\\[6pt]
	A_{ij}&=-w_{ij}^n,
	\quad j\in\mathcal{N}(i),
	\label{eq:matrix-offdiagonal}
	\\[6pt]
	b_i&=\frac{\bar{\phi}_i^n}{\Delta\tau}
	-\frac{r_i^nG_i^n}
	{(1+\alpha^{n+1}\Delta\tau)f^{n+1}}.
	\label{eq:matrix-rhs}
\end{align}
The large sparse linear system \eqref{eq:linear-system} is solved using an algebraic multigrid (AMG) method \cite{brandt1986algebraic}.
Once the updated cell averages are obtained, $r_i^{n+1}$ is recovered from \eqref{eq:sav-r-update}. 

Under homogeneous Dirichlet boundary conditions, the SAV update is unconditionally energy-stable (Section~\ref{section:energy-stability}), motivating the use of large global pseudo-time steps in the reported computations. The nonsingular M-matrix structure of $A$ yields a discrete comparison principle that guarantees nonnegative cell averages of the physical wall distance under the pseudo-time step restriction given in Section~\ref{section:m-matrix-bound-preservation}. The residual-based artificial viscosity preserves the formal order of spatial accuracy, as shown in Section~\ref{section:accuracy-preserving}. As the Eikonal residuals $G_i^n$ tend to zero in every cell, the face weights $w_{ij}^n$ also tend to zero, so the artificial-viscosity term vanishes in the limit and the scheme recovers the original discrete Eikonal equation.

%
%

Although this work employs a second-order finite-volume reconstruction in Section \ref{section:wls-reconstruction}, the SAV formulation proposed in this paper is independent of the choice of reconstruction scheme. The method can therefore be extended to higher-order finite-volume schemes by adapting the upwind cell selection criterion to accommodate their enlarged reconstruction stencils. The upwind treatment described in \cite{mor2025central} provides guidance for this extension.

\subsection{Boundary conditions}\label{section:boundary-condition}

The physical wall-distance field satisfies a homogeneous Dirichlet condition on solid walls and a zero-normal-gradient condition on symmetry planes. At far-field boundaries, the distance is extrapolated from the interior. With these boundary treatments, the discrete SAV system may admit the spurious equilibrium $\bar{\phi}_i=r_i=0$ in every cell. The corresponding constant field has zero gradient and therefore violates the steady-state Eikonal equation $\lVert\nabla\phi\rVert=1$. This spurious equilibrium can hinder convergence to the physical wall-distance field.

To exclude the all-zero state, the discrete system is formulated in terms of the shifted field $\tilde{\phi}=\phi+1$, for which the wall condition becomes $\tilde{\phi}|_{\Gamma}=1$. The physical wall distance is recovered as $\phi=\tilde{\phi}-1$ and is nonnegative whenever $\tilde{\phi}\ge1$. Since $\nabla\tilde{\phi}=\nabla\phi$, the two fields yield the same Eikonal residual. For notational simplicity, in the discrete implementation described below, $\phi$ denotes the shifted field.

At each boundary face of a cell $\Omega_i$, both the polynomial reconstruction and the artificial-viscosity discretization require a location and a value to replace the centroid and cell average of the missing neighbor. These data are specified according to the boundary type.

\begin{figure}
	\centering
	\subfigure[Wall]{%
		\includegraphics[width=0.3\linewidth]{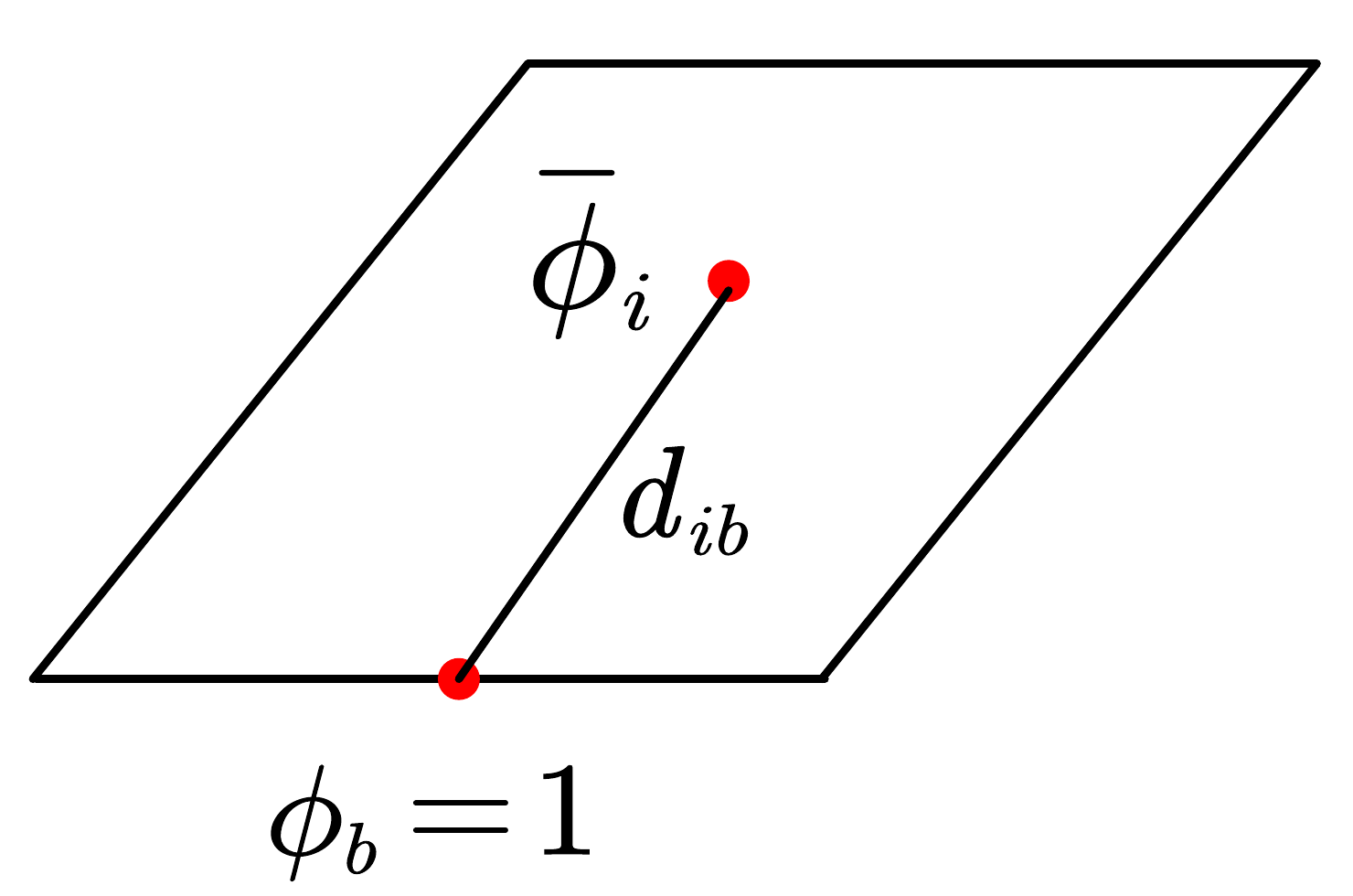}
		 \label{fig:wall-bc}
	}
	\hfill
	\subfigure[Far field]{%
		\includegraphics[width=0.3\linewidth]{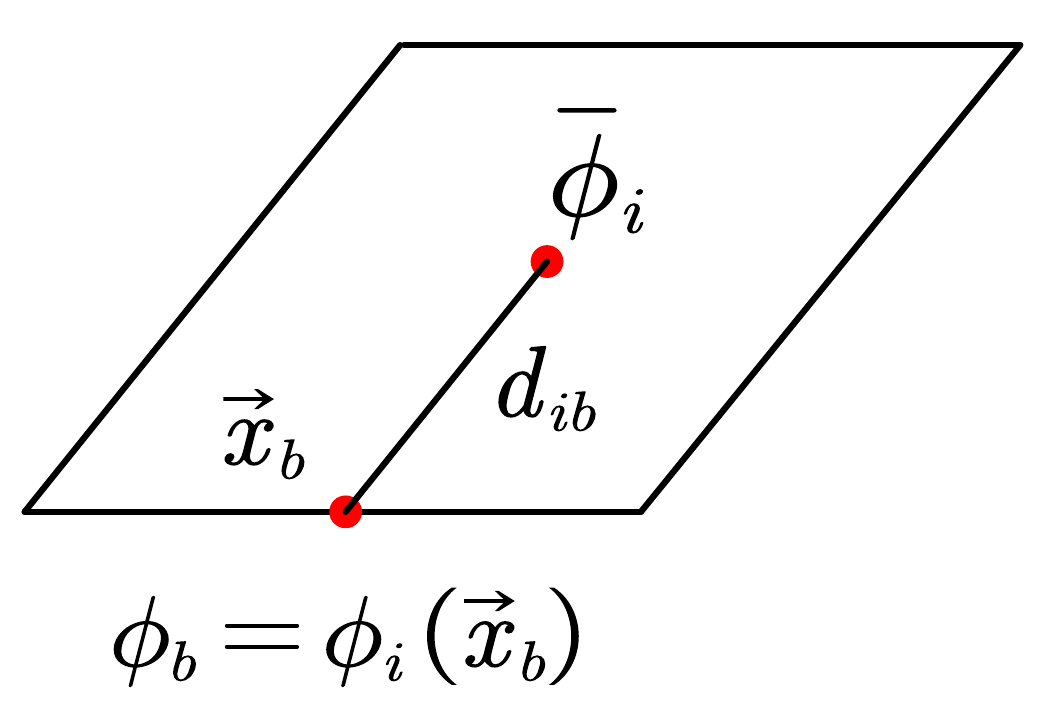}
		 \label{fig:far-bc}
	}
	\hfill
	\subfigure[Symmetry]{%
		\includegraphics[width=0.3\linewidth]{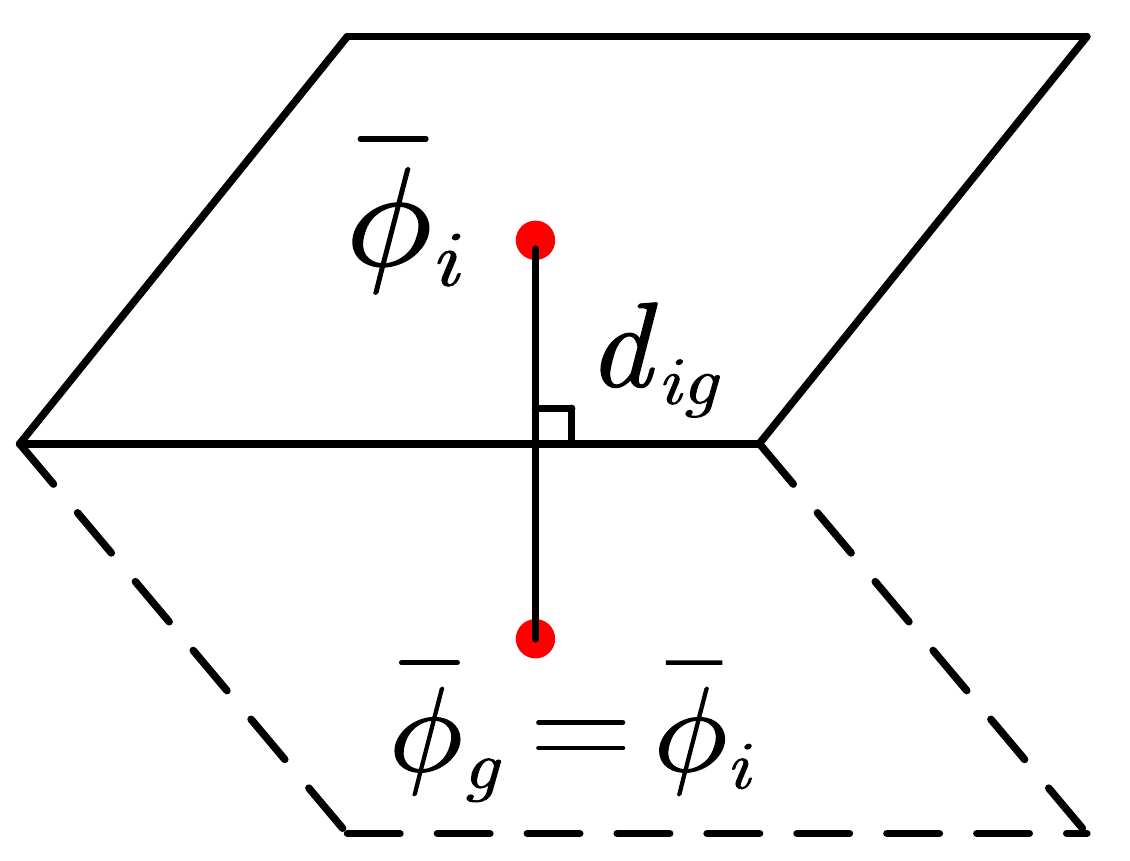}
		\label{fig:sym-bc}
	}
	\caption{Boundary condition treatment.}
	\label{fig:bounary-condition-treatment}
\end{figure}

\begin{itemize}
	\item \textit{Solid wall.} The center $\vec{x}_b$ of the boundary face serves as the boundary point, where the shifted wall value $\phi_b=1$ is prescribed (Figure~\ref{fig:wall-bc}).
	
	\item \textit{Far field.} The boundary point is again the face center $\vec{x}_b$. The value at this point is extrapolated from the reconstruction polynomial in $\Omega_i$ at the preceding pseudo-time step, giving $\phi_b^{n+1}=\phi_i^n(\vec{x}_b)$ (Figure~\ref{fig:far-bc}).
	
	\item \textit{Symmetry plane.} A ghost cell $\Omega_g$ is constructed by reflecting $\Omega_i$ across the symmetry plane. The ghost-cell centroid serves as the neighboring point, and its cell average is set to $\bar{\phi}_g=\bar{\phi}_i$ (Figure~\ref{fig:sym-bc}).
\end{itemize}

\subsection{Implementation} \label{sec:implementation}

	The SAV-FV procedure for wall-distance computation is summarized in Algorithm~\ref{alg:sav-fv-wall-distance}. 
	The algorithm can be easily integrated into existing unstructured finite-volume codes, providing a convenient tool for wall-distance computation in large-scale simulations of wall-bounded turbulent flows.
	
	\begin{algorithm}[t]
		\caption{SAV-FV algorithm for wall-distance computation}
		\label{alg:sav-fv-wall-distance}
		\small
		\begin{algorithmic}[1]
			\Require Mesh,  parameters $c$, $\beta$ and $\alpha_\phi$, global
			pseudo-time step schedule \(\{\Delta\tau^n\}\), max iteration $ite\_max$, convergence criterion $\varepsilon_{\nabla \phi}$
			\Ensure Cell-averaged physical wall distances \(\{\bar d_i\}\)
			
			\State Set \(n\gets0\), \(\tau^0\gets0\),
			\(r_i^0\gets1\), and \(\bar{\phi}_{\min}\gets1+10^{-8}\).
			
			\State Compute the initial cell averages of the physical wall-distance $\left\{\bar{\phi}_i^0\right\}$ using the Euclidean distance from the origin.
			
			\State Shift the wall distance 
			\(\bar{\phi}_i^0 \gets \bar{\phi}_i^0 + 1\) in each cell.
			
			
			
			
			
			\While{ $L_1^n > \varepsilon_{\nabla \phi}$ and $n < ite\_max$}
			
			\State Reconstruct the linear polynomials \(\{\phi_i^n\}\) using upwind weighted least squares with the weights defined in \eqref{eq:final-weight}.
			
			\State Compute the cell-averaged Eikonal residual
			\(\bar G_i^n\gets\lVert\nabla \phi_i^n\rVert-1\).
			
			\State Compute the residual norms \(L_1^n\gets \sum_i|\Omega_i|\,|\bar G_i^n| / \sum_i|\Omega_i|\) and \(L_\infty^n\gets\max_i|\bar G_i^n|\).
			
			\State Compute the bounded cell residuals $\left\{G_i^n\right\}$ using
			\eqref{eq:regularized-G}.
			\State Compute the diffusion weights $\left\{w^n_{ij}\right\}$ using
			\eqref{eq:vis-weight}.
			\State Set \(\tau^{n+1}\gets\tau^n+\Delta\tau_n\), then evaluate \(f^{n+1}\) and \(\alpha^{n+1}\)
			at \(\tau^{n+1}\).
			
			\State Assemble the system matrix \(A\) and the right-hand-side vector \(\boldsymbol b\) using \eqref{eq:linear-sav-cell}.
			
			\State Solve \(A\bar{\boldsymbol\phi}^{\,n+1,*}=\boldsymbol b\) using the AMG solver.
			
			\State Enforce the lower bound by setting
			\(\bar{\phi}_i^{n+1}\gets
			\max\!\left(\bar{\phi}_i^{n+1,*},\bar{\phi}_{\min}\right)\).
			
			\State Recover \(r_i^{n+1}\) from \eqref{eq:sav-r-update} using the corrected cell average \(\bar{\phi}_i^{n+1}\).
			
			\State Recover the cell-averaged physical wall distance by setting
			\(\bar d_i\gets\bar{\phi}_i^{n+1}-1\).
			
			
			\State Set \(n\gets n+1\).
			

			\EndWhile
			
			\State \Return $\left\{\bar d_i \right\}$
		\end{algorithmic}
	\end{algorithm}

\section{Properties of the numerical scheme} \label{section:Analysis}
This section collects the main algebraic properties of the SAV-FV scheme. The analysis is written for the fully discrete system \eqref{eq:sav-phi-update}--\eqref{eq:sav-r-update}. Let
\begin{equation}
\left(u,v\right)_h=\sum_i |\Omega_i|u_i v_i,\qquad
\|u\|_h^2=\left(u,u\right)_h,
\label{eq:discrete-inner-product}
\end{equation}
and denote the implicit viscosity operator by
\begin{equation}
D_i^n(v)=\sum_{j\in\mathcal{N}(i)}w_{ij}^n(v_j-v_i).
\label{eq:viscosity-operator}
\end{equation}

\subsection{A priori accuracy preservation}\label{section:accuracy-preserving}

Let $h_i=\max_{j\in\mathcal{N}(i)}d_{ij}$. Assume that the reconstruction is $k$-th order accurate on a shape-regular local stencil, and further assume that $\bar{\phi}_i^{n}$ also maintains $k$-th order accuracy, i.e. 
\begin{equation}
	\bar{\phi}_i^{n}=\bar{\phi}_{exact}+\mathcal{O}(h_i^{k}),
\end{equation}
where $\mathcal{O}(h_i^{k})$ means there exists a constant $C>0$ which is independent of $h_i$, such that $|\bar{\phi}_i^{n}-\bar{\phi}_{exact}| \le Ch_i^{k}$.
Then the reconstructed gradient is $(k-1)$th-order accurate, and the cell residual satisfies
\begin{equation}
\bar{G}_i^n=\bar{G}_i(\phi_{\text{exact}})+\mathcal{O}(h_i^{k-1})=\mathcal{O}(h_i^{k-1}).
\label{eq:Gbar-order}
\end{equation}
The regularized residual in \eqref{eq:regularized-G} obeys
\begin{equation}
|G_i^n|
=\frac{|\bar{G}_i^n|}{\beta \tilde d_i^{-1}+|\bar{G}_i^n|}
\le \frac{h_i}{\beta}|\bar{G}_i^n|,
\label{eq:G-order-bound}
\end{equation}
and hence
\begin{equation}
G_i^n=\mathcal{O}(h_i^k)
\quad \text{whenever}\quad
\bar{G}_i^n=\mathcal{O}(h_i^{k-1}).
\label{eq:G-order}
\end{equation}
Therefore the factor $\beta \tilde d_{i}^{-1}$ in \eqref{eq:regularized-G} helps to promote the accuracy order of the residual used by the SAV update.

The artificial viscosity has the same a priori scaling. From \eqref{eq:vis-weight},
\begin{equation}
w_{ij}^n=\frac{(|\hat{G}_i^n|+|\hat{G}_j^n|)|S_{ij}|}{2|\Omega_i|},
\qquad
\hat{G}_i^n=\frac{G_i^n}{1+|G_i^n|},
\label{eq:w-order}
\end{equation}
so that $|\hat{G}_i^n|\le |G_i^n|$ and
\begin{equation}
w_{ij}^n=\mathcal{O}\left((h_i^k+h_j^k)\frac{|S_{ij}|}{|\Omega_i|}\right).
\label{eq:w-scaling}
\end{equation}
For a smooth solution, $\bar{\phi}_j^{n+1}-\bar{\phi}_i^{n+1}=\mathcal{O}(d_{ij})$, while $|S_{ij}|/|\Omega_i|=\mathcal{O}(h_i^{-1})$ on a shape-regular stencil. Hence
\begin{equation}
D_i^n(\bar{\phi}^{n+1})
=\sum_{j\in\mathcal{N}(i)}w_{ij}^n
(\bar{\phi}_j^{n+1}-\bar{\phi}_i^{n+1})
=\mathcal{O}(h_i^k).
\label{eq:viscosity-order}
\end{equation}
Thus the added viscosity is of the same formal order as $G_i^n$ and does not reduce the order of the finite-volume discretization. 
Then we have 
\begin{equation}
	\bar{\phi}_i^{n+1}= \bar{\phi}_i^{n}+ \Delta \tau \left[D_i^n(\bar{\phi}^{n+1})
	-\frac{r_i^{n+1}}{f^{n+1}} G_i^n\right]
	=\bar{\phi}_{exact}+ \mathcal{O}(h_i^k).
	\label{eq:accuracy-order}
\end{equation}
Since $\bar{\phi}_i^{n}$ is $k$-th order accurate, the updated value $\bar{\phi}_i^{n+1}$ retains the same $\mathcal{O}(h_i^k)$ precision, provided that $\Delta \tau \sim \mathcal{O}(h_i^q)$ for $q \ge 0$. 
Therefore, the overall accuracy of the proposed scheme is preserved.

As the Eikonal residual tends to zero in every cell, the rescaled residuals and the associated viscosity weights also vanish:
\begin{equation}
	G_i^n\rightarrow 0,
	\qquad
	\hat{G}_i^n\rightarrow 0,
	\qquad
	w_{ij}^n\rightarrow 0.
	\label{eq:vanishing-viscosity-analysis}
\end{equation}
Consequently, the artificial-viscosity term vanishes in the limit, and the converged solution satisfies the discrete form of the original steady-state Eikonal equation.


\subsection{Energy stability under homogeneous boundary conditions} \label{section:energy-stability}

Consider the homogeneous Dirichlet boundary condition $\phi|_{\Gamma}=0$. For an interior face $S_{ij}$, define
\begin{equation}
m_{ij}^n=|\Omega_i|w_{ij}^n
=\frac{(|\hat{G}_i^n|+|\hat{G}_j^n|)|S_{ij}|}{2}
=|\Omega_j|w_{ji}^n=m_{ji}^n.
\label{eq:face-symmetric-weight}
\end{equation}
For a boundary face $S_{ib}$ with zero prescribed value, let
\begin{equation}
m_{ib}^n=|\Omega_i|w_{ib}^n\ge 0.
\label{eq:boundary-weight}
\end{equation}
Then the viscosity operator satisfies the discrete summation-by-parts (SBP) identity
\begin{equation}
\begin{split}
\left(D^n v,v\right)_h
&=\sum_i |\Omega_i|v_i\sum_{j\in\mathcal{N}(i)}w_{ij}^n(v_j-v_i) \\
&=-\sum_{S_{ij}\in\mathcal{F}_h^{\rm int}}
m_{ij}^n(v_i-v_j)^2
-\sum_{S_{ib}\in\mathcal{F}_h^{\rm D}}m_{ib}^n v_i^2
\le 0 .
\end{split}
\label{eq:diffusion-negative}
\end{equation}
Here $\mathcal{F}_h^{\rm int}$ and $\mathcal{F}_h^{\rm D}$ denote the sets of interior faces and boundary faces, respectively.

Multiplying \eqref{eq:sav-phi-update} by $|\Omega_i|\bar{\phi}_i^{n+1}$, and \eqref{eq:sav-r-update} by $|\Omega_i|r_i^{n+1}$, summing over all cells, and moving all terms to the left gives
\begin{equation}
\begin{split}
&\left(\frac{\bar{\phi}^{n+1}-\bar{\phi}^{n}}{\Delta\tau},
\bar{\phi}^{n+1}\right)_h
+\left(\frac{r^{n+1}-r^{n}}{\Delta\tau},r^{n+1}\right)_h
+\alpha^{n+1}\|r^{n+1}\|_h^2 \\
&\quad
-\left(D^n\bar{\phi}^{n+1},\bar{\phi}^{n+1}\right)_h
+\sum_i |\Omega_i|\frac{r_i^{n+1}G_i^n\bar{\phi}_i^{n+1}}{f^{n+1}}
-\sum_i |\Omega_i|\frac{\bar{\phi}_i^{n+1}G_i^n r_i^{n+1}}{f^{n+1}}
=0 .
\end{split}
\label{eq:energy-add}
\end{equation}
The two SAV coupling terms cancel exactly. Using
\begin{equation}
2a(a-b)=a^2-b^2+(a-b)^2,
\label{eq:energy-identity}
\end{equation}
with the discrete inner product, define the modified discrete energy
\begin{equation}
E^n=\|\bar{\phi}^{n}\|_h^2+\|r^{n}\|_h^2.
\label{eq:modified-energy}
\end{equation}
Equations \eqref{eq:energy-add}--\eqref{eq:energy-identity} yield the exact balance
\begin{equation}
\begin{split}
&E^{n+1}-E^n
+\|\bar{\phi}^{n+1}-\bar{\phi}^{n}\|_h^2
+\|r^{n+1}-r^{n}\|_h^2
+2\Delta\tau\alpha^{n+1}\|r^{n+1}\|_h^2 \\
&\quad
+2\Delta\tau\sum_{S_{ij}\in\mathcal{F}_h^{\rm int}}
m_{ij}^n(\bar{\phi}_i^{n+1}-\bar{\phi}_j^{n+1})^2
+2\Delta\tau\sum_{S_{ib}\in\mathcal{F}_h^{\rm D}}
m_{ib}^n(\bar{\phi}_i^{n+1})^2=0 .
\end{split}
\label{eq:energy-balance}
\end{equation}
Since $\alpha^{n+1}=c/(\tau^{n+1}+c)^2>0$ and $m_{ij}^n,m_{ib}^n\ge 0$, all terms after $E^{n+1}-E^n$ are non-negative. 
Therefore
\begin{equation}
E^{n+1}\le E^n,
\qquad \forall\,\Delta\tau>0.
\label{eq:unconditional-energy-stability}
\end{equation}
The SAV-FV scheme is thus unconditionally energy stable under homogeneous Dirichlet boundary conditions. 

To exclude the spurious all-zero state during the SAV update, we apply a constant shift to the distance field and impose the wall boundary condition $\phi|_{\Gamma}=\phi_b$, with $\phi_b=1$ in this work. This nonhomogeneous boundary condition is compatible with unconditional energy stability because the viscosity operator satisfies
\begin{equation}
		\left(D^n\bar{\phi}^{n+1},\bar{\phi}^{n+1}\right)_h
		= -\sum_{S_{ij}\in\mathcal{F}_h^{\mathrm{int}}}
		m_{ij}^n
		\left(\bar{\phi}_i^{n+1}-\bar{\phi}_j^{n+1}\right)^2
		-\sum_{S_{ib}\in\mathcal{F}_h^{\mathrm{D}}}
		m_{ib}^n\bar{\phi}_i^{n+1}
		\left(\bar{\phi}_i^{n+1}-\phi_b\right)
		\le0.
\end{equation}
The inequality follows from the nonnegativity of the viscosity weights and the bound $\bar{\phi}_i^{n+1}\ge\phi_b>0$, which is preserved during the SAV update, as shown in Section \ref{section:m-matrix-bound-preservation}.


For non-Dirichlet boundary conditions, the discrete energy estimate generally takes the form $E^{n+1}-E^n=E_b-\tilde{E}_{\mathrm{int}}$, where $E_b$ denotes the boundary contribution and $\tilde{E}_{\mathrm{int}}$ collects the nonnegative dissipation terms,
\begin{equation}
	\tilde{E}_{int}=\|\bar{\phi}^{n+1}-\bar{\phi}^{n}\|_h^2
	+\|r^{n+1}-r^{n}\|_h^2
	+2\Delta\tau\alpha^{n+1}\|r^{n+1}\|_h^2+2\Delta\tau\sum_{S_{ij}\in\mathcal{F}_h^{\rm int}}
	m_{ij}^n(\bar{\phi}_i^{n+1}-\bar{\phi}_j^{n+1})^2.
	\label{eq:t_E_int}
\end{equation}
For symmetry faces, imposing zero normal flux yields a zero contribution to $E_b$. For other boundary treatments, such as far-field extrapolation, establishing $E_b-\tilde{E}_{\mathrm{int}}\le0$ requires an additional estimate bounding the boundary contribution in terms of the interior dissipation. Deriving such an estimate for general configurations remains challenging. The numerical experiments nevertheless demonstrate stable behavior of the SAV-FV scheme across the test cases considered, including those involving complex geometries, high aspect ratio grids, and relatively large pseudo-time steps.


\subsection{M-matrix structure and bound preservation} \label{section:m-matrix-bound-preservation}

We analyze the structure of the matrix $A$ in \eqref{eq:linear-system}. From \eqref{eq:matrix-diagonal}--\eqref{eq:matrix-offdiagonal}, we have
\begin{equation}
A_{ii}>0,
\qquad
A_{ij}=-w_{ij}^n\le 0 \quad (j\ne i),
\label{eq:z-matrix}
\end{equation}
and
\begin{equation}
A_{ii}-\sum_{j\ne i}|A_{ij}|
=\frac{1}{\Delta\tau}
+\frac{\Delta \tau (G_i^n)^2}
{(1+\alpha^{n+1}\Delta \tau)(f^{n+1})^2}
+\sum_{b\in\mathcal{B}(i)}w_{ib}^n
>0 .
\label{eq:diagonal-dominance}
\end{equation}
Here, $\mathcal{B}(i)$ denotes the set of physical boundary faces contributing to the diagonal entry for cell $i$, and symmetry faces are excluded from $\mathcal{B}(i)$ since their diffusive flux is zero.
Hence $A$ is a strictly diagonally dominant $Z$-matrix and therefore a nonsingular M-matrix \cite{giorgi2022nonsingular}. In particular,
\begin{equation}
A^{-1}\ge 0 .
\label{eq:inverse-positive}
\end{equation}

For the shifted unknown introduced in Section \ref{section:boundary-condition}, the wall value is $1$. Let $\boldsymbol{1}=(1,\ldots,1)^T$ and introduce the scaled system
\begin{equation}
\widetilde{A}\bar{\boldsymbol{\phi}}^{n+1}=\widetilde{\boldsymbol{b}},
\qquad
\widetilde{A}=\Delta\tau A,
\qquad
\widetilde{\boldsymbol{b}}=\Delta\tau\boldsymbol{b}.
\label{eq:scaled-system}
\end{equation}
Because $\widetilde{A}^{-1}\ge0$, the comparison condition
\begin{equation}
\widetilde{\boldsymbol{b}}-\widetilde{A}\boldsymbol{1}\ge 0 .
\label{eq:comparison-condition}
\end{equation}
implies the lower bound $\bar{\boldsymbol{\phi}}^{n+1}\ge\boldsymbol{1}$, and ensures the positivity of the physical distance field, given by $\bar{\boldsymbol{\phi}}^{n+1}-\boldsymbol{1}$. 
The row-wise form of the comparison condition
\eqref{eq:comparison-condition} is
\begin{equation}
	d_i^n
	-\frac{\Delta\tau r_i^nG_i^n}{(1+\alpha^{n+1}\Delta\tau)f^{n+1}}
	-\frac{\Delta\tau^2(G_i^n)^2}{(1+\alpha^{n+1}\Delta\tau)(f^{n+1})^2}+\Delta\tau\sum_{b \in \mathcal B(i)}
	w_{ib}^n\bigl(\phi_b^{n+1}-1\bigr)
	\ge 0,
	\label{eq:bound-condition-cell-bc}
	\end{equation}
where $d_i^n=\bar{\phi}_i^n-1$, and $\phi_b^{n+1}$ is the prescribed or explicitly extrapolated value of the shifted distance at boundary face $b$.  
At a solid wall, $\phi_b^{n+1}=1$, so the corresponding contribution vanishes. At far-field boundaries, where the physical wall distance is positive, the shifted boundary values are expected to satisfy $\phi_b^{n+1}\ge1$. Whenever this bound holds, the corresponding contributions are nonnegative since $w_{ib}^n\ge0$.
Therefore, a more restrictive sufficient condition for bound preservation is
\begin{equation}
	d_i^n
	-\frac{\Delta\tau r_i^nG_i^n}
	{(1+\alpha^{n+1}\Delta\tau)f^{n+1}}
	-\frac{\Delta\tau^2(G_i^n)^2}
	{(1+\alpha^{n+1}\Delta\tau)(f^{n+1})^2}
	\ge 0.
	\label{eq:bound-condition-cell}
\end{equation}
Equivalently, after multiplying by $1+\alpha^{n+1}\Delta\tau>0$, a sufficient local time-step condition is
\begin{equation}
	\frac{(G_i^n)^2}{(f^{n+1})^2} \Delta\tau^2
	+\left[\frac{r_i^nG_i^n}{f^{n+1}}-\alpha^{n+1}d_i^n\right]\Delta\tau
	-d_i^n\le 0 .
	\label{eq:quadratic-bound-condition}
\end{equation}
Thus, if $d_i^n > 0$ and there comes
\begin{equation}
0<\Delta\tau\le \Delta\tau_i^{\rm bp},
\label{eq:bp-step}
\end{equation}
with
\begin{equation}
	\Delta\tau_i^{\rm bp}
	=\frac{-q_i^n+\left[(q_i^n)^2+4p_i^n d_i^n\right]^{1/2}}{2p_i^n},
	\qquad
	p_i^n=\frac{(G_i^n)^2}{(f^{n+1})^2},
	\qquad
	q_i^n=\frac{r_i^nG_i^n}{f^{n+1}}-\alpha^{n+1}d_i^n,
	\label{eq:bp-step-root}
\end{equation}
for all cells with $p_i^n>0$, then \eqref{eq:comparison-condition} holds, and when $p_i^n=0$, condition \eqref{eq:bp-step} reduces to arbitrary $\Delta\tau \ge 0$ according to algebraic condition \eqref{eq:quadratic-bound-condition}, since $\alpha^{n+1}$ and $d_i^n$ are positive.
Thus, when $G_i^n \rightarrow 0$, we have $\Delta\tau_i^{\rm bp} \rightarrow \infty$. Given that $|G_i^n| < 1$ and is bounded, it follows that $\Delta\tau_i^{\rm bp}$ admits a positive lower bound, the details are as follows.

The maximum allowable pseudo-time step $\Delta\tau_i^{\rm bp}$ can be written as
\begin{equation}
	\Delta\tau_i^{\rm bp}
	=\frac{-q_i^n+\left[(q_i^n)^2+4p_i^n d_i^n\right]^{1/2}}{2p_i^n}
	= \frac{2d_i^n}{q_i^n+\left[(q_i^n)^2+4p_i^n d_i^n\right]^{1/2}}.
	\label{eq:tau-bp}
\end{equation}
By the definition of $f^{n+1}$,  $\alpha^{n+1}$ and $G_i^n$, we have $f^{n+1} \ge e^{-1}, \; \alpha^{n+1}>0, \; |G_i^n|<1$, then we get
\begin{equation}
	p_i^n<e^2, \qquad q_i^n <e C_r,
	\label{eq:tau-bp-ineq}
\end{equation}
and the lower bound of $\Delta\tau_i^{\rm bp}$ is given by
\begin{equation}
	\Delta\tau_i^{\rm bp} > \frac{2d_i^n}{e C_r+\left[(e C_r)^2+4e^2 d_i^n\right]^{1/2}}
	=\frac{e^{-1}\left(\sqrt{(C_r)^2+4 d_i^n}-C_r\right)}{2},
	\label{eq:tau-bp-ineq2}
\end{equation}
where $C_r$ denotes the bound of $r_i^n$. This bound depends solely on the initial computational conditions, as the energy satisfies $E^n \le E^0$. Additionally, the lower bound of the time step remains strictly positive because $d_i^n > 0$ always holds. So the positivity is preserved as long as $\Delta \tau \le e^{-1}\left(\sqrt{(C_r)^2+4 d_i^n}-C_r\right)/2$. Given that $d_i^n$ is the mean wall distance of cell $i$ and is of the same order as the mesh scale $\Delta x$ in the near-wall region, it follows that $\Delta \tau \sim d_i^n/(eC_r) \sim \mathcal{O}(\Delta x)$, as $d_i^n \ll C_r$ in the near-wall region.
 
The pseudo-time step restriction \eqref{eq:bp-step} ensures that $\widetilde{\boldsymbol{b}}\ge\widetilde{A}\boldsymbol{1}$ componentwise. Since $\widetilde{A}$ is a nonsingular M-matrix, multiplication by its entrywise nonnegative inverse preserves this inequality and yields
\begin{equation}
	\bar{\boldsymbol{\phi}}^{n+1}
	= \widetilde{A}^{-1}\widetilde{\boldsymbol{b}}
	\ge \widetilde{A}^{-1}\widetilde{A}\boldsymbol{1}
	= \boldsymbol{1}.
	\label{eq:lower-bound-proof}
\end{equation}
By this discrete comparison principle, the SAV-FV scheme preserves the lower bound $\bar{\phi}_i\ge 1$ for the shifted distance field in every cell under the stated pseudo-time step restriction.

\begin{figure} 
	\centering
	\includegraphics[width=0.6\linewidth]{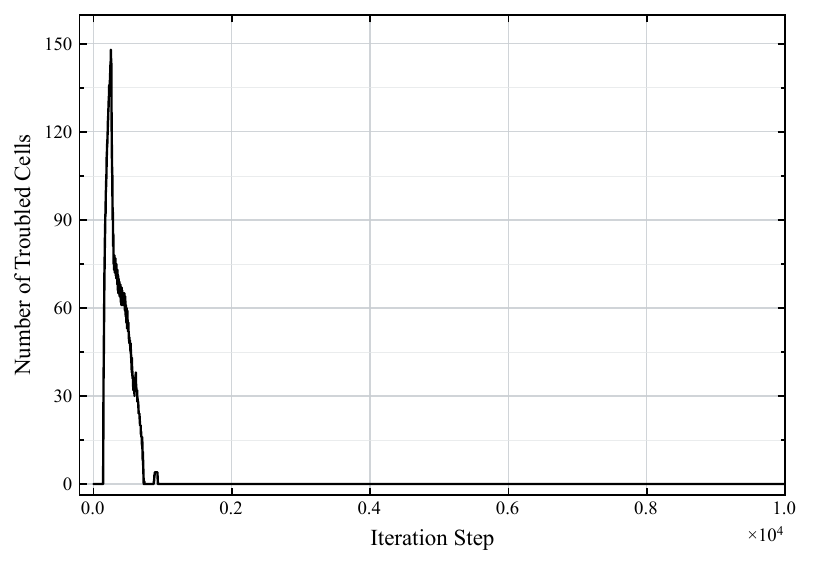}
	\caption{Number of cells requiring the lower-bound correction at each pseudo-time step in the flat-plate test case.}
	\label{fig:pp-cell-BL}
\end{figure}

\textbf{Remark:} 
In this subsection, we show that the SAV-FV system matrix is a nonsingular M-matrix. Together with the bound on the right-hand side, its discrete comparison principle ensures that the update preserves the lower bound $\bar{\phi}_i\ge1$ for the shifted cell averages under the pseudo-time step restriction \eqref{eq:bp-step}. This sufficient condition requires impractically small pseudo-time steps.
For computational efficiency, the numerical experiments use a larger global pseudo-time step $\Delta\tau$ and enforce a lower bound on the shifted cell averages after each SAV update.
Let $\bar{\phi}_i^{n+1,*}$ denote the shifted cell average in cell $i$ before truncation. The corrected value is
\[
\bar{\phi}_i^{n+1}
= \max\!\left(\bar{\phi}_i^{n+1,*},\,\bar{\phi}_{\min}\right),
\qquad \bar{\phi}_{\min}=1+10^{-8}.
\]
This correction ensures that the physical wall-distance average is at least $10^{-8}$ in every cell,
and the scalar auxiliary variable $r^{n+1}_i$ is then updated from the corrected value $\bar{\phi}_i^{n+1}$ via \eqref{eq:sav-r-update}.
At the large global pseudo-time steps used in the numerical experiments reported in Section~\ref{section:numerical_results}, only the flat-plate case in Section~\ref{section:flat-plate} required truncation. Figure \ref{fig:pp-cell-BL} shows the number of cells requiring correction at each pseudo-time step. In this case, truncation was active during the early iterations and became inactive as the solution approached convergence.

\subsection{Analysis of parameter effects on convergence} \label{section:convergence-factors}

\begin{figure} 
	\centering
	\includegraphics[width=0.6\linewidth]{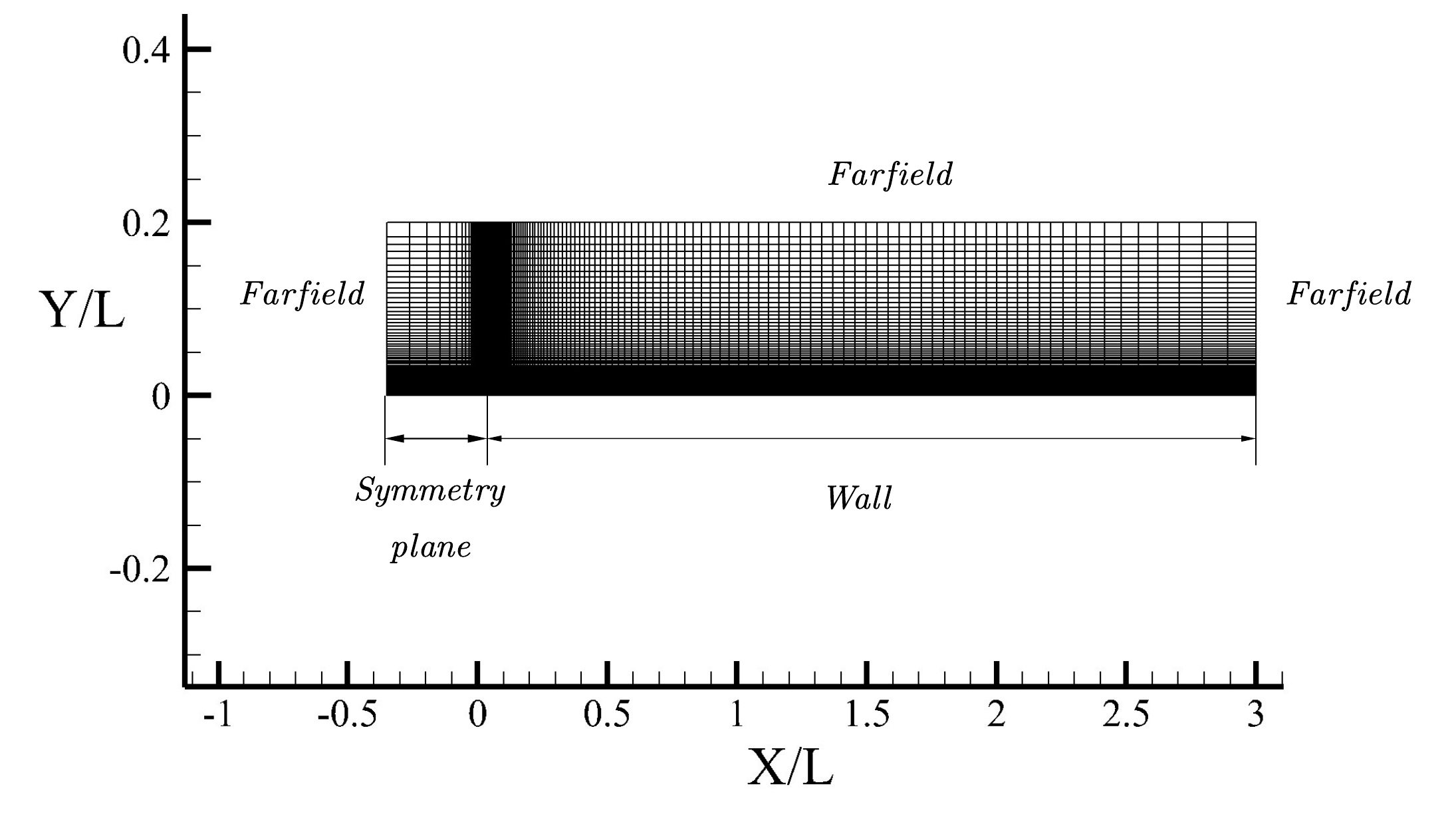}
	\caption{Computational mesh and boundary conditions for the extended flat-plate convergence study.}
	\label{fig:convergence-BL-geometry}
\end{figure}

This subsection examines how the upwind weighting parameter $\alpha_{\phi}$, the initial wall-distance field, geometric singularities, and cell aspect ratios affect the convergence of the SAV-FV scheme. We use the extended flat-plate test case described in Section~\ref{section:flat-plate}, with a reference length of $L=1$ and a global pseudo-time step of $\Delta\tau=2$. Figure~\ref{fig:convergence-BL-geometry} shows the reference mesh and boundary conditions. The mesh contains 123,328 cells and has a first-layer height of $10^{-6}L$ at the wall. Unless otherwise specified, the residual regularization parameter is set to $\beta=10$.


\subsubsection{Upwind weighting parameter}

\begin{figure} 
	\centering
	\subfigure[$L_1(\nabla \phi)$ \label{fig:convergence-alpha-L1}]{%
		\includegraphics[width=0.48\linewidth]{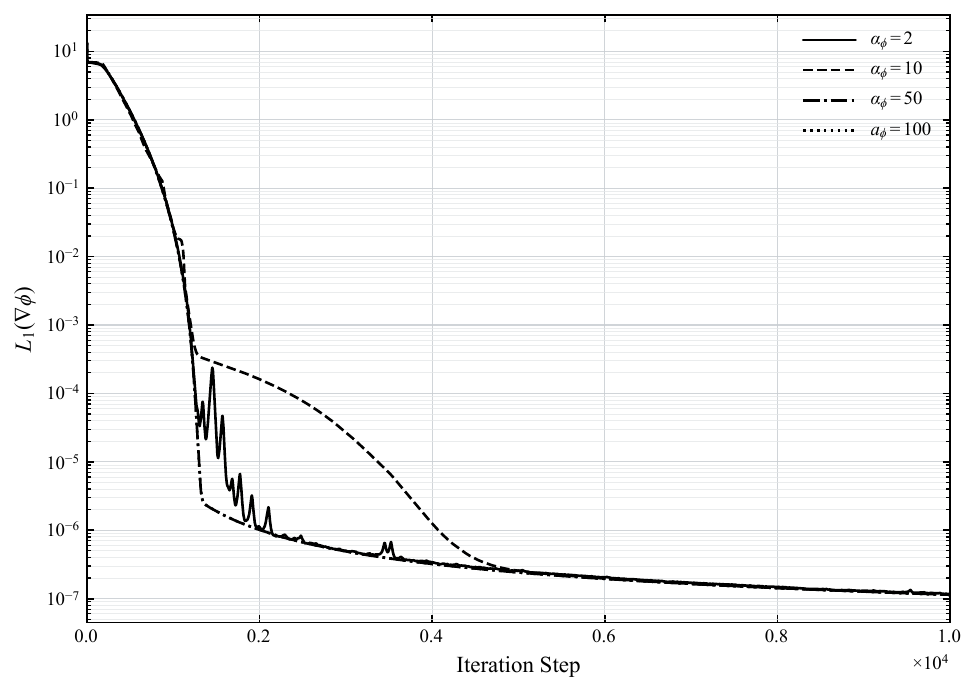}%
	}
	\hfill
	\subfigure[$L_{\infty}(\nabla \phi)$ \label{fig:convergence-alpha-Linf}]{%
		\includegraphics[width=0.48\linewidth]{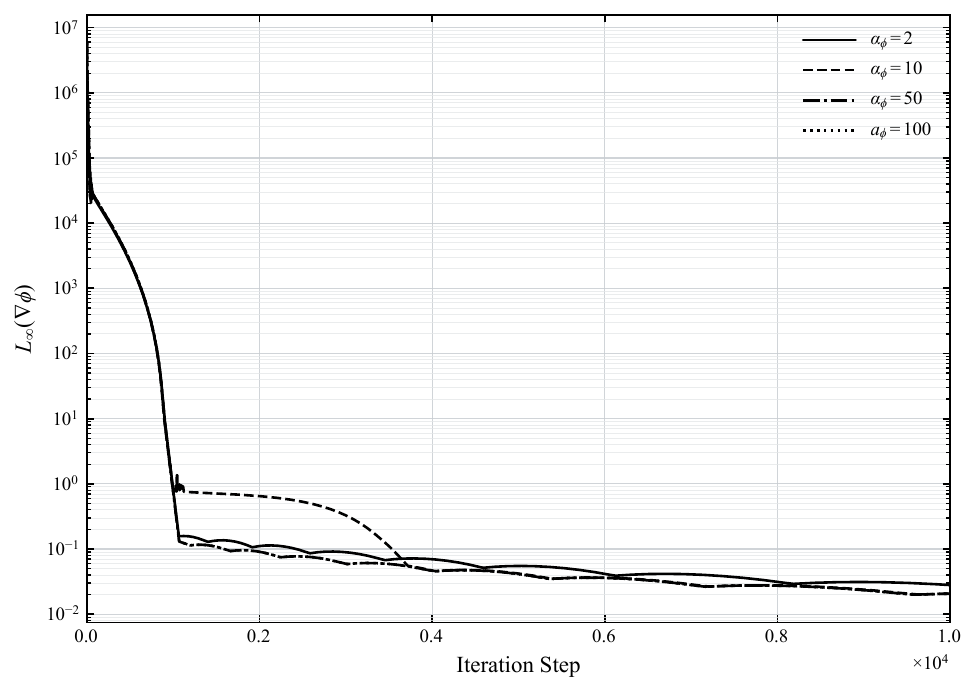}%
	}
	\caption{Convergence histories for different upwind weighting parameters $\alpha_{\phi}$.}
	\label{fig:convergence-alpha}
\end{figure}

The influence of $\alpha_{\phi}$ is first examined using $\alpha_{\phi}=2$, 10, 50, and 100. The initial field is the Euclidean distance from the origin with the unit shift. Figure \ref{fig:convergence-alpha} compares the corresponding histories of the $L_1(\nabla \phi)$ and $L_{\infty}(\nabla \phi)$ residuals, which are defined in Section \ref{section:numerical_results}.
As shown in Figure \ref{fig:convergence-alpha}, the convergence rate depends non-monotonically on $\alpha_{\phi}$, and the two curves for $\alpha_{\phi}=50$ and $\alpha_{\phi}=100$ coincide, indicating that once $\alpha_{\phi}$ becomes sufficiently large, the computational results would not change anymore. The smaller and larger values produce similar convergence curves, whereas an intermediate value, particularly $\alpha_{\phi}=10$ in this comparison, delays the residual reduction during the early iterations. These differences become small in the later stage, and all choices approach the same converged result. Thus, the influence of $\alpha_{\phi}$ on convergence is mainly confined to the initial transient.

\begin{figure} 
	\centering
	\subfigure[$\alpha_{\phi}=1$
	\label{fig:convergence-alpha-sdy-1}]{%
		\includegraphics[
		height=0.39\textwidth,
		trim=0bp 57.45bp 200bp 215.05bp,
		clip
		]{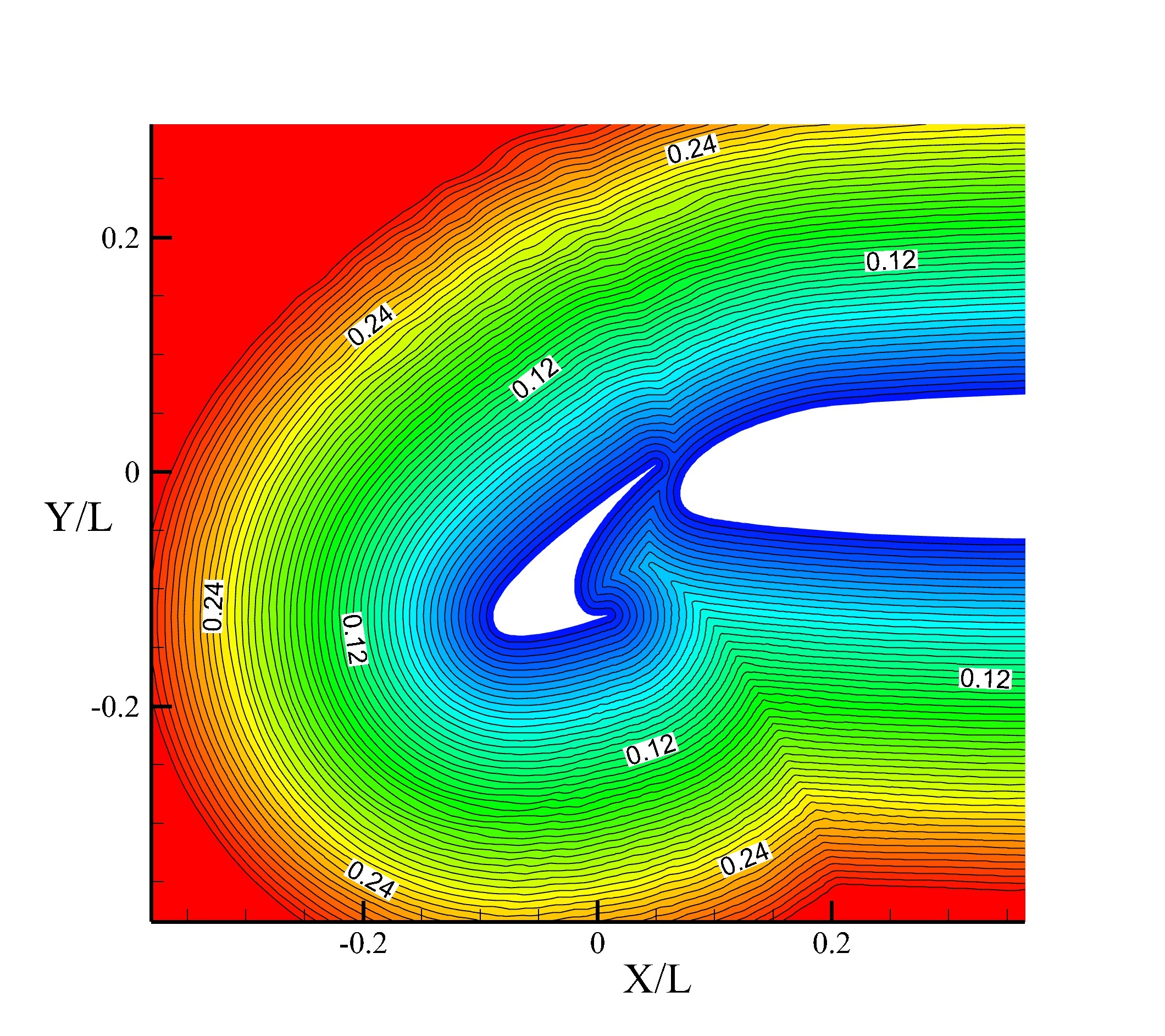}%
	}
	\hfill
	\subfigure[$\alpha_{\phi}=10$
	\label{fig:convergence-alpha-sdy-10}]{%
		\includegraphics[
		height=0.39\textwidth,
		trim=0bp 25.34bp 0bp 11.32bp,
		clip
		]{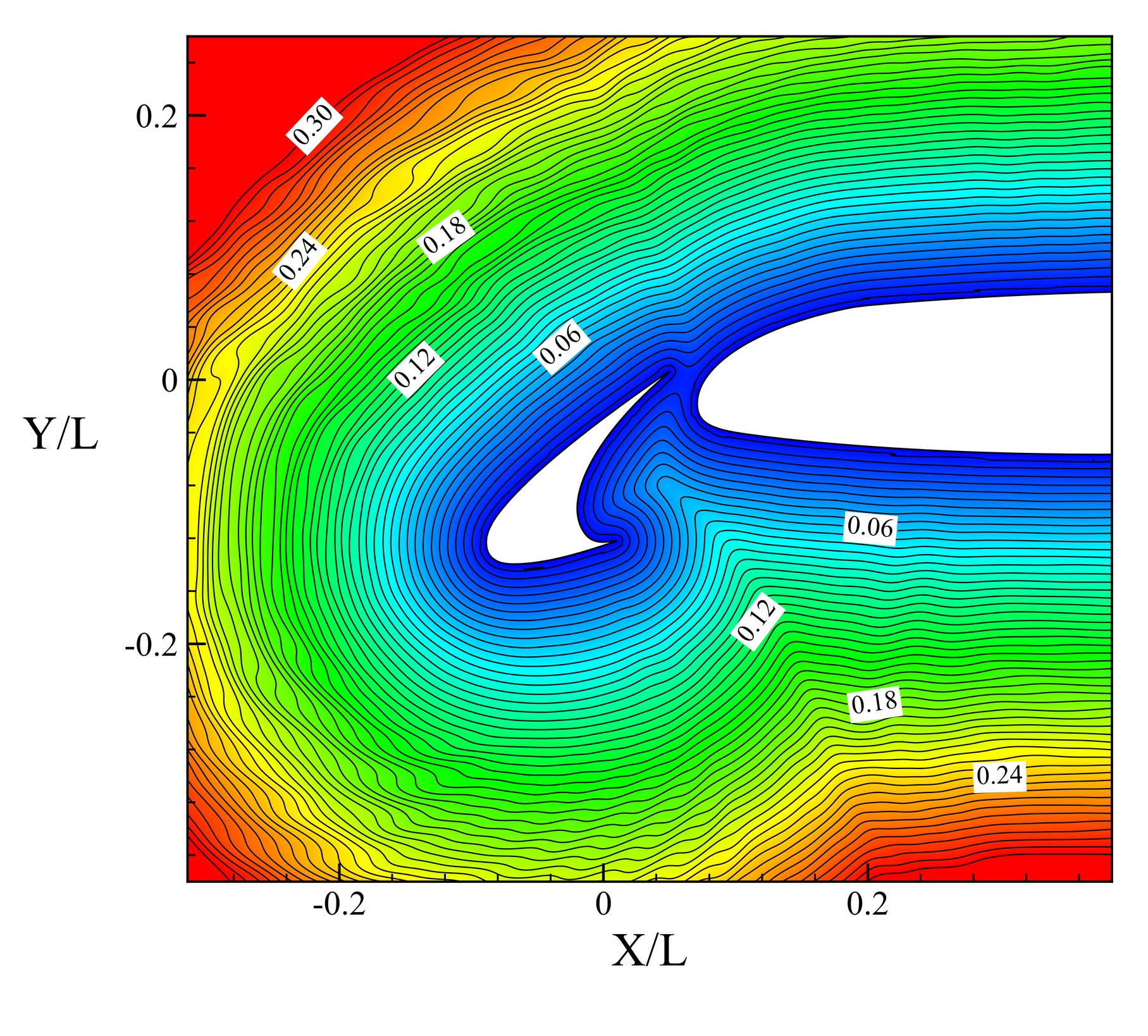}%
	}
	\caption{Wall-distance contours near the slat of the
		three-element airfoil for different values of $\alpha_{\phi}$.}
	\label{fig:convergence-alpha-sdy}
\end{figure}

However, increasing $\alpha_{\phi}$ excessively can amplify numerical disturbances generated at geometric singularities and produce pronounced oscillations. This behavior is particularly evident for complex configurations such as the three-element airfoil considered in Section \ref{section:30P-30N}.
Figure \ref{fig:convergence-alpha-sdy} compares the wall-distance contours near the slat for $\alpha_{\phi}=1$ and 10. The contours obtained with $\alpha_{\phi}=1$ are visibly smoother, whereas those for $\alpha_{\phi}=10$ exhibit more pronounced local oscillations. This comparison supports the observation that a larger $\alpha_{\phi}$ can amplify singularity-induced numerical oscillations in complex geometries.

Conversely, smaller values reduce the propagation of singularity-induced oscillations but provide only a limited upwind bias. In this situation, the required upwind information relies mainly on the increase in grid spacing from the wall toward the far field. A small $\alpha_{\phi}$ can therefore be effective when the near-wall mesh is sufficiently fine and has a suitable growth rate away from the wall. For meshes with insufficient near-wall resolution or weak grid-spacing growth, particularly uniform meshes, a sufficiently large $\alpha_{\phi}$ is required to supply the necessary upwind bias.

Based on these observations, $1\le\alpha_{\phi}\le10$ is recommended as a practical safe range, subject to the mesh requirements discussed above. The default choice $\alpha_{\phi}=10$ provides sufficient upwinding without excessively amplifying singularity-induced oscillations. Determining an optimal value of $\alpha_{\phi}$ for different geometries and meshes remains an open question and will be addressed in future work.

\subsubsection{Initial distance field}

\begin{figure} 
	\centering
	\subfigure[$L_1(\nabla \phi)$ \label{fig:convergence-initial-L1}]{%
		\includegraphics[width=0.48\linewidth]{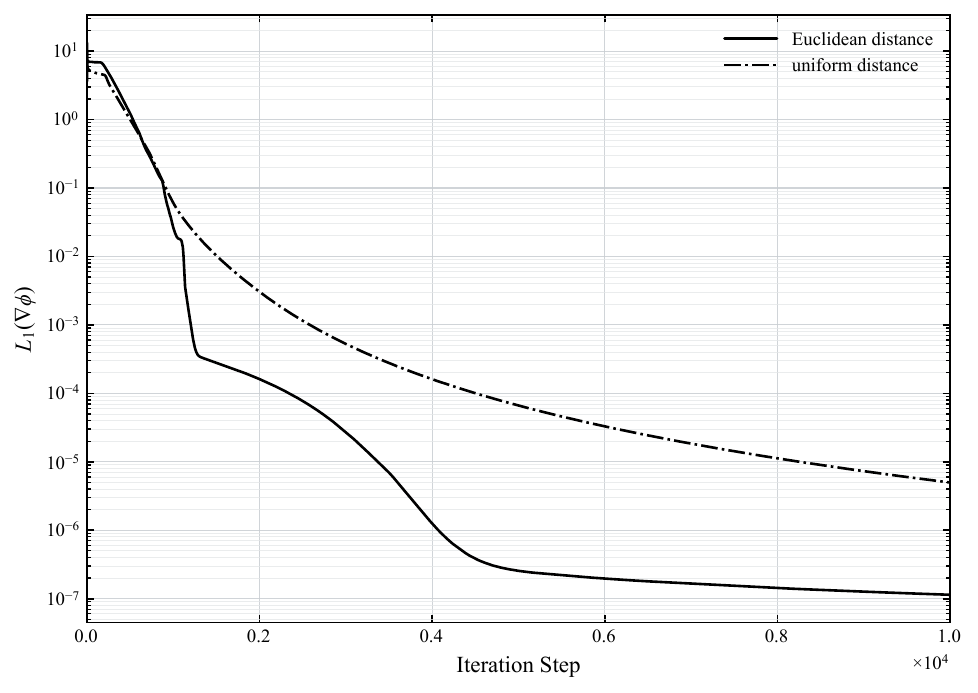}%
	}
	\hfill
	\subfigure[$L_{\infty}(\nabla \phi)$ \label{fig:convergence-initial-Linf}]{%
		\includegraphics[width=0.48\linewidth]{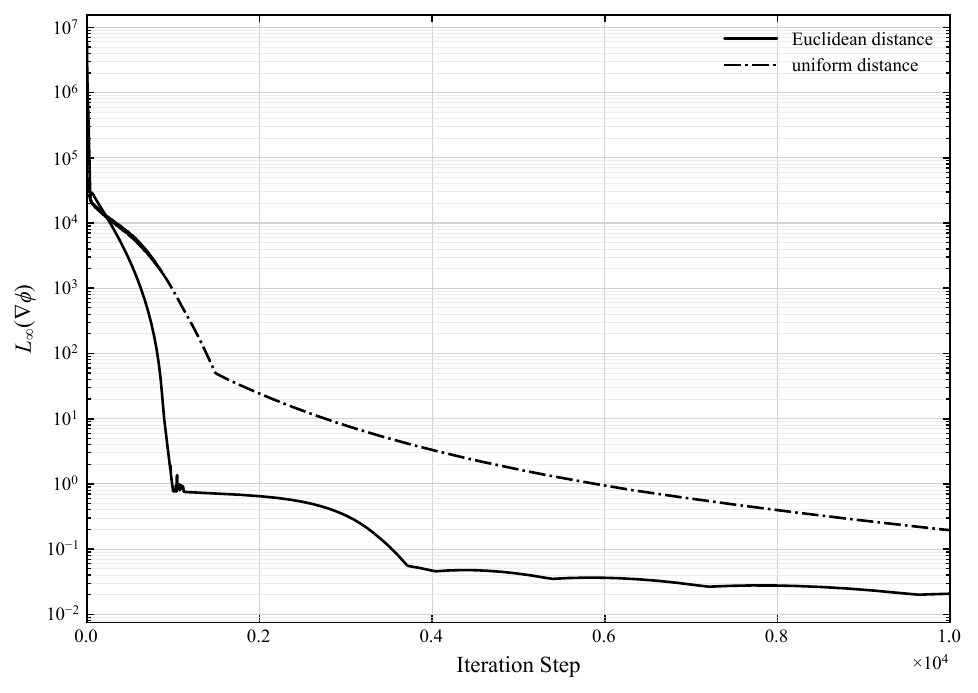}%
	}
	\caption{Convergence histories for uniform and Euclidean initial distance fields with $\alpha_{\phi}=10$.}
	\label{fig:convergence-initial}
\end{figure}

The influence of the initial field is examined on the same mesh and with the same boundary conditions, using $\alpha_{\phi}=10$. In Figure \ref{fig:convergence-initial}, the curves labeled ``uniform distance'' use the constant initial value $\tilde{\phi}^0=2$, whereas those labeled ``Euclidean distance'' use $\tilde{\phi}^0=\sqrt{x^2+y^2}+1$. These values include the unit offset introduced in Section \ref{section:boundary-condition}, the corresponding physical initial distances are therefore 1 and $\sqrt{x^2+y^2}$, respectively.

Figure \ref{fig:convergence-initial} shows that the Euclidean initialization yields a faster reduction in both residuals. After the initial decay, its residual curves remain below those obtained with the uniform initial field and reach low residual levels in fewer iterations. This behavior is reasonable because the Euclidean initial field generally provides a closer approximation to the final wall-distance distribution than a spatially uniform value. Consequently, less adjustment of the distance field is required during the pseudo-time iteration.

\subsubsection{Singularities and aspect-ratio of grids}
\label{sec:aspect-ratio}

\begin{figure}
	\centering
	\includegraphics[width=0.6\linewidth]{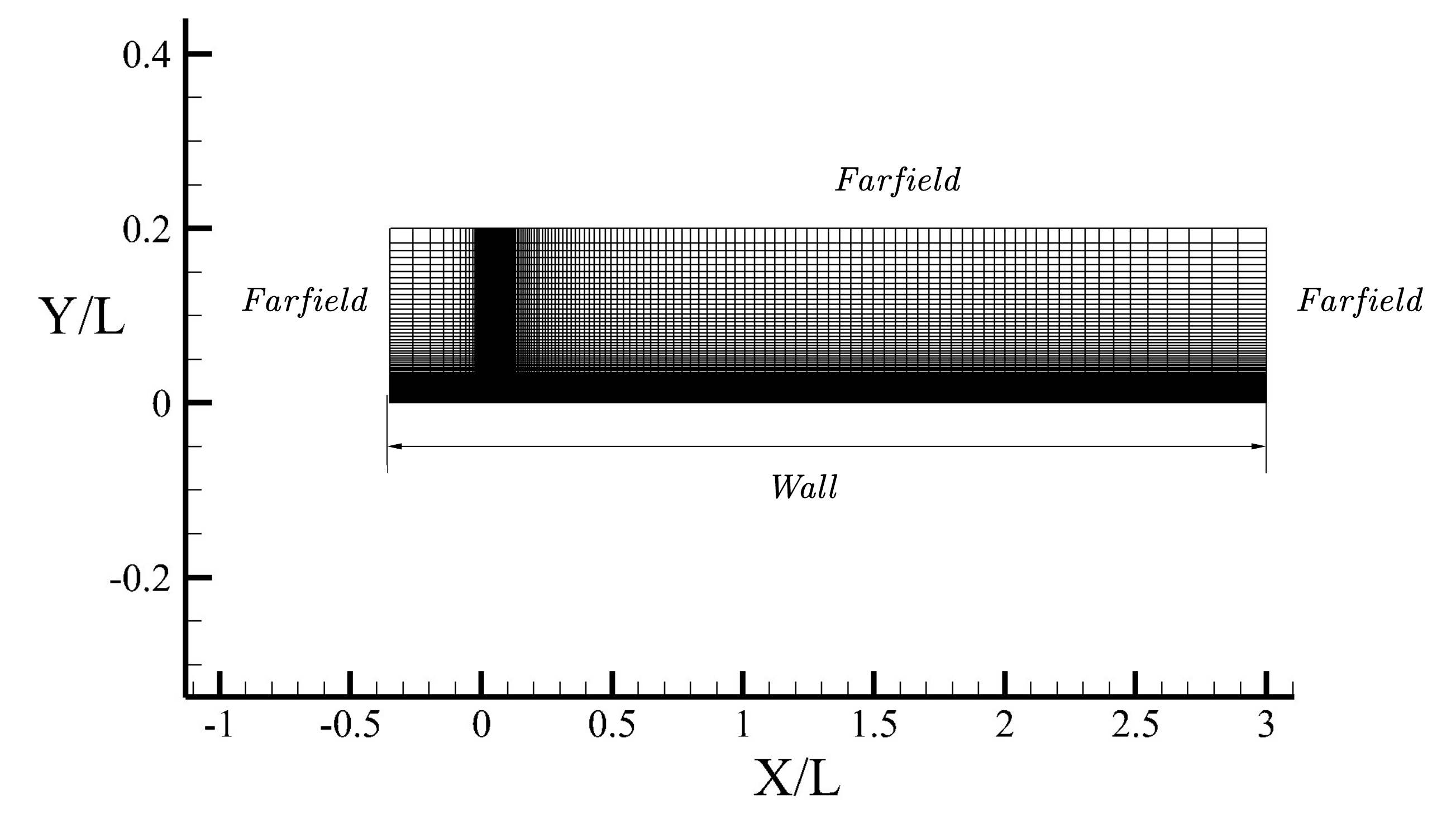}
	\caption{Computational mesh and boundary conditions for the ``wall\_all'' configuration.}
	\label{fig:convergence-BL-wall-all}
\end{figure}

\begin{figure} 
	\centering
	\subfigure[$L_1(\nabla \phi)$ \label{fig:convergence-mesh-L1}]{%
		\includegraphics[width=0.48\linewidth]{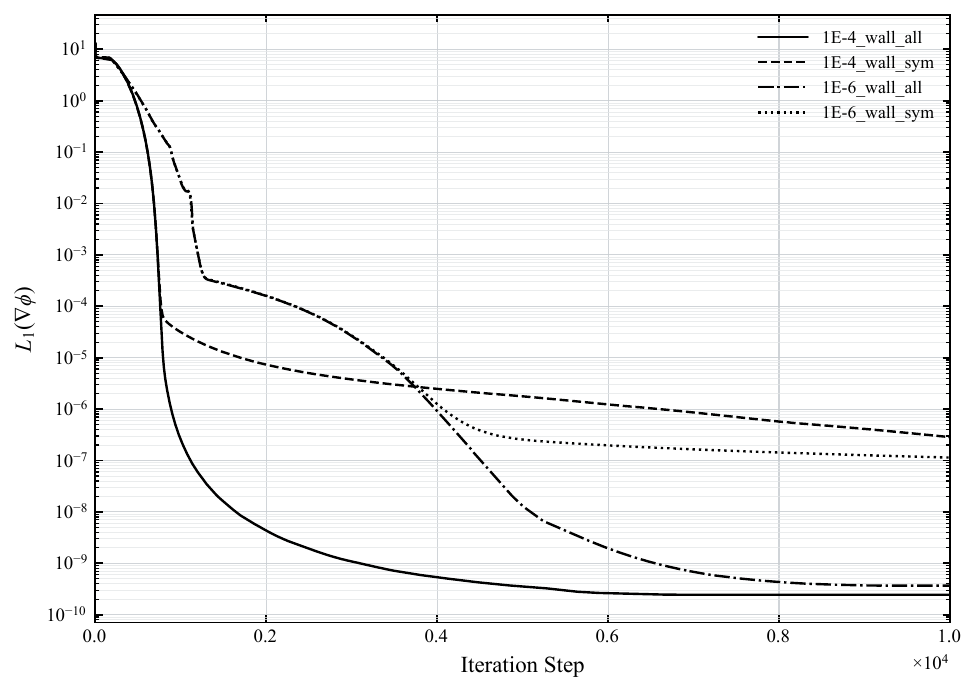}%
	}
	\hfill
	\subfigure[$L_{\infty}(\nabla \phi)$ \label{fig:convergence-mesh-Linf}]{%
		\includegraphics[width=0.48\linewidth]{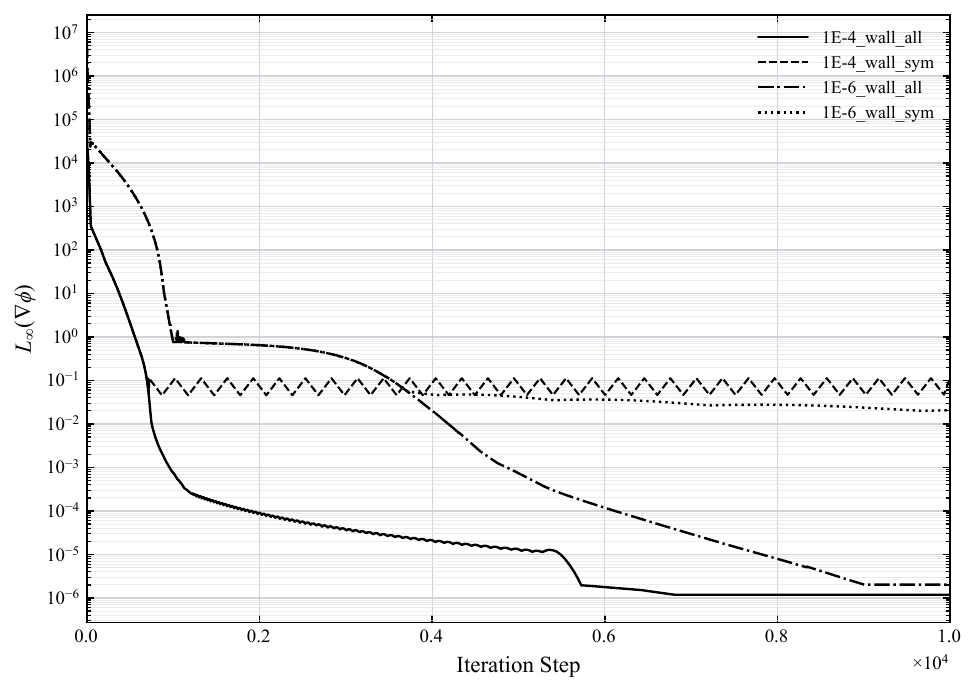}%
	}
	\caption{Convergence histories for different boundary conditions and first-layer grid heights.}
	\label{fig:convergence-mesh}
\end{figure}

The final comparison examines the leading-edge singularity and grid stretching, with $\alpha_{\phi}=10$ and the Euclidean initial field. In Figure \ref{fig:convergence-mesh}, labels containing ``wall\_sym'' use the boundary conditions in Figure \ref{fig:convergence-BL-geometry}, whereas ``wall\_all'' replaces the symmetry-plane segment by a wall, as shown in Figure \ref{fig:convergence-BL-wall-all}. 
This replacement removes the wall–symmetry junction at the original leading edge and its associated wall-distance singularity.
The far-field boundaries are retained, and both configurations use the same mesh. 

The labels ``1E-6'' and ``1E-4'' denote first-layer heights of $10^{-6}L$ and $10^{-4}L$, respectively. Both meshes contain the same number of cells and differ in their first-layer spacing and the resulting stretching, with the $10^{-6}L$ mesh having substantially larger near-wall cell aspect ratios. This comparison therefore distinguishes the influence of the boundary-induced singularity from that of the grid aspect ratio without changing the cell count.

As shown in Figure \ref{fig:convergence-mesh}, at a fixed first-layer height, the initial rapid-decay stages occur over comparable iteration ranges for the two boundary configurations, whereas their terminal residual levels differ substantially. In particular, the ``wall\_all'' cases reach much lower residual levels, while the leading-edge singularity in the ``wall\_sym'' cases limits further residual reduction. These results suggest that, for this test, the singularity mainly affects the attainable residual magnitude and has a comparatively weaker influence on the convergence speed.

In both boundary configurations, reducing the first-layer height from $10^{-4}L$ to $10^{-6}L$ markedly slows the decay of the residual. For the cases examined, this comparison suggests that cell aspect ratio has a greater influence on the convergence rate than the presence of the leading-edge singularity. Future work will focus on accelerating convergence on meshes with high cell aspect ratios while preserving the stability and accuracy of the SAV-FV scheme.


\section{Numerical results} \label{section:numerical_results}
Five test cases are simulated on three-dimensional hexahedral meshes. For the two-dimensional cases, simulations are performed using hexahedral meshes with five layers of grid points in the spanwise direction. 
The benchmark suite comprises (i) a quarter-cylinder, (ii) an extended flat-plate domain, (iii) a NACA0012 airfoil, (iv) a multi-element airfoil, and (v) the three-dimensional ONERA M6 wing. We use the quarter-cylinder test case to verify the spatial order of accuracy of the SAV-FV scheme.
These test cases cover complex geometries and are performed on
high-aspect-ratio meshes used in practical RANS simulations,
thereby providing a rigorous validation of the proposed scheme.
Several general remarks and notations are established prior to proceeding:

\begin{itemize}
	\item Hereafter, unless otherwise specified, the default values for both $\beta$ and $\alpha_{\phi}$ are set to 10;
	\item For all test cases, the initial distance field is defined as the Euclidean distance from the origin plus a unit offset, i.e., $\sqrt{x^2+y^2+z^2}+1$ in three dimensions and $\sqrt{x^2+y^2}+1$ in two dimensions, where the unit offset arises from the computational distance field introduced in Section~\ref{section:boundary-condition};
	\item The initial value of the scalar auxiliary variable $r$ is set to 1 ;
	\item The linear system \eqref{eq:linear-system} is solved for $\bar{\boldsymbol{\phi}}^{n+1}$ using an algebraic multigrid (AMG) method \cite{brandt1986algebraic}. The solver stops when the residual falls to $10^{-8}$ of its initial value or after 50 iterations, whichever occurs first;
	\item Let $R(\vec{x})=\lVert\nabla\phi(\vec{x})\rVert-1$ denote the pointwise Eikonal residual. 
	Convergence of the pseudo-time iteration is assessed using both the volume-averaged absolute residual $L_1(\nabla\phi)=\lvert\Omega\rvert^{-1}\sum_i\int_{\Omega_i}\lvert R(\vec{x})\rvert\,dV$ and the maximum absolute residual $L_\infty(\nabla\phi)=\max_{\vec{x}\in\Omega}\lvert R(\vec{x})\rvert$.
	\item The numerical solution represents the shifted distance field defined in Section~\ref{section:boundary-condition}. Subtracting $1$ from this solution yields the physical wall-distance field.
	\item The relative error is defined as $\mathrm{Err} = |\phi - \phi_{\text{exact}}|/\phi_{\text{exact}}$, where $\phi_{\text{exact}}$ denotes the exact wall distance. 
	In the numerical computations, $\phi_{\text{exact}}$ is obtained using a search algorithm based on the alternating digital tree (ADT) data structure \cite{boger2001efficient}, which yields exact distances to the piecewise-linear wall boundaries represented by the mesh used in this study.
\end{itemize}

\subsection{Quarter cylinder} \label{section:quarter-cylinder}

\begin{figure} 
	\centering
	\subfigure[Mesh and boundary conditions\label{fig:quarter-grids}]{%
		\includegraphics[width=0.45\linewidth]{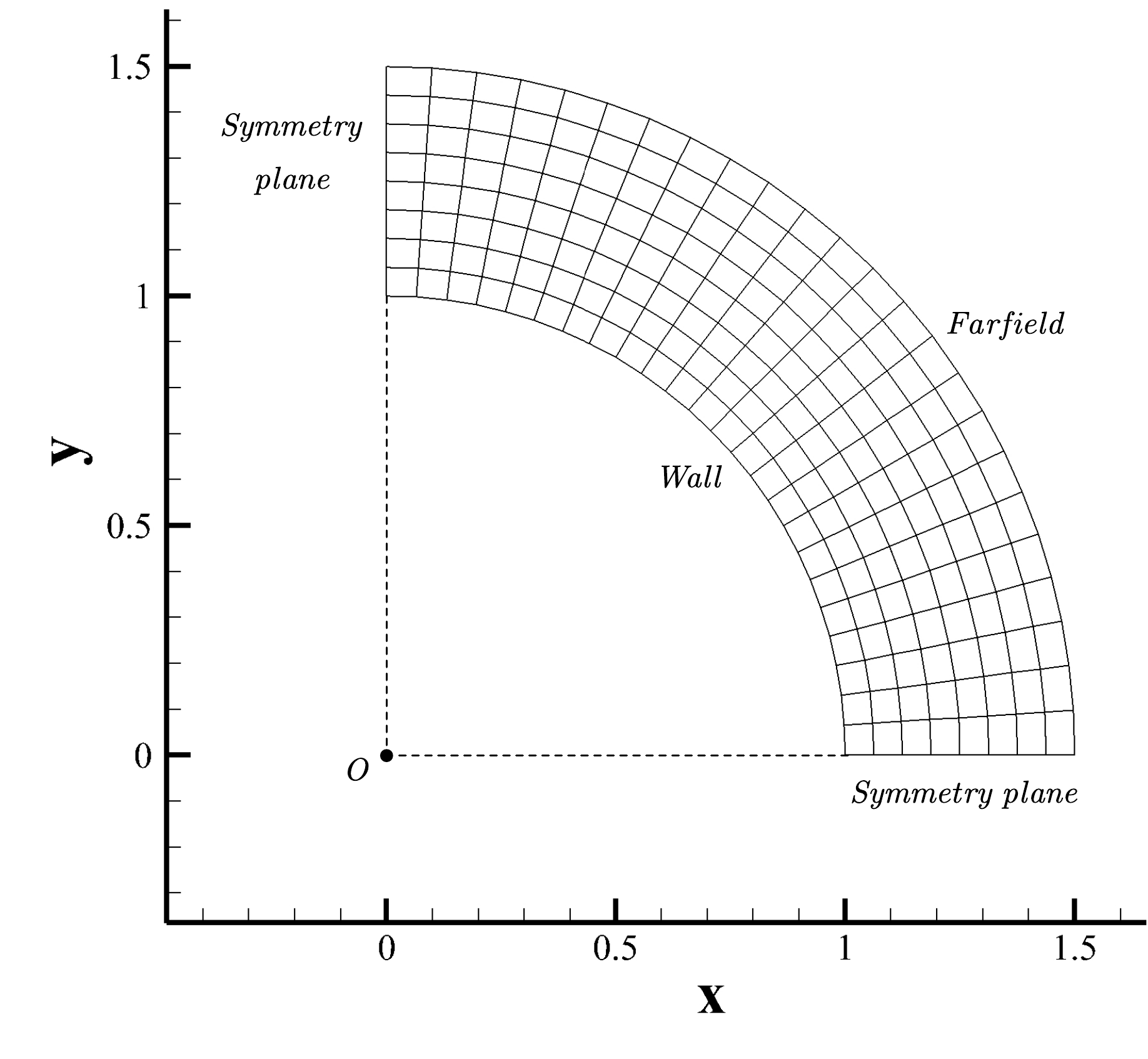}%
	}
	\hfill
	\subfigure[Wall distance distribution\label{fig:quarter-dist}]{%
		\includegraphics[width=0.54\linewidth]{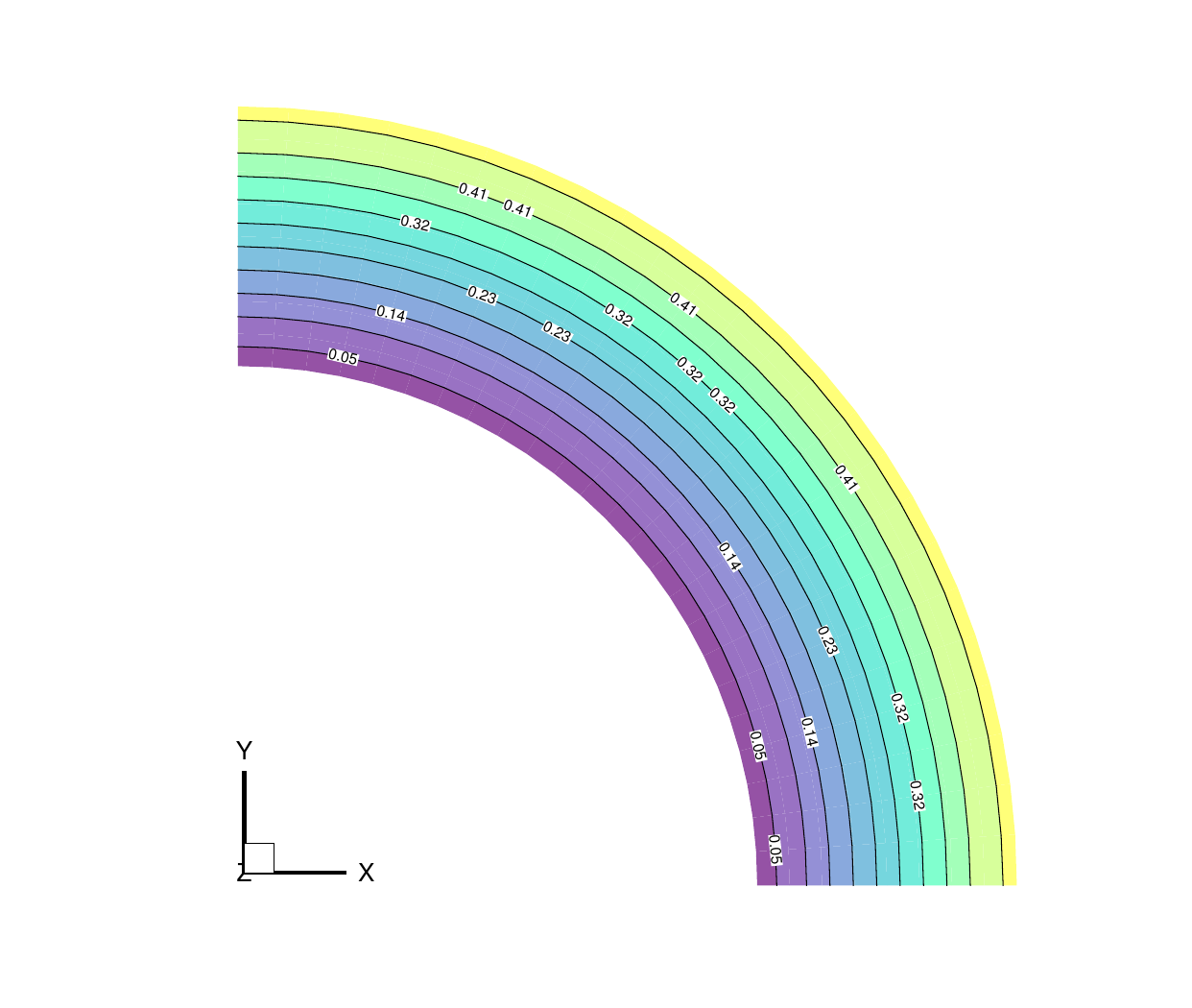}%
	}
	\caption{Results of the quarter-cylinder on $8 \times 24$ cells.}
	\label{fig:quarter-cylinder-geometry}
\end{figure}

The quarter-cylinder test assesses whether the proposed SAV-FV scheme achieves its designed second-order spatial accuracy for the steady-state Eikonal solution. 
Figure~\ref{fig:quarter-grids} shows the boundary conditions and the coarsest mesh, which contains $8$ cells in the radial direction and $24$ along the circular arc. Finer meshes are generated by uniformly reducing the grid spacing in both directions while holding all other simulation settings fixed. The observed order of convergence is estimated from the wall-distance errors on successive meshes.

Figure~\ref{fig:quarter-dist} shows the computed wall-distance field on the coarsest mesh. The contours follow the curvature of the wall and remain smooth near the symmetry boundaries. The absence of visible spurious oscillations in these regions is consistent with the intended stabilizing effect of the vanishing artificial viscosity.

\begin{table}[htbp]
	\centering
	\caption{Wall-distance errors and orders of convergence.}
	\label{tab:accuracy order}
	\begin{tabular}{c c c c c}
	\toprule
		Mesh & \(L^1\) error & Order & \(L^\infty\) error & Order \\
		\midrule
		$8 \times 24$ & \(5.59 \times 10^{-4}\) & -- & \(9.62 \times 10^{-4}\) & -- \\
		$16 \times 48$ & \(1.44 \times 10^{-4}\) & 1.96 & \(2.46 \times 10^{-4}\) & 1.97 \\
		$32 \times 96$ & \(3.64 \times 10^{-5}\) & 1.98 & \(6.19 \times 10^{-5}\) & 1.99 \\
		$64 \times 192$ & \(9.07 \times 10^{-6}\) & 2.00 & \(1.55 \times 10^{-5}\) & 2.00 \\
		$128 \times 384$ & \(2.27 \times 10^{-6}\) & 2.00 & \(3.89 \times 10^{-6}\) & 1.99 \\
		\bottomrule
	\end{tabular}
\end{table}

For the quarter-cylinder geometry, the exact wall distance is $\phi_{\mathrm{exact}}(x,y)=\sqrt{x^2+y^2}-1$. Since $\phi_i$ represents the shifted distance field, the error at the cell centroid $\vec{x}_i$ is $e_i=\bigl(\phi_i(\vec{x}_i)-1\bigr)-\phi_{\mathrm{exact}}(\vec{x}_i)$. The volume-weighted mean absolute error and the maximum absolute error are defined by
\begin{equation}
	L^1(\phi)
	= \frac{\sum_i \lvert\Omega_i\rvert\,\lvert e_i\rvert}
	{\sum_i \lvert\Omega_i\rvert},
	\qquad
	L^\infty(\phi)
	= \max_i \lvert e_i\rvert.
\end{equation}
Table~\ref{tab:accuracy order} reports both errors and their observed orders of convergence under mesh refinement. The results confirm the expected second-order spatial accuracy of the SAV-FV scheme for this test case.


\subsection{Extended flat plate} \label{section:flat-plate}

The extended flat-plate configuration tests the robustness and accuracy of the SAV-FV scheme on highly stretched meshes representative of those used in simulations of turbulent boundary layers over flat plates. The reference length is $L=1$, and the global pseudo-time step is $\Delta\tau=2$. The computational domain extends upstream of the plate, allowing the test to assess whether the scheme suppresses spurious oscillations near the leading-edge singularity.
The region above the plate provides a direct test of whether the scheme reproduces the exact physical wall-distance field $\phi=y$. The linear polynomial reconstruction can represent this field exactly. Wherever it does so, $\lVert\nabla\phi\rVert=1$ and the cell residual vanishes. The artificial-viscosity weight is therefore zero on any interior face shared by two cells that reproduce the linear profile.

\begin{figure} 
	\centering
	\includegraphics[width=0.7\linewidth]{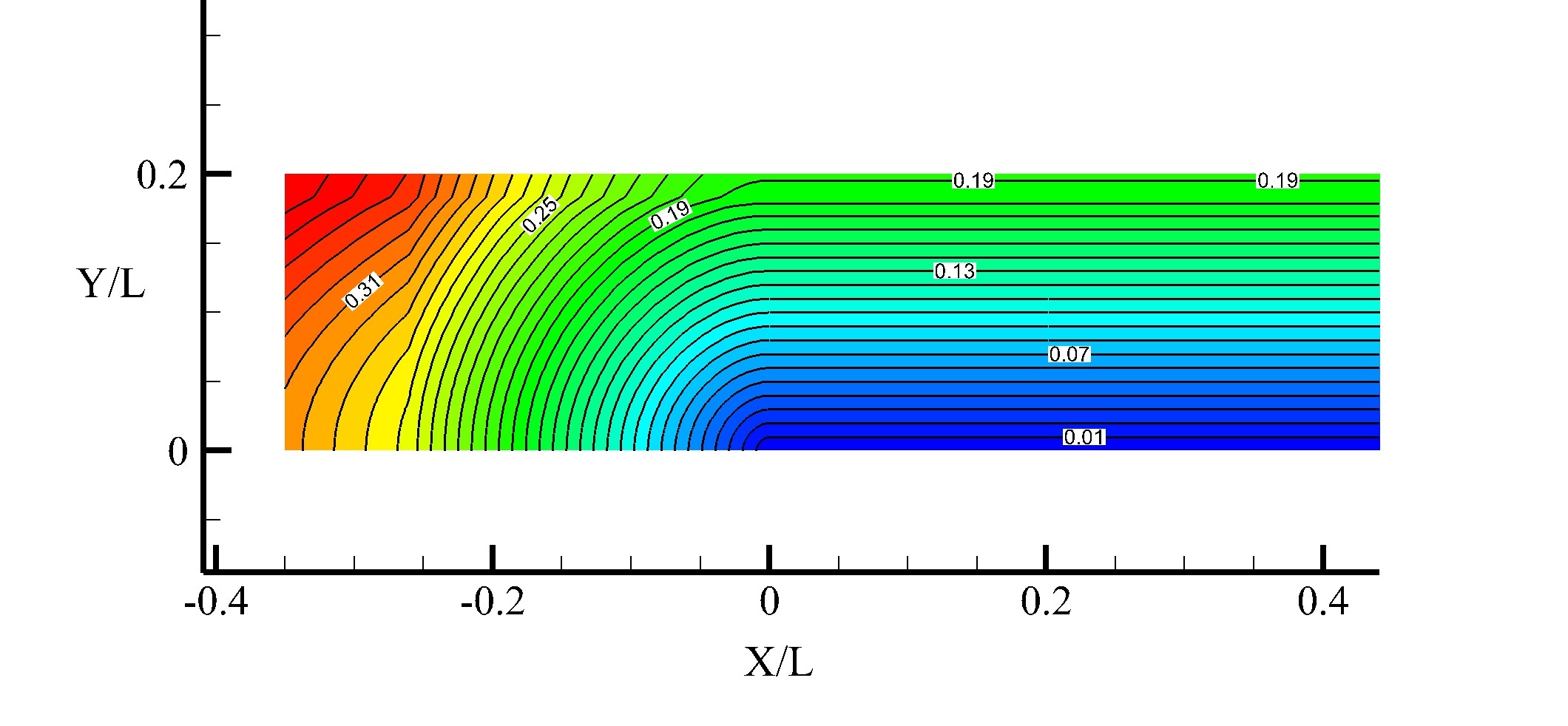}
	\caption{Computed wall distance of the extended flat plate.}
	\label{fig:BL-dist}
\end{figure}

The extended flat-plate case is computed on a strongly anisotropic mesh containing 123,328 cells. The first-layer height is $10^{-6}L$, and the maximum cell aspect ratio, of order $10^5$, occurs near the downstream end of the plate. 
Figure~\ref{fig:convergence-BL-geometry} shows the computational mesh and boundary conditions. Figure \ref{fig:BL-dist} shows the computed wall-distance field. The field varies smoothly near the leading-edge singularity, with no visible spurious oscillations. Away from the leading edge, the wall-distance contours are nearly parallel to the plate, as expected for the exact profile $\phi=y$.

\begin{figure} 
	\centering
	\subfigure[$L_1(\nabla \phi)$  \label{fig:BL-L1}]{%
		\includegraphics[width=0.48\linewidth]{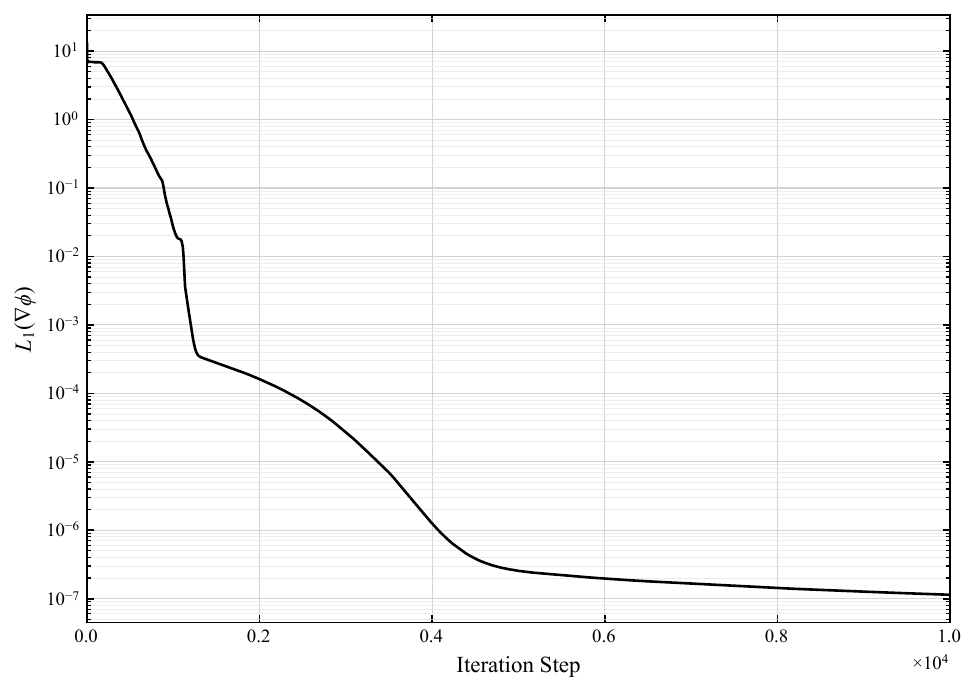}%
	}
	\hfill
	\subfigure[$L_{\infty}(\nabla \phi)$  \label{fig:BL-Linf}]{%
		\includegraphics[width=0.48\linewidth]{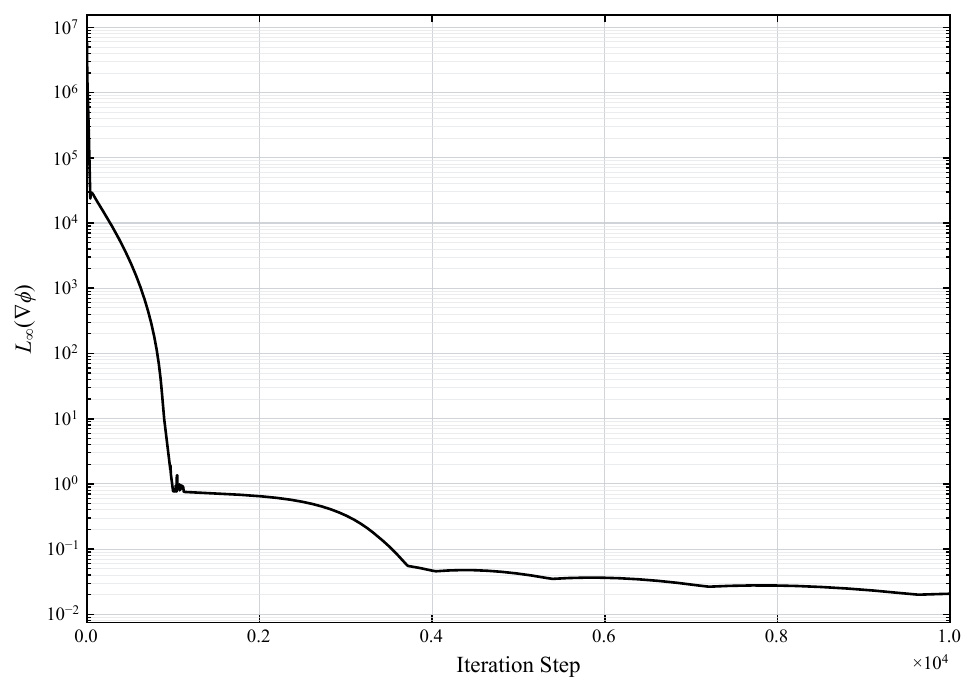}%
	}
	\caption{Convergence history of the extended flat plate.}
	\label{fig:BL-Res}
\end{figure}

Figure~\ref{fig:BL-Res} shows the convergence histories of the Eikonal residual norms. The volume-averaged norm $L_1(\nabla\phi)$ measures convergence across the domain, whereas $L_\infty(\nabla\phi)$ captures the largest local residual, including residuals in thin near-wall cells that may contribute little to the volume average. Both norms decrease rapidly during the initial pseudo-time iterations and then level off approximately eight orders of magnitude below their respective initial values.

\begin{figure} 
	\centering
	\subfigure[Relative error near singular point \label{fig:BL-Err-1}]{%
		\includegraphics[width=0.48\linewidth]{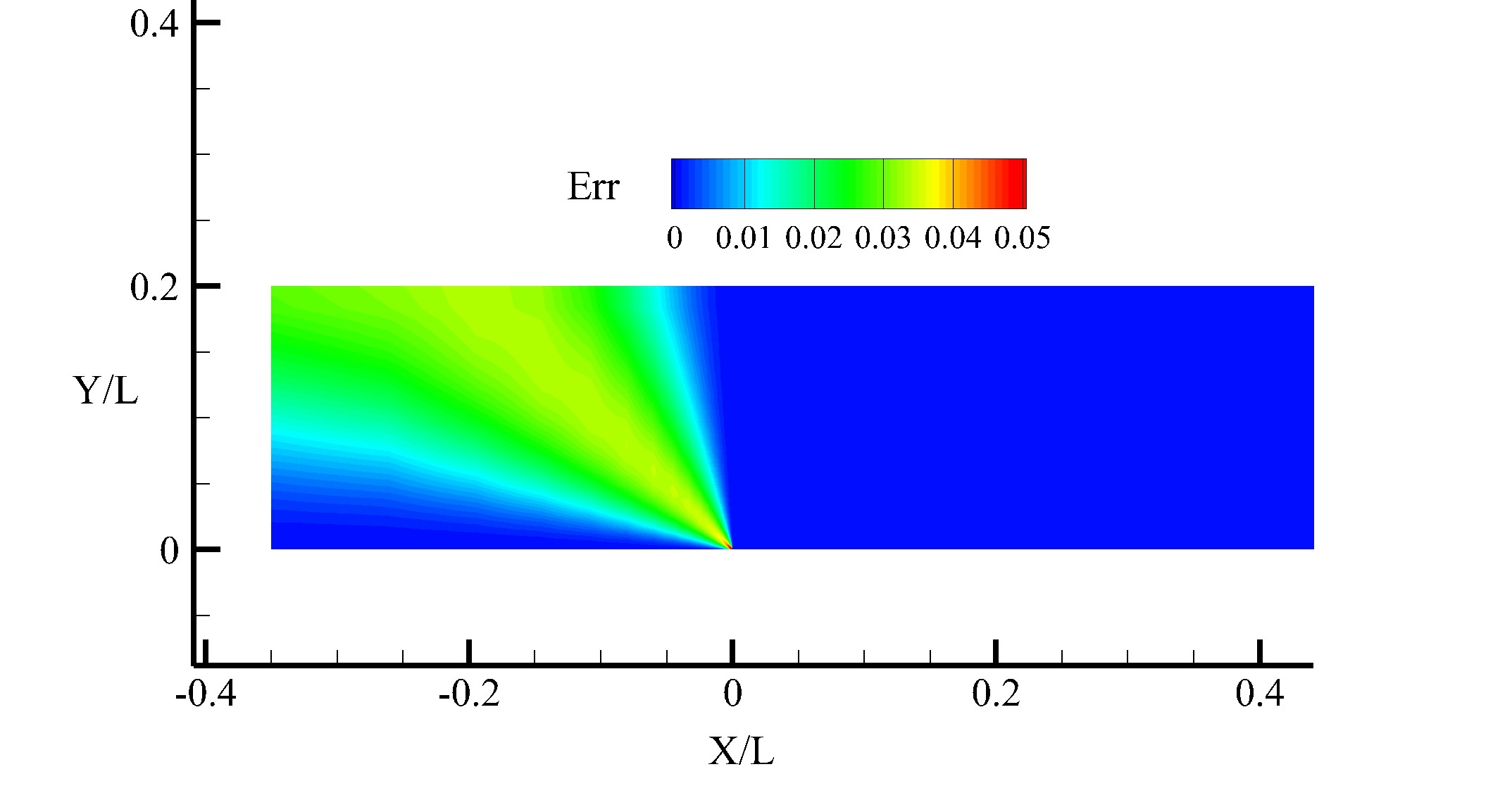}%
	}
	\hfill
	\subfigure[Absolute error in the boundary-layer region \label{fig:BL-Err-2}]{%
		\includegraphics[width=0.48\linewidth]{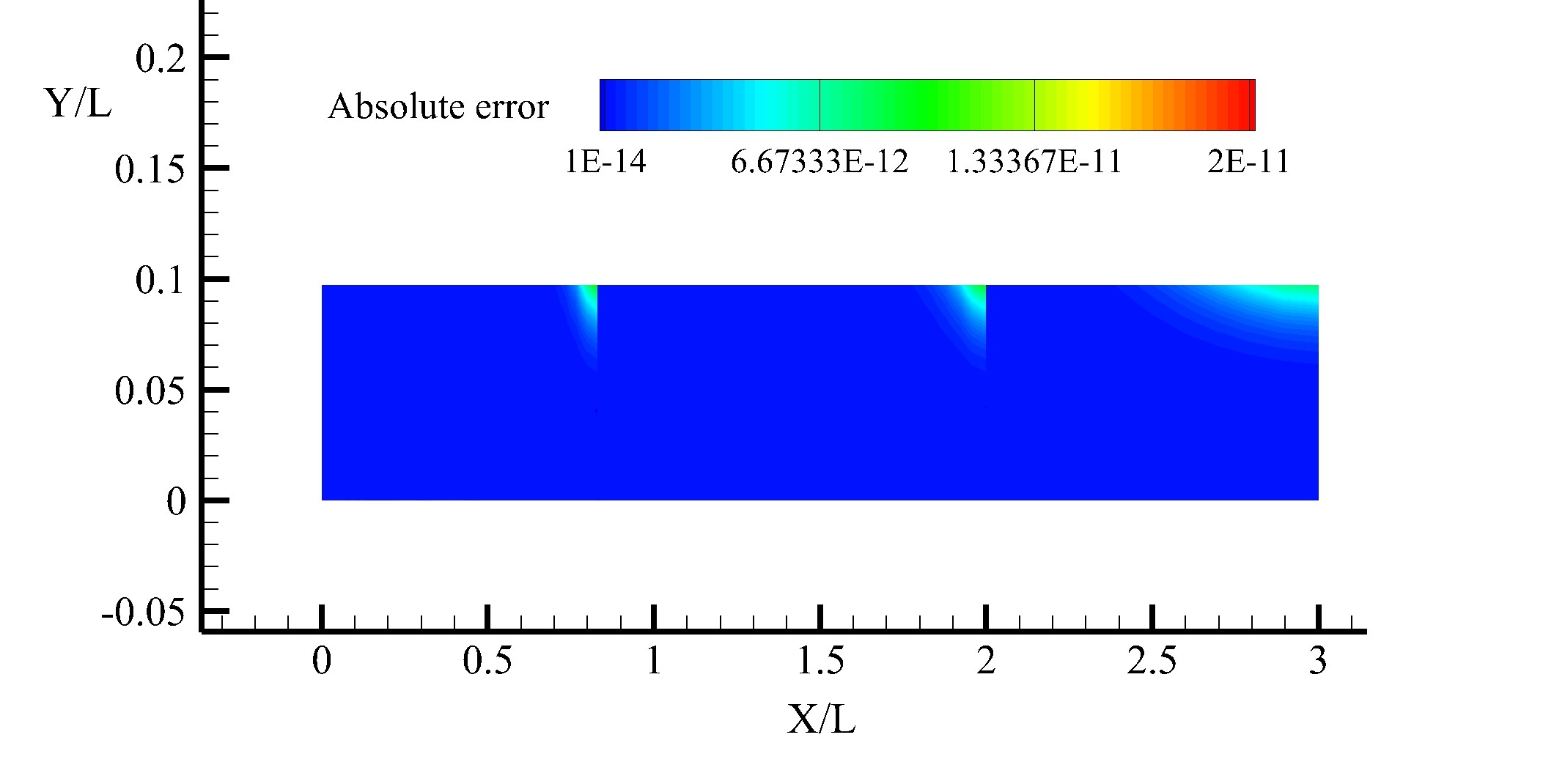}%
	}
	\caption{Wall distance errors of the extended flat-plate.}
	\label{fig:BL-Err}
\end{figure}

Figure~\ref{fig:BL-Err-1} shows the distribution of the relative wall-distance error. The largest errors are concentrated near the leading-edge singularity, where the exact distance field is nonsmooth. The relative error remains below $5\%$ over most of the domain. Figure~\ref{fig:BL-Err-2} shows the absolute error in the strip $0\le y\le0.1$, which covers the entire boundary-layer region. Over most of this strip above the plate, the SAV-FV scheme reproduces the linear wall-distance profile $\phi=y$ with an absolute error of order $10^{-14}$.

\subsection{NACA0012 airfoil} \label{section:naca0012}
\begin{figure} 
	\centering
	\subfigure[Boundary conditions \label{fig:0012-bc}]{%
		\includegraphics[width=0.48\linewidth]{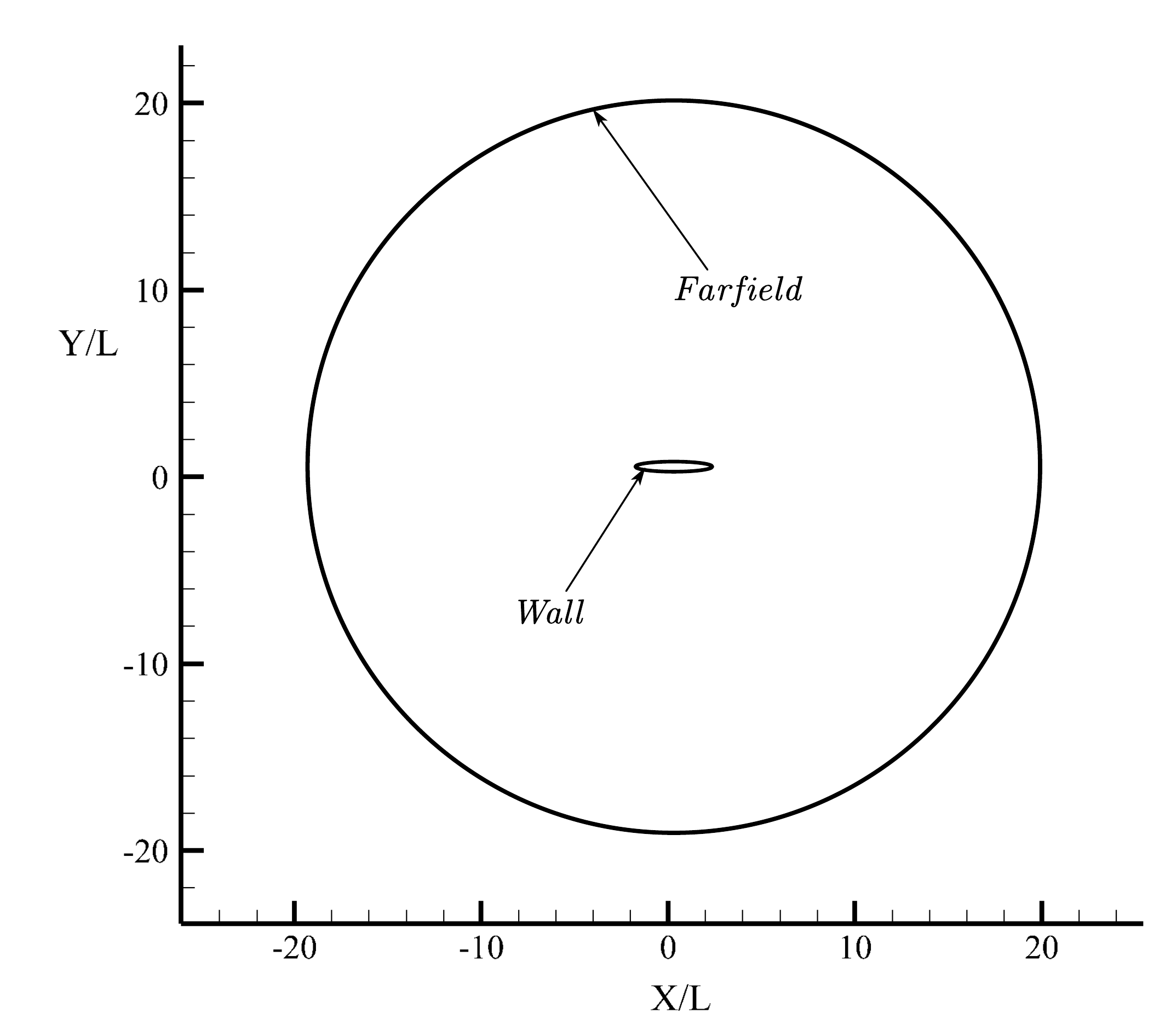}%
	}
	\hfill
	\subfigure[Mesh near the wall \label{fig:0012-grid}]{%
		\includegraphics[width=0.48\linewidth]{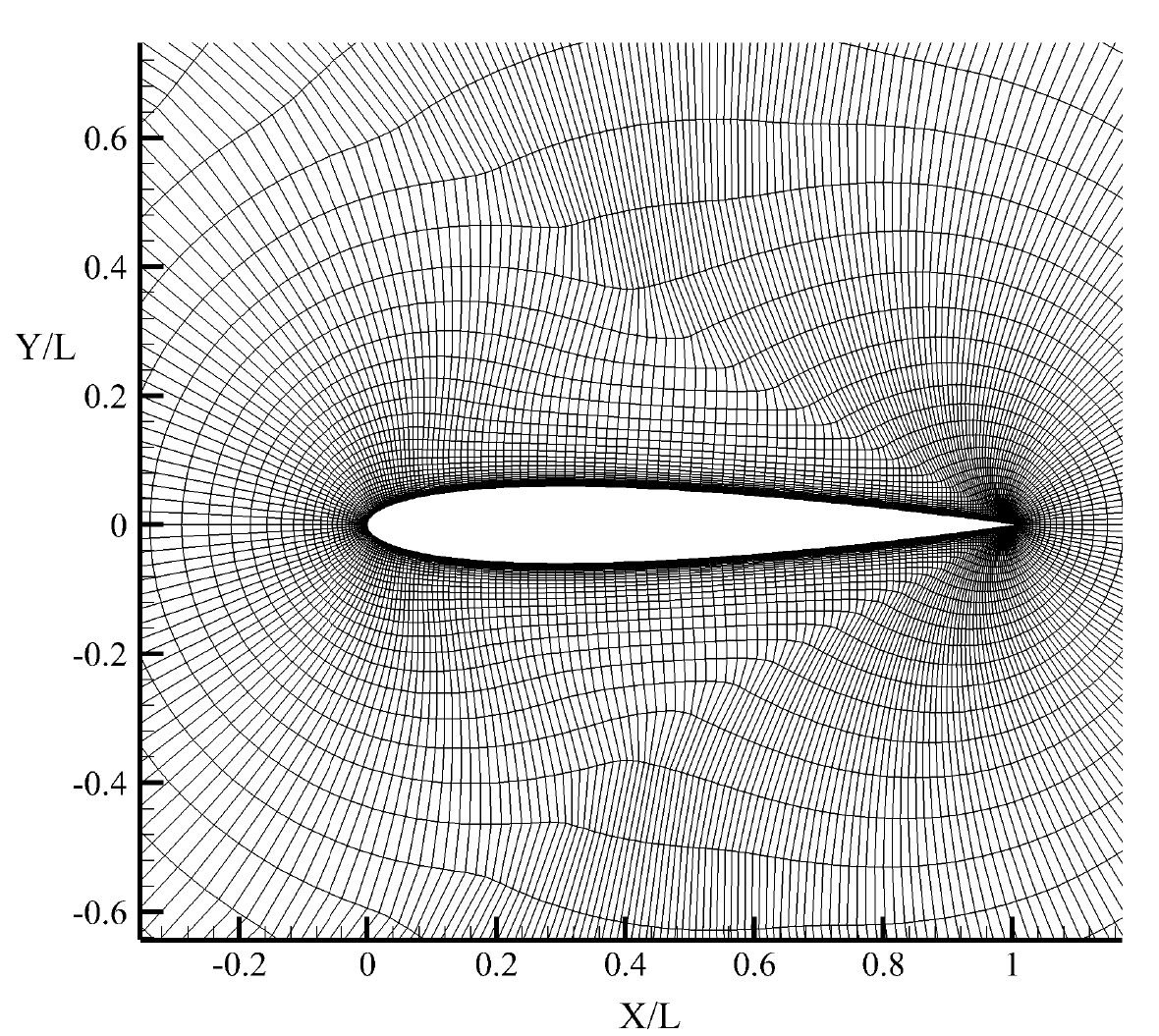}%
	}
	\caption{NACA0012 simulation setup.}
	\label{fig:0012-geo}
\end{figure}

The NACA0012 airfoil test assesses the accuracy and robustness of the SAV-FV scheme on a curved geometry with a sharp trailing edge. Figure~\ref{fig:0012-geo} shows the computational domain and boundary conditions. The airfoil surface is treated as a wall, and a far-field condition is imposed at the outer boundary. The stretched, body-fitted mesh contains 194,560 cells, with a first-layer height of $10^{-6}L$ and near-wall cell aspect ratios of order $10^4$, where $L$ denotes the chord length. Figure~\ref{fig:0012-grid} shows an enlarged view of the mesh. The global pseudo-time step is $\Delta\tau=2$.

\begin{figure} 
	\centering
	\subfigure[Wall distance distribution \label{fig:0012-dist}]{%
		\includegraphics[width=0.45\linewidth]{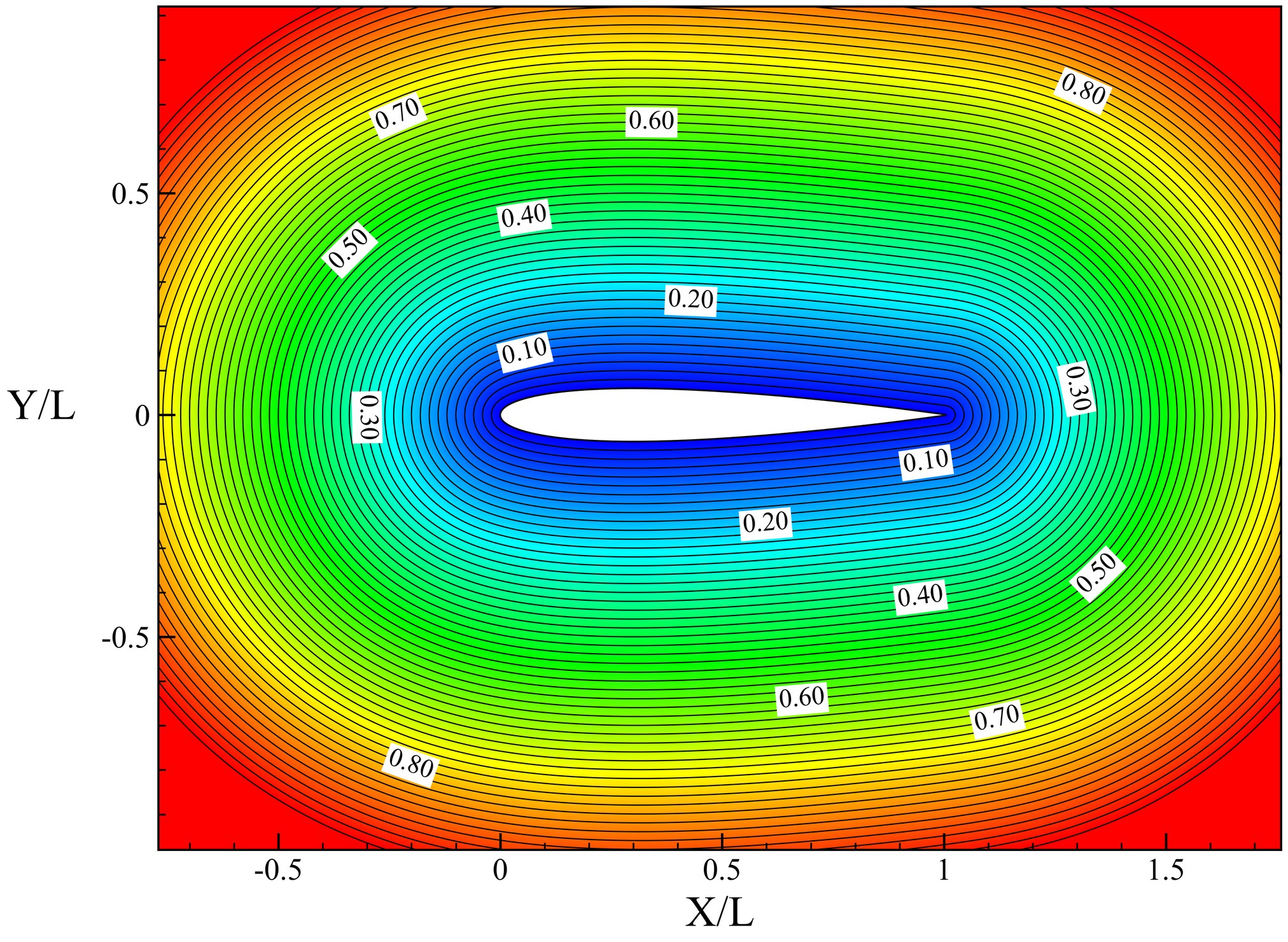}%
	}
	\hfill
	\subfigure[Wall distance distribution near tail \label{fig:0012-tail}]{%
		\includegraphics[width=0.46\linewidth]{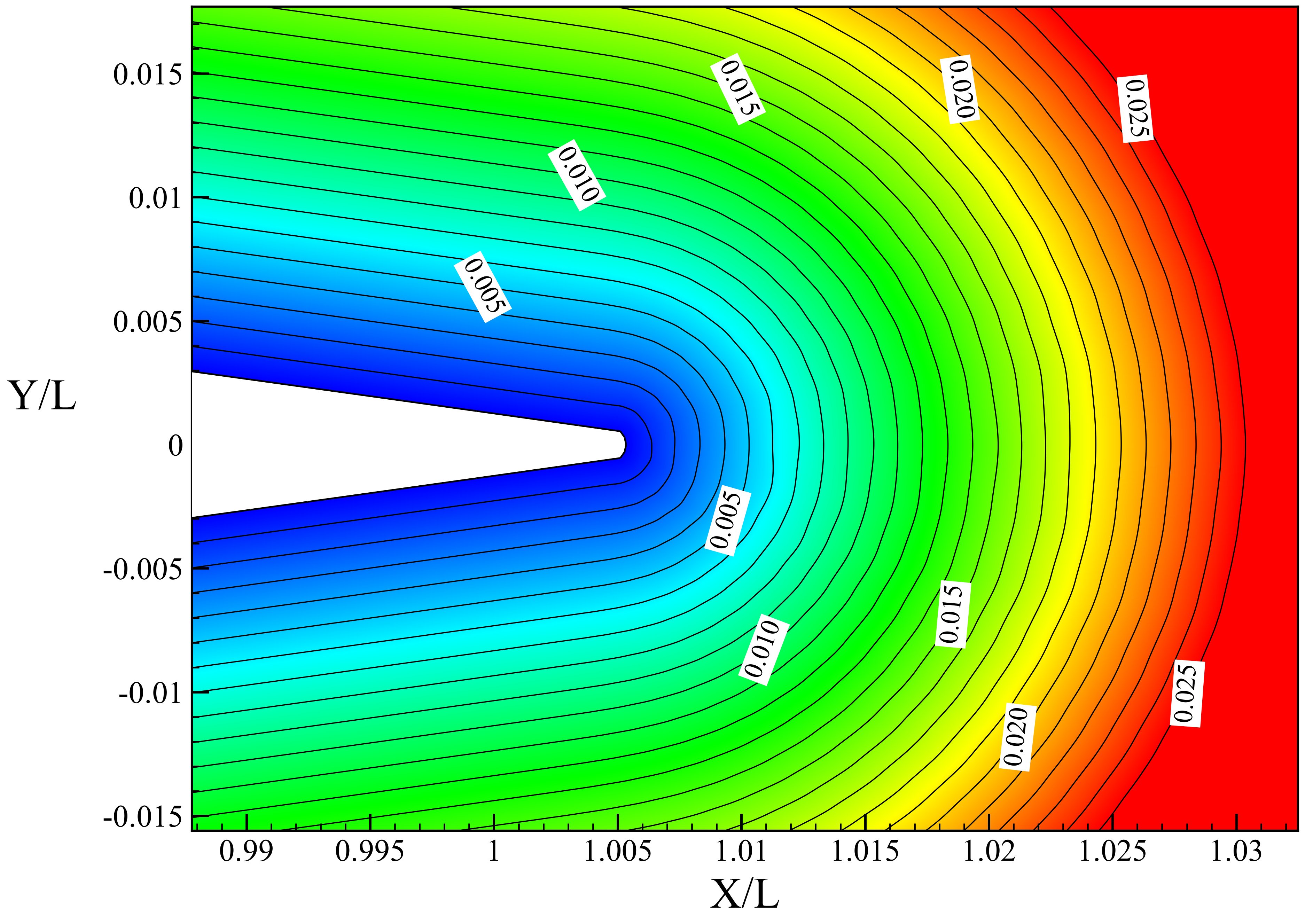}%
	}
	\hfill
	\subfigure[Error overall \label{fig:0012-err-1}]{%
		\includegraphics[width=0.5\linewidth]{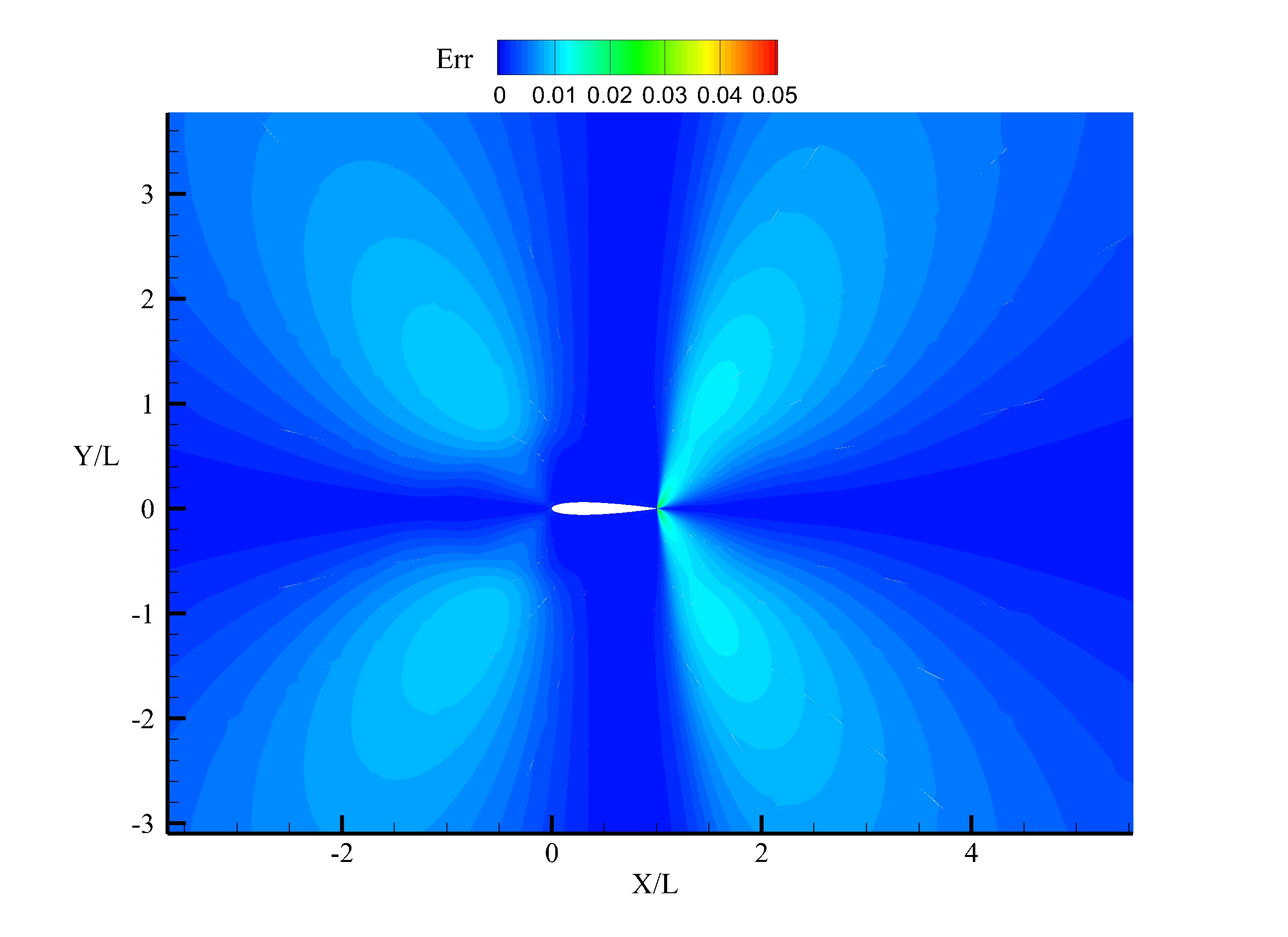}%
	}
	\hfill
	\subfigure[Error near tail \label{fig:0012-err-2}]{%
		\includegraphics[width=0.45\linewidth]{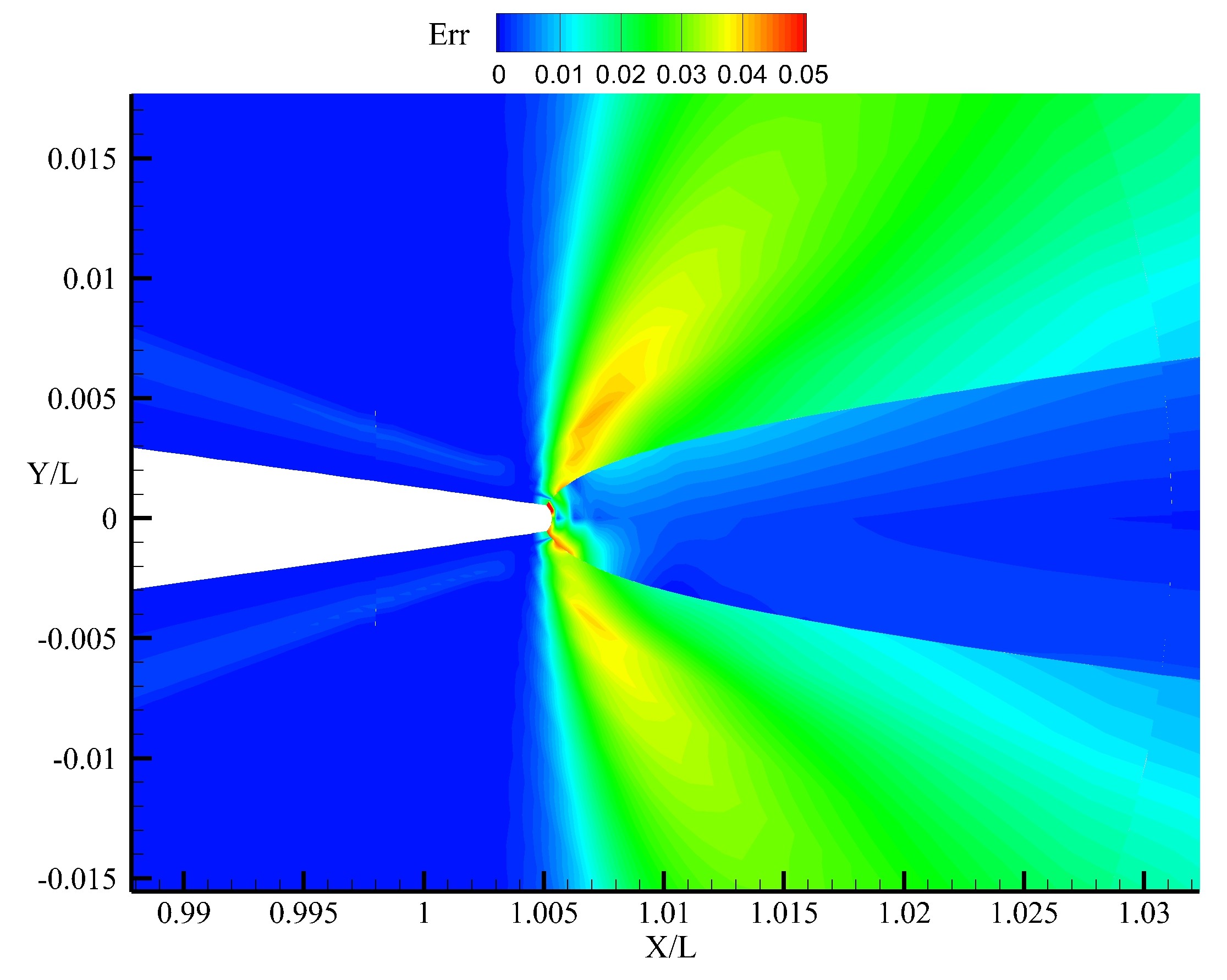}%
	}
	\caption{Results of the NACA0012 airfoil.}
	\label{fig:0012-result}
\end{figure}

Figure~\ref{fig:0012-result} shows the computed wall-distance contours and the corresponding relative-error distribution. The contours follow the airfoil surface smoothly and remain regular throughout the domain. In particular, the enlarged view in Figure~\ref{fig:0012-tail} shows no visible spurious oscillations near the trailing-edge singularity.
Figure~\ref{fig:0012-err-1} shows that the relative wall-distance error is below $2\%$ over nearly the entire domain. The largest errors are confined to a small neighborhood of the trailing edge, as shown in Figure~\ref{fig:0012-err-2}. Near this sharp feature of the piecewise-linear boundary, the wall-distance gradient can vary abruptly across a reconstruction stencil, reducing local accuracy. Even in the enlarged trailing-edge region, the relative error remains below $5\%$ over nearly the entire area shown.

\begin{figure}
	\centering
	\subfigure[$L_1(\nabla \phi)$  \label{fig:0012-L1}]{%
		\includegraphics[width=0.48\linewidth]{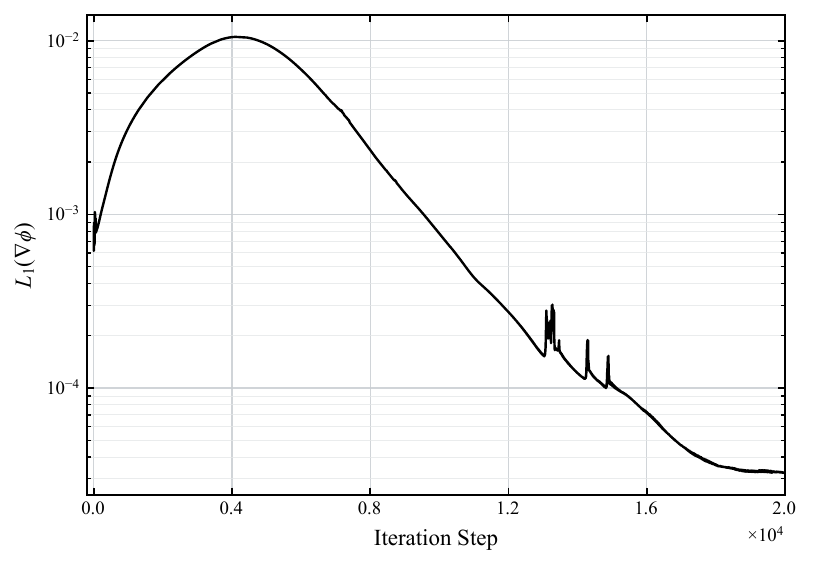}%
	}
	\hfill
	\subfigure[$L_{\infty}(\nabla \phi)$  \label{fig:0012-Linf}]{%
		\includegraphics[width=0.48\linewidth]{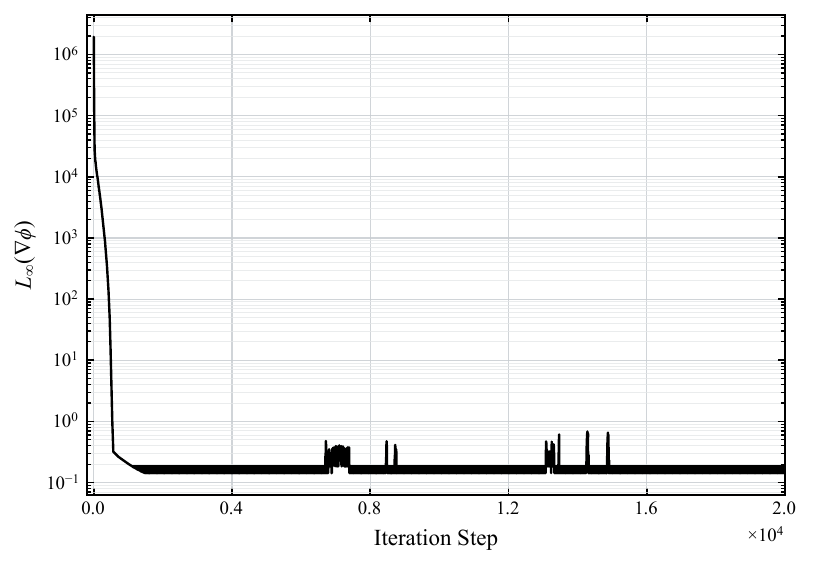}%
	}
	\caption{Convergence history of the NACA0012 airfoil.}
	\label{fig:0012-Res}
\end{figure}

As shown in Figure~\ref{fig:0012-Res}, the maximum residual $L_\infty(\nabla\phi)$ decreases by approximately seven orders of magnitude before reaching a plateau. The volume-averaged residual $L_1(\nabla\phi)$ decreases more modestly and remains low after the initial transient. Similar convergence histories are observed in the multi-element airfoil and ONERA M6 wing cases. This difference between the two norms is consistent with the choice of the initial field as the Euclidean distance from the origin $O$. The initial reconstructed gradient magnitude is already close to unity in far-field cells, which occupy a large fraction of the domain. The initial volume-averaged residual is therefore small, and substantial reductions in local residuals, including those in thin near-wall cells, can have little effect on $L_1(\nabla\phi)$. Near geometric singularities, nonsmooth wall-distance gradients are more difficult to reconstruct, and persistent local residuals can limit further reductions in both norms.


\subsection{30P-30N multi-element airfoil} \label{section:30P-30N}

\begin{figure}
	\centering
	\subfigure[Near slat element \label{fig:sdy-grid-slat}]{%
		\includegraphics[width=0.3\linewidth]{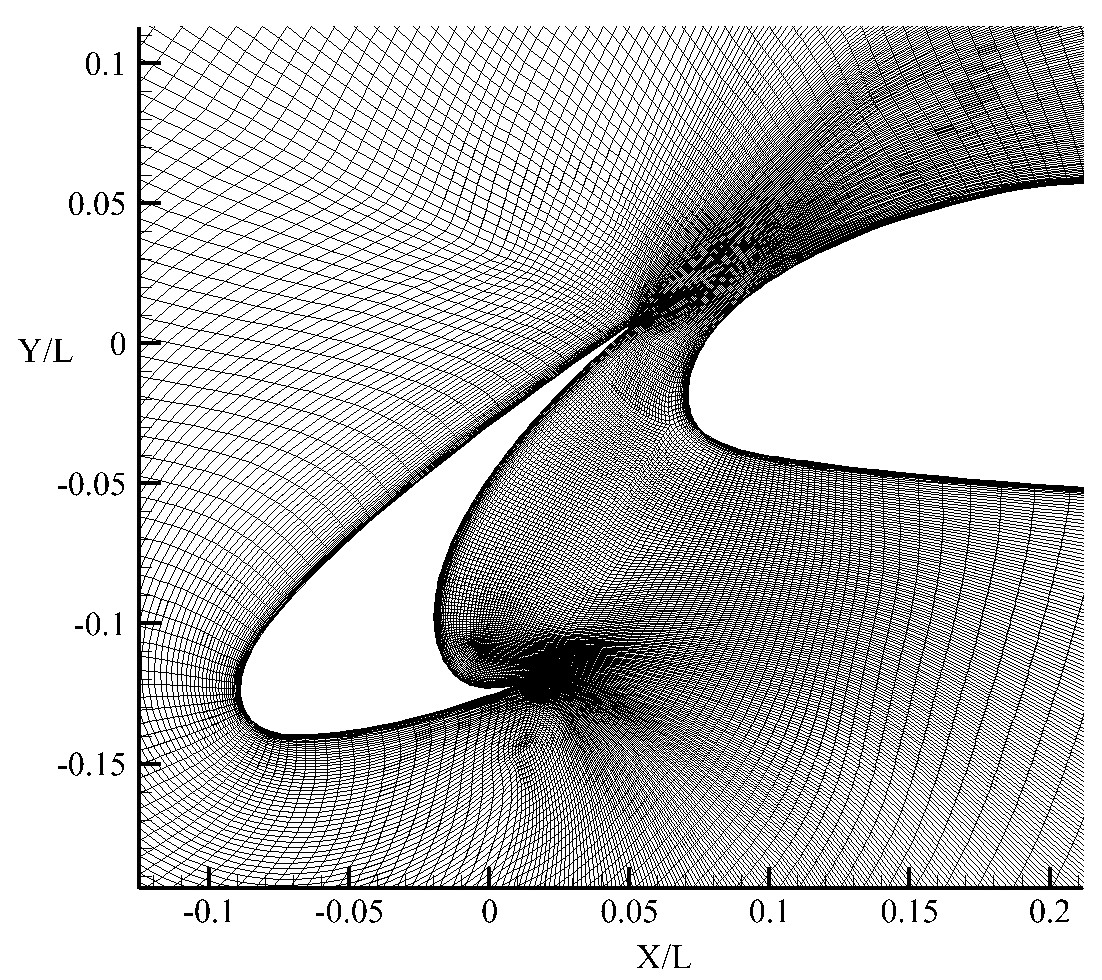}%
	}
	\hfill
	\subfigure[General view \label{fig:sdy-grid}]{%
		\includegraphics[width=0.3\linewidth]{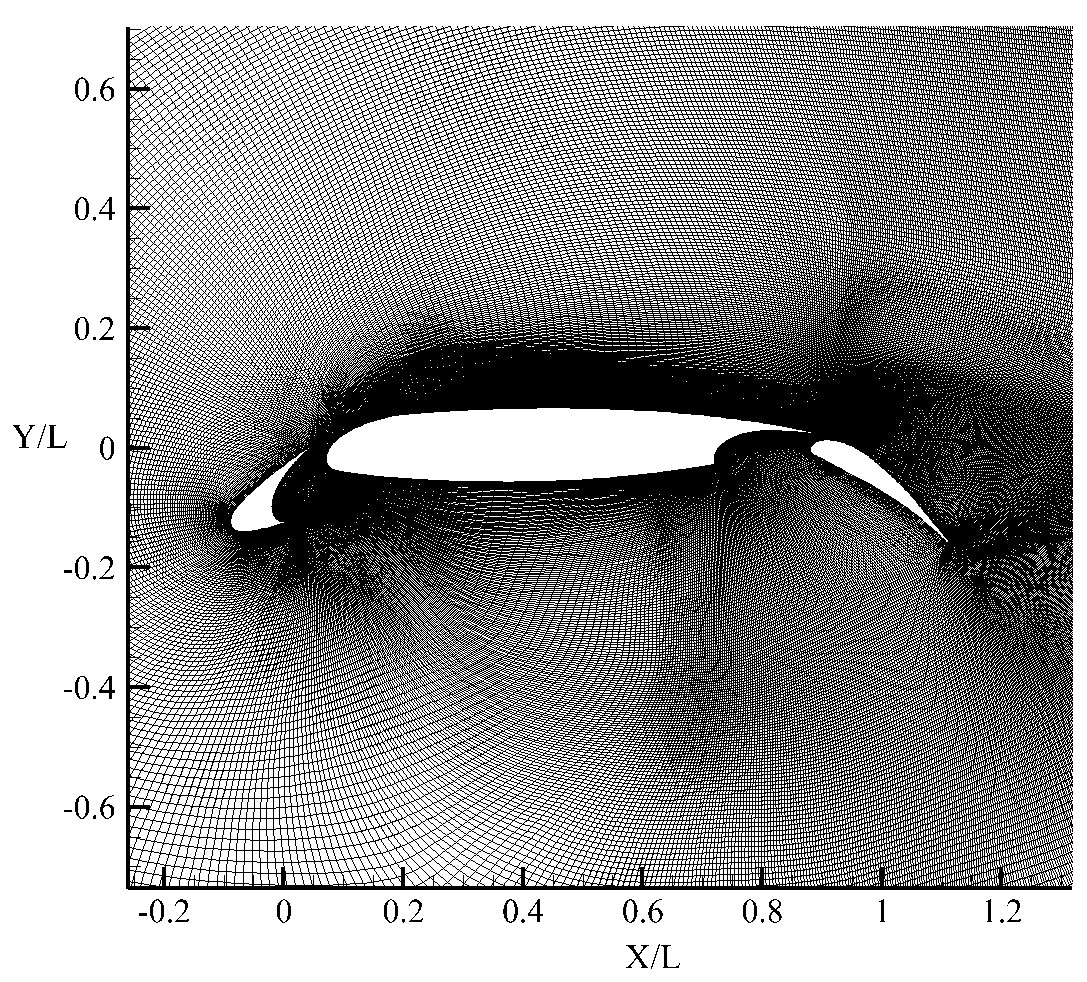}%
	}
	\hfill
	\subfigure[Near flap element \label{fig:sdy-grid-flap}]{%
		\includegraphics[width=0.3\linewidth]{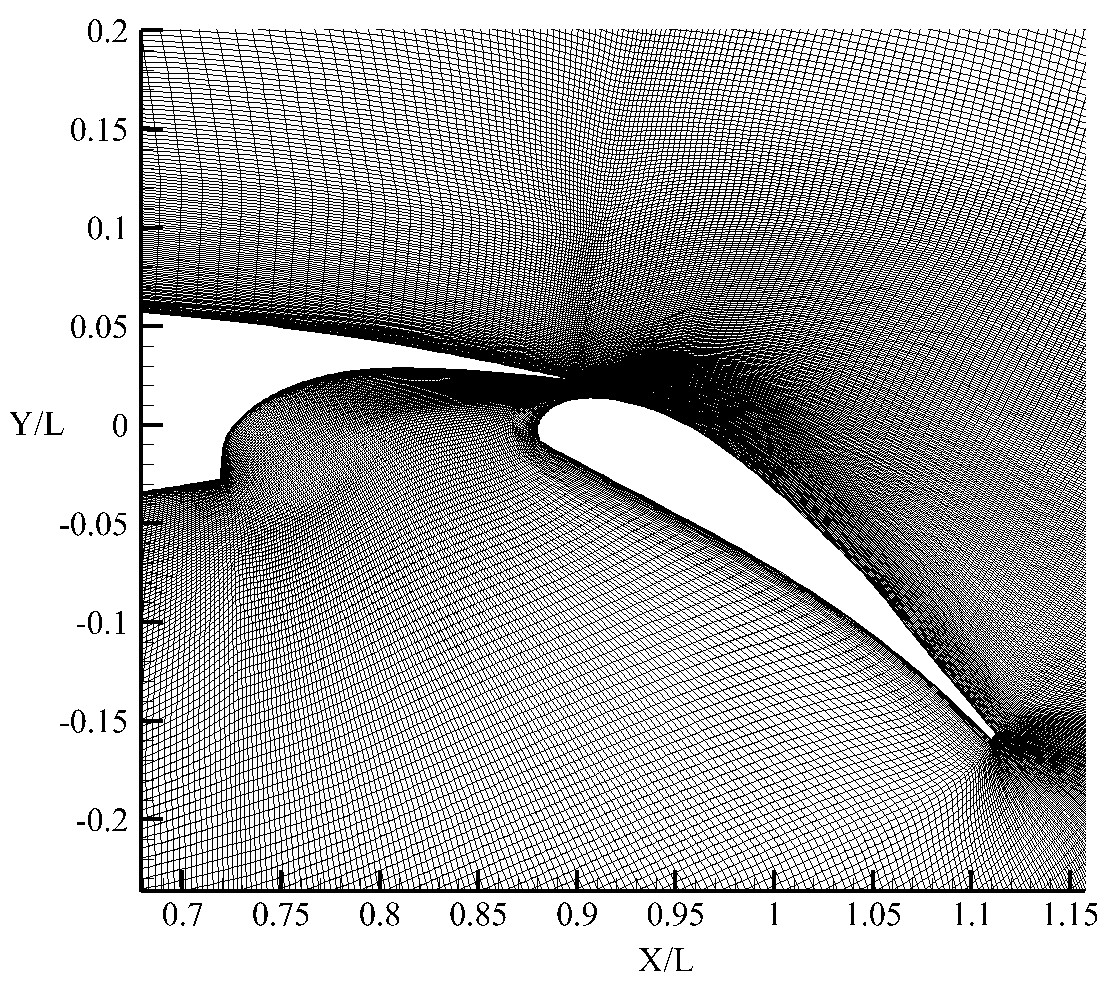}%
	}
	\caption{Computational mesh of the 30P-30N multi-element airfoil.}
	\label{fig:sdy-grids}
\end{figure}

The 30P--30N multi-element airfoil is used to assess the performance of the SAV-FV scheme in computing wall distances for multibody configurations.
The O-type computational domain has an outer radius of $17L$, with the reference length set to $L=1$. As in the NACA0012 case in Section~\ref{section:naca0012}, the airfoil surfaces are treated as walls, and a far-field condition is imposed at the outer boundary. The mesh shown in Figure.~\ref{fig:sdy-grids} contains 815,200 hexahedral cells, with a first-layer height of $10^{-5}L$ and near-wall cell aspect ratios of approximately $10^3$. The global pseudo-time step is $\Delta\tau=2$.

The wall-distance reconstruction must account for geometric singularities near the leading and trailing edges of the slat, main element, and flap. Additional difficulties arise along the medial axes between elements, where two or more wall points are equally close. As the nearest wall changes from one element to another across a medial axis, the distance gradient changes discontinuously. Resolving these features on strongly anisotropic cells places particular demands on the reconstruction. To limit the propagation of numerical oscillations from geometric singularities, the upwind weighting parameter is set to $\alpha_{\phi}=1$, so that only the geometric weights enter the reconstruction.

%

\begin{figure}
	\centering
	\subfigure[$L_1(\nabla \phi)$  \label{fig:sdy-L1}]{%
		\includegraphics[width=0.48\linewidth]{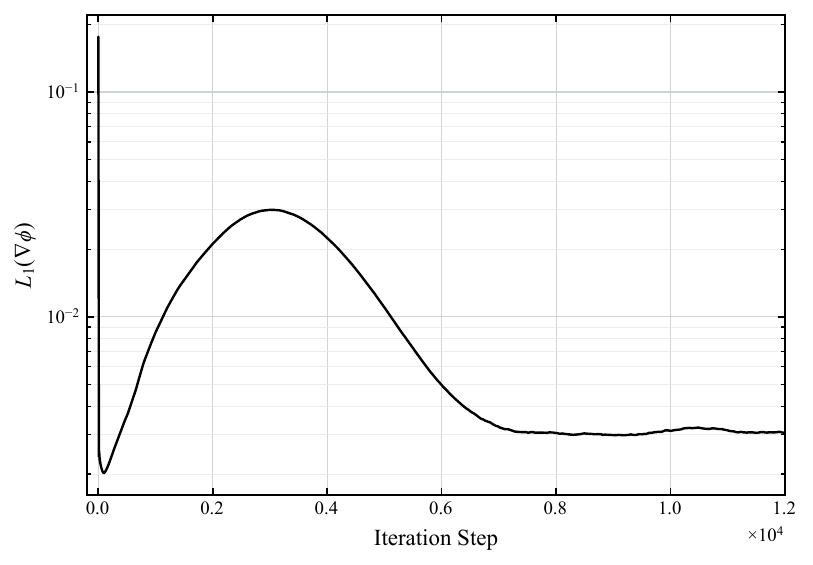}%
	}
	\hfill
	\subfigure[$L_{\infty}(\nabla \phi)$  \label{fig:sdy-Linf}]{%
		\includegraphics[width=0.48\linewidth]{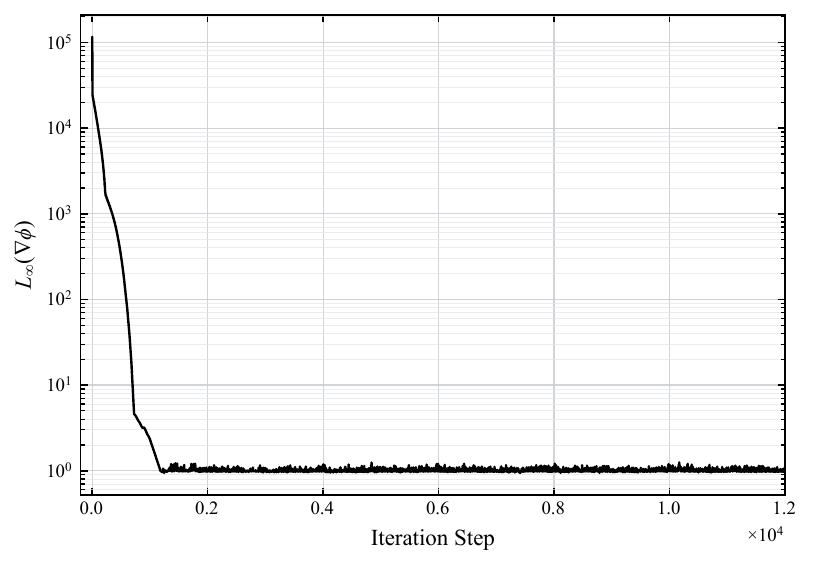}%
	}
	\caption{Convergence history of the 30P-30N multi-element airfoil.}
	\label{fig:sdy-Res}
\end{figure}

The residual histories for the 30P--30N airfoil in Figure~\ref{fig:sdy-Res} follow the same overall pattern as those for the NACA0012 airfoil in Figure~\ref{fig:0012-Res}. The maximum residual $L_\infty(\nabla\phi)$ decreases rapidly by approximately five orders of magnitude before leveling off. The volume-averaged residual $L_1(\nabla\phi)$ decreases more modestly and then fluctuates slightly about a slowly varying level. These fluctuations are consistent with the difficulty of reconstructing the wall-distance gradient near geometric singularities and across the medial axes between elements. Both residual histories indicate stable behavior over the pseudo-time iterations shown.

\begin{figure}
	\centering
	\subfigure[Near slat element \label{fig:sdy-dist-slat}]{%
		\includegraphics[width=0.3\linewidth]{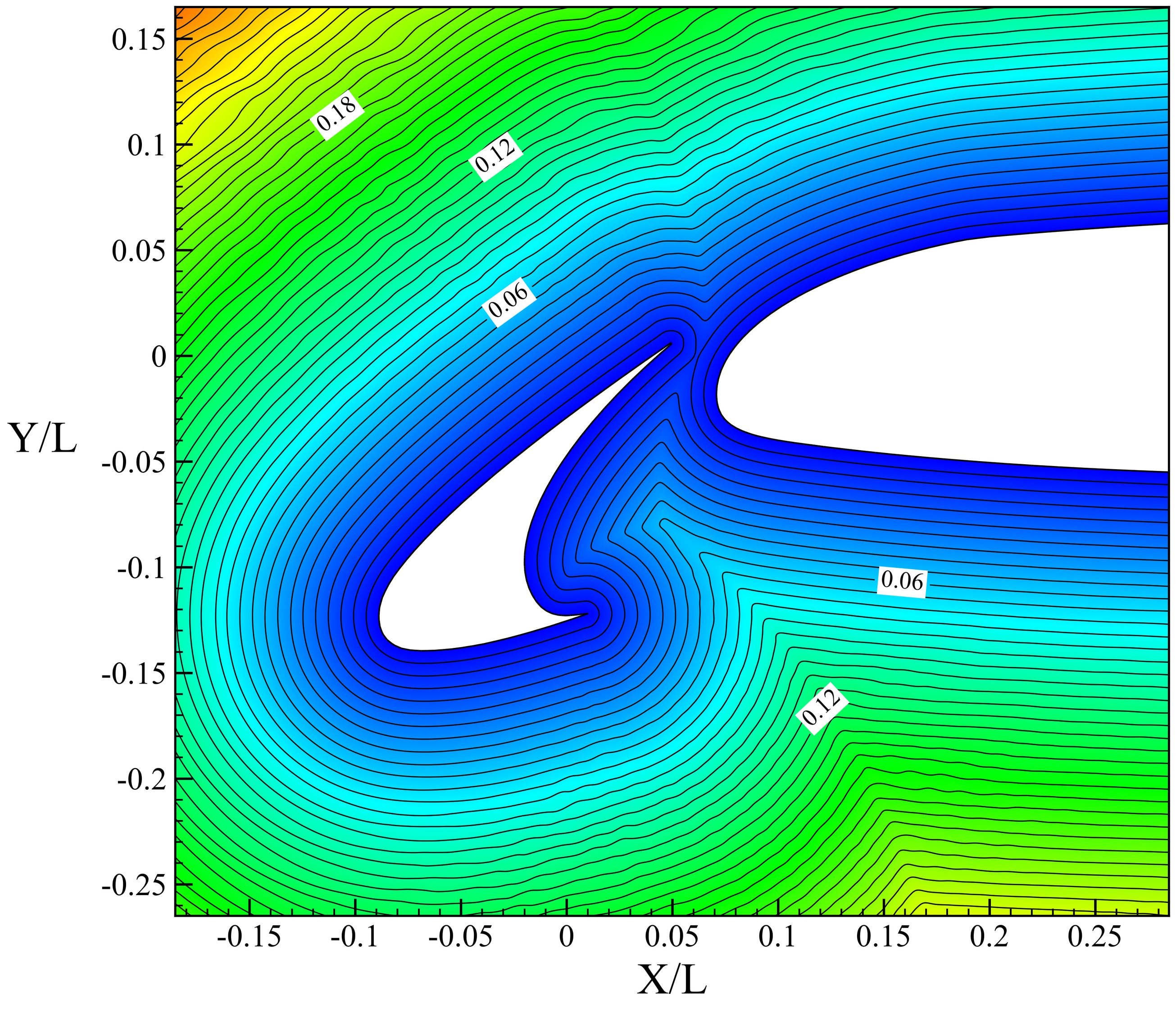}%
	}
	\hfill
	\subfigure[General view \label{fig:sdy-dist}]{%
		\includegraphics[width=0.3\linewidth]{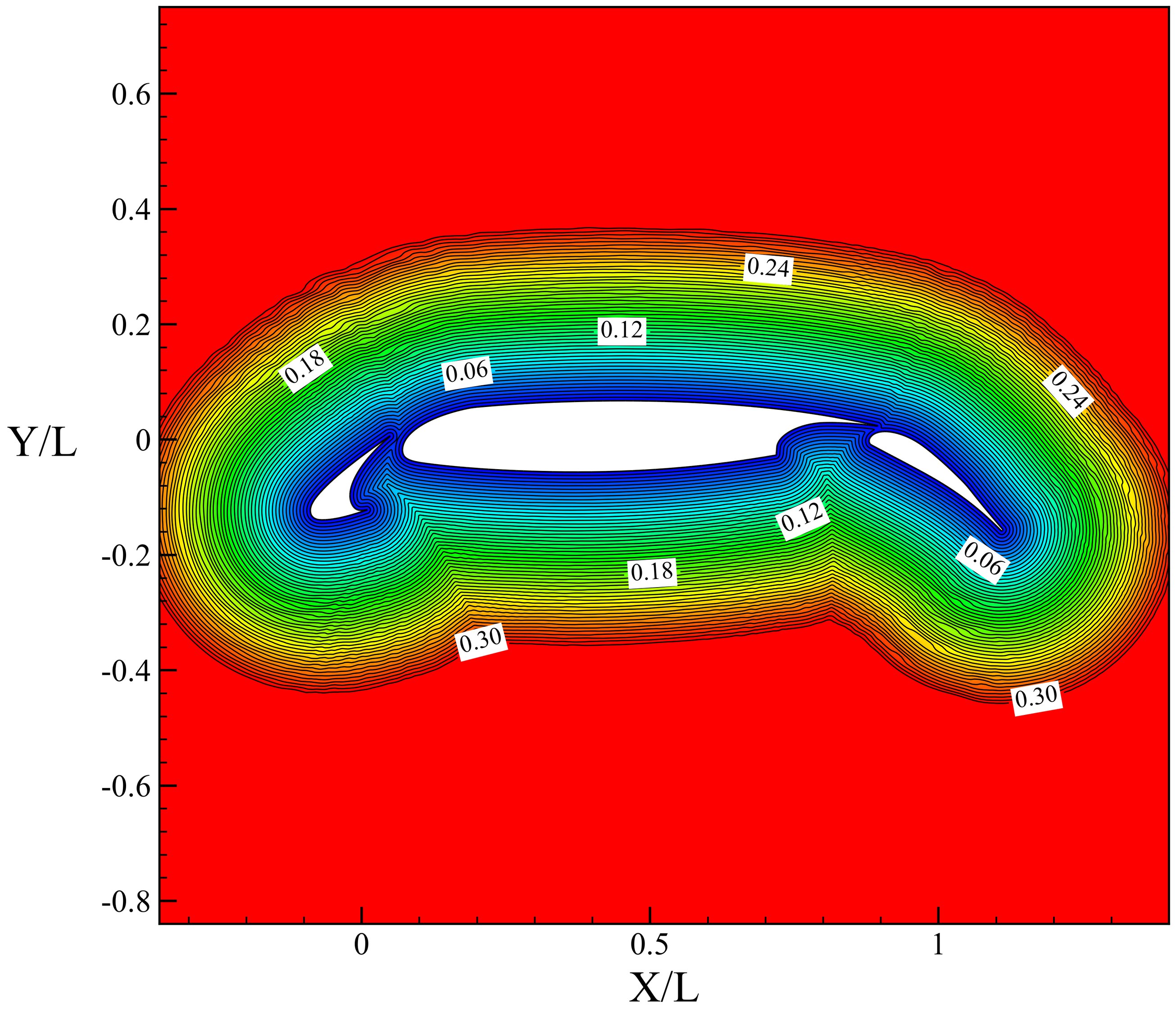}%
	}
	\hfill
	\subfigure[Near flap element \label{fig:sdy-dist-flap}]{%
		\includegraphics[width=0.3\linewidth]{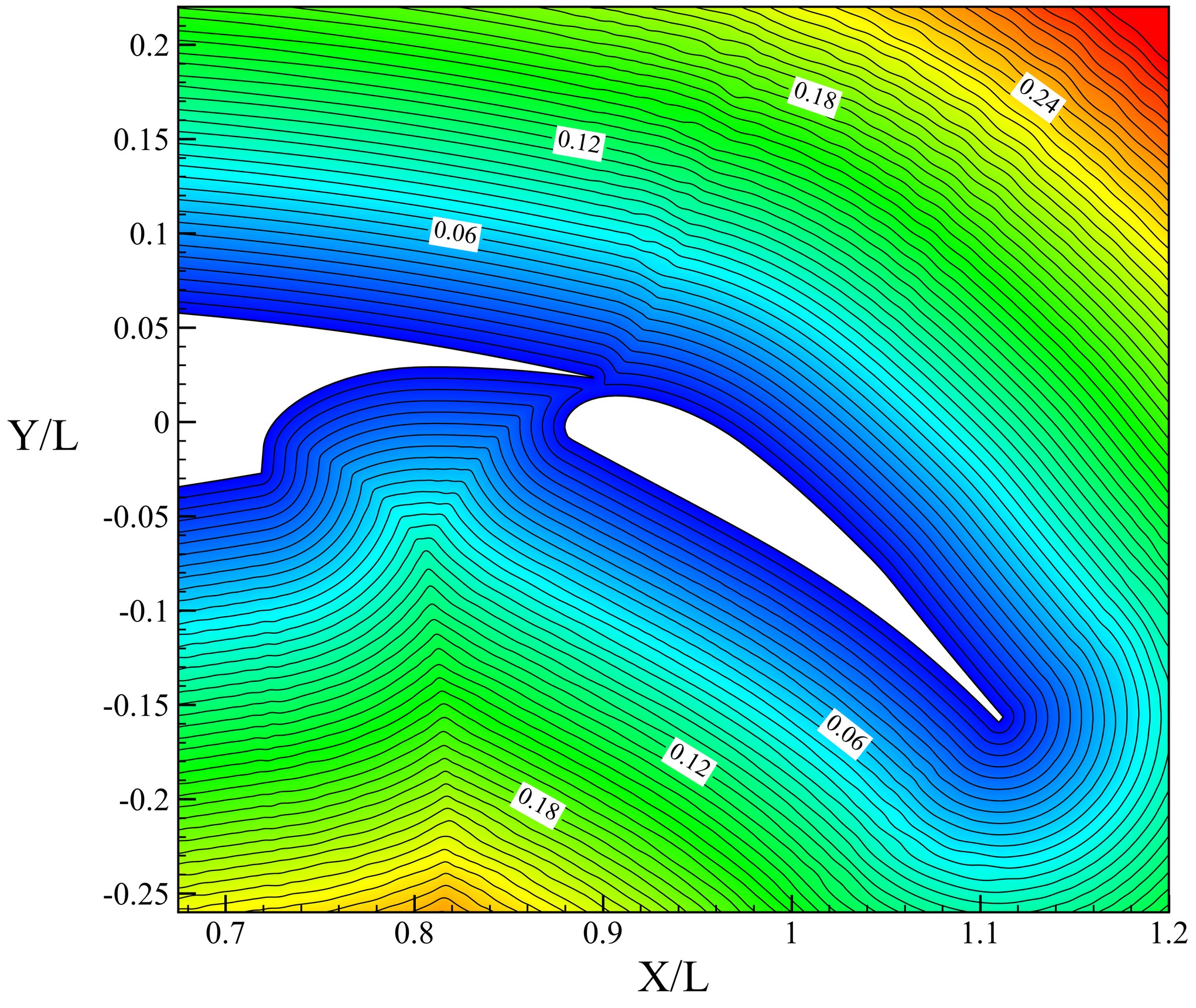}%
	}
	\caption{Computed wall distance of the multi-element airfoil.}
	\label{fig:sdy-distance}
\end{figure}

Figure \ref{fig:sdy-distance} presents the computed wall-distance field. The contours exhibit regular and smooth distributions around all three elements, including the regions adjacent to the leading- and trailing-edge singularities of the slat, main airfoil, and flap. No spurious oscillation is observed near these singular points, indicating that the proposed scheme effectively suppresses singularity-induced numerical oscillations in a complex multi-element configuration.

\begin{figure}
	\centering
	\subfigure[Near slat element \label{fig:sdy-Err-slat}]{%
		\includegraphics[width=0.3\linewidth]{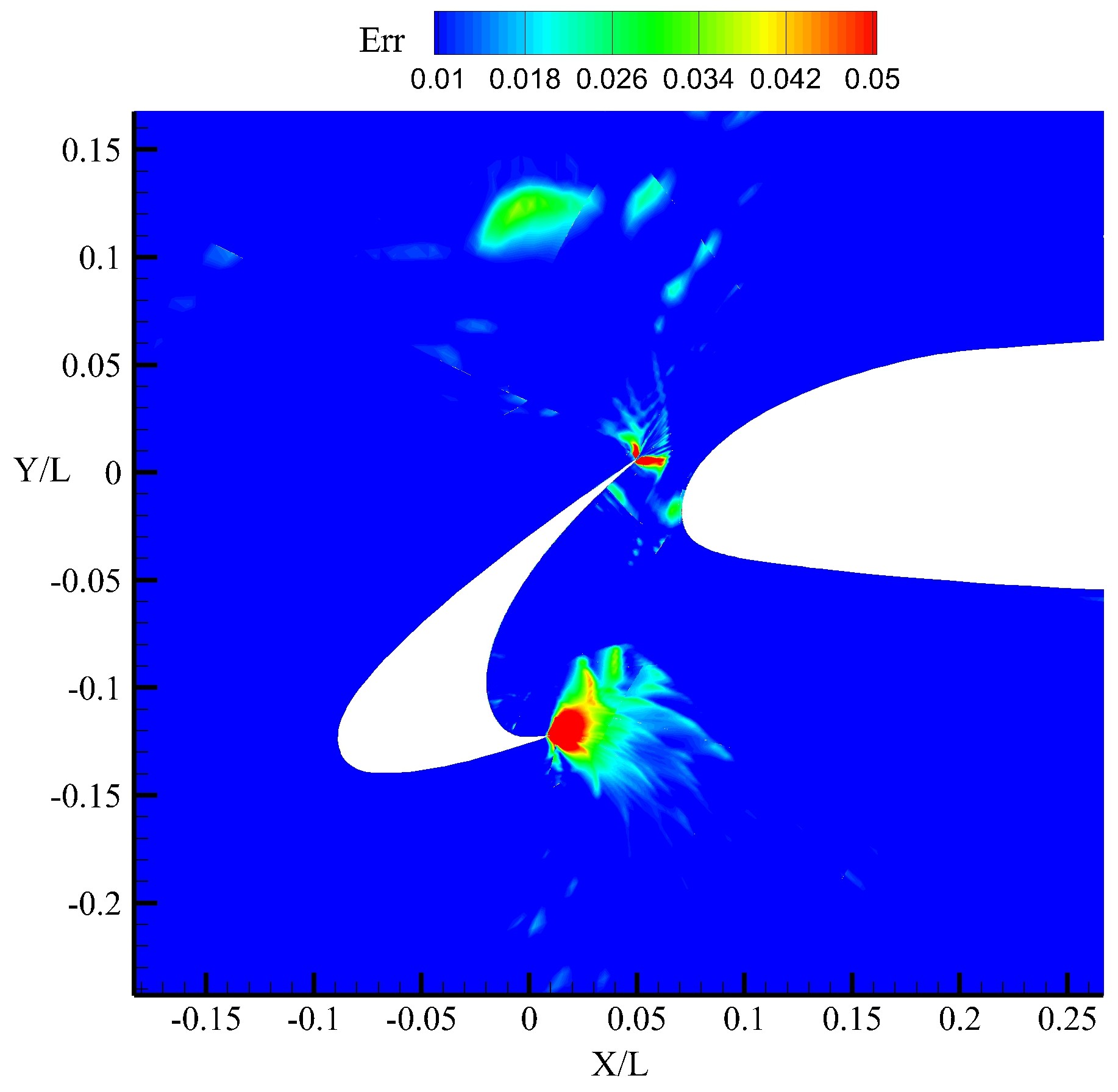}%
	}
	\hfill
	\subfigure[General view \label{fig:sdy-Err}]{%
		\includegraphics[width=0.3\linewidth]{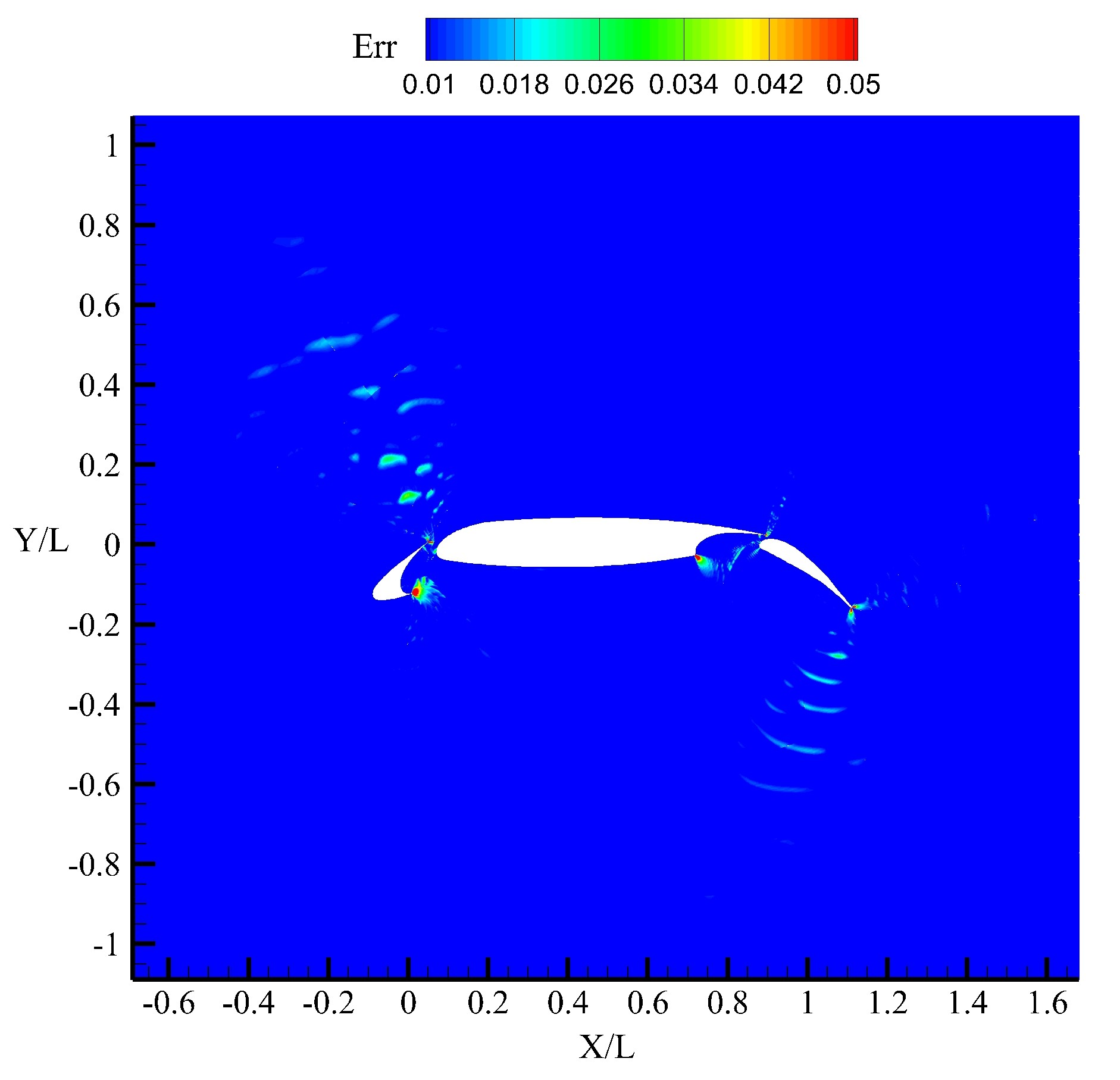}%
	}
	\hfill
	\subfigure[Near flap element \label{fig:sdy-Err-flap}]{%
		\includegraphics[width=0.3\linewidth]{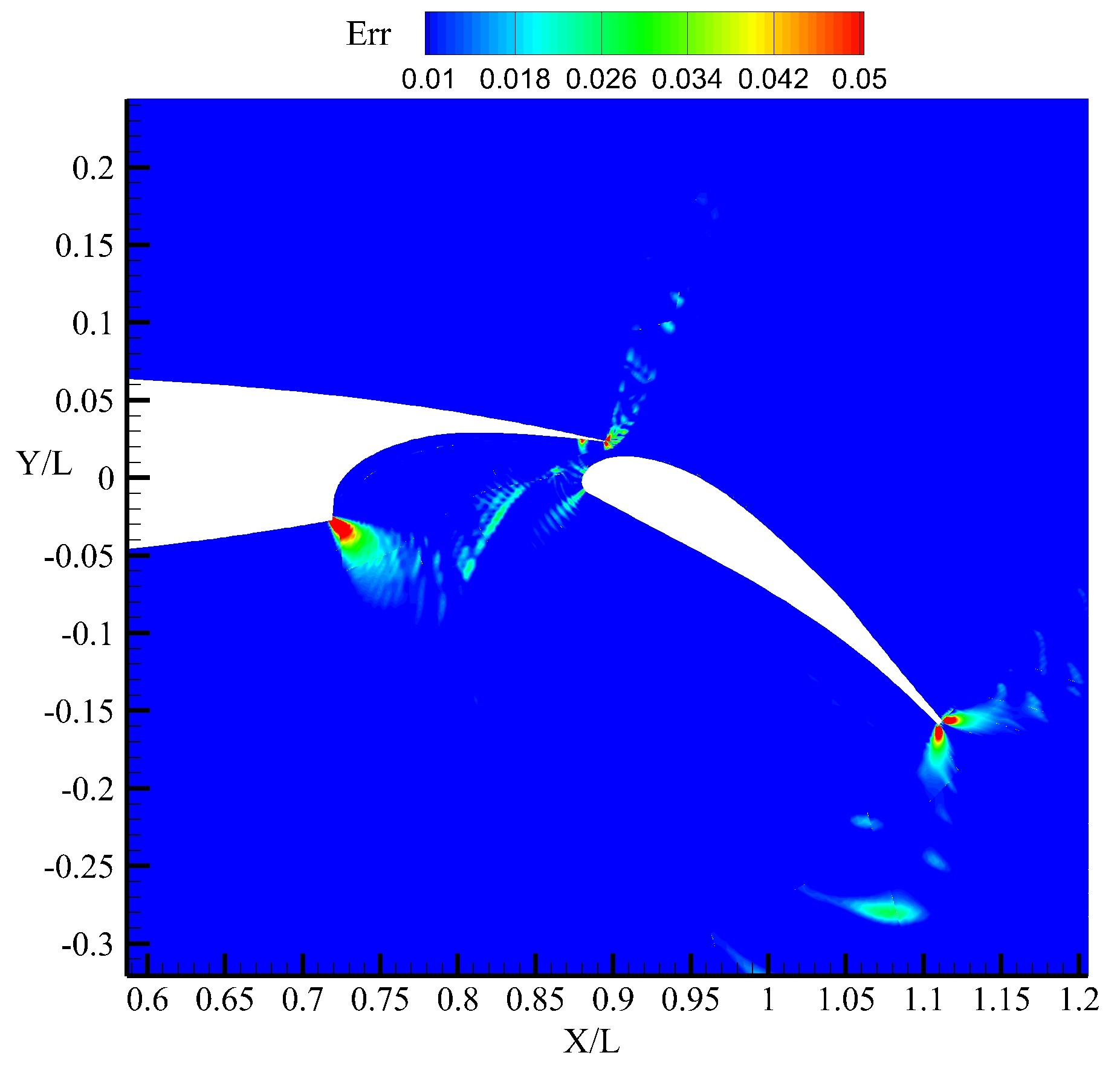}%
	}
	\caption{Relative error distribution of the multi-element airfoil.}
	\label{fig:sdy-Errs}
\end{figure}

The relative-error distributions are shown in Figure \ref{fig:sdy-Errs}. The relative errors in the vast majority of the computational domain, including the near-wall region, are approximately 1\%, whereas values exceeding 5\% are restricted to small neighborhoods of the geometric singularities. The error contours also reveal that local errors generated at these singular points propagate outward along the grid lines and gradually decay to a negligible level. Thus, despite the presence of multiple geometric singularities and discontinuous distance gradients, the numerical scheme produces a non-oscillatory wall-distance field while limiting and damping the associated error propagation. These results demonstrate the robustness and effectiveness of the SAV-FV scheme for wall-distance computation on complex multi-element geometries.

\subsection{ONERA M6 wing} \label{section:M6}

\begin{figure}
	\centering
	\subfigure[Boundary conditions \label{fig:M6-bc}]{%
		\includegraphics[width=0.57\linewidth]{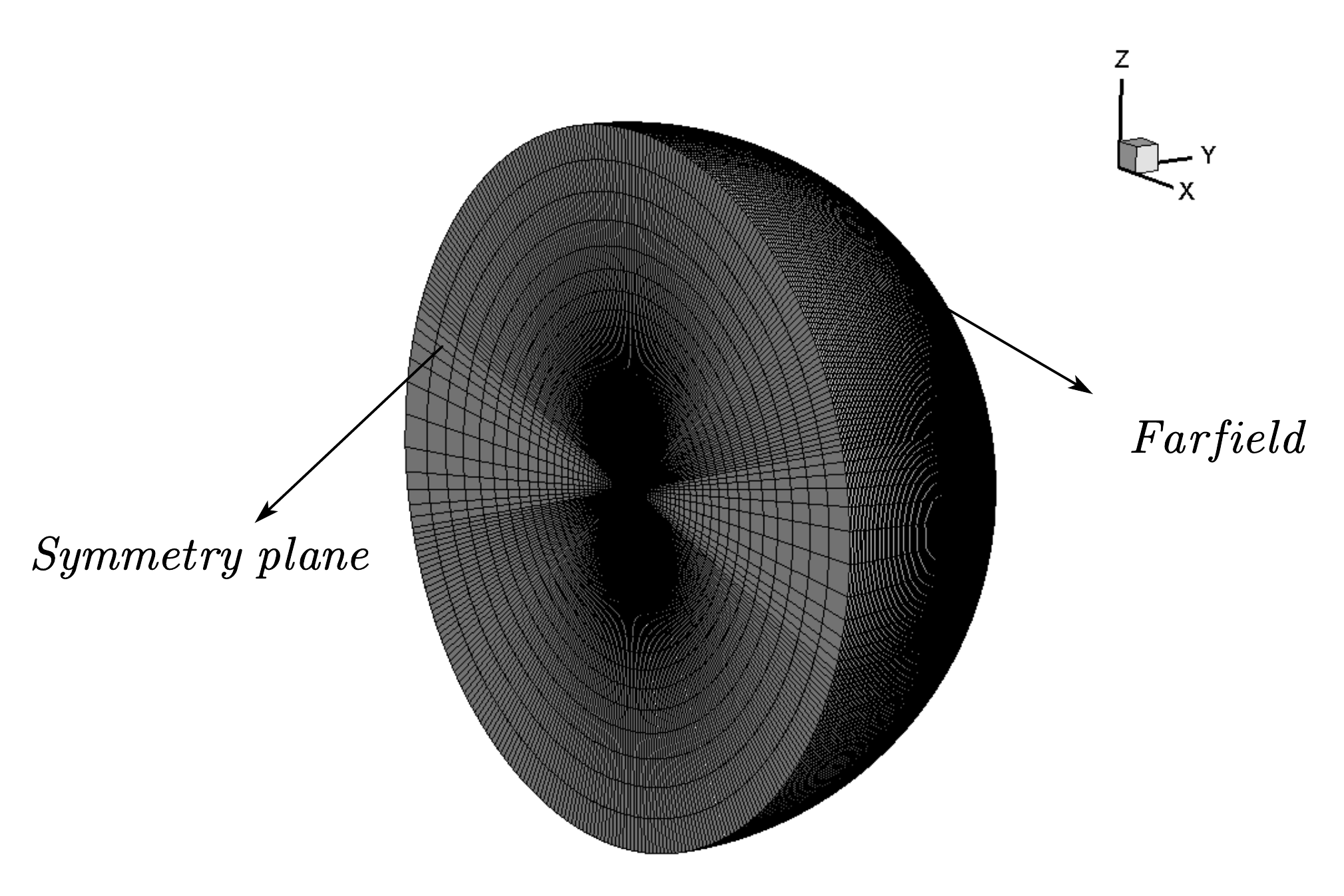}%
	}
	\hfill
	\subfigure[Mesh near the wall \label{fig:M6-grid}]{%
		\includegraphics[width=0.38\linewidth]{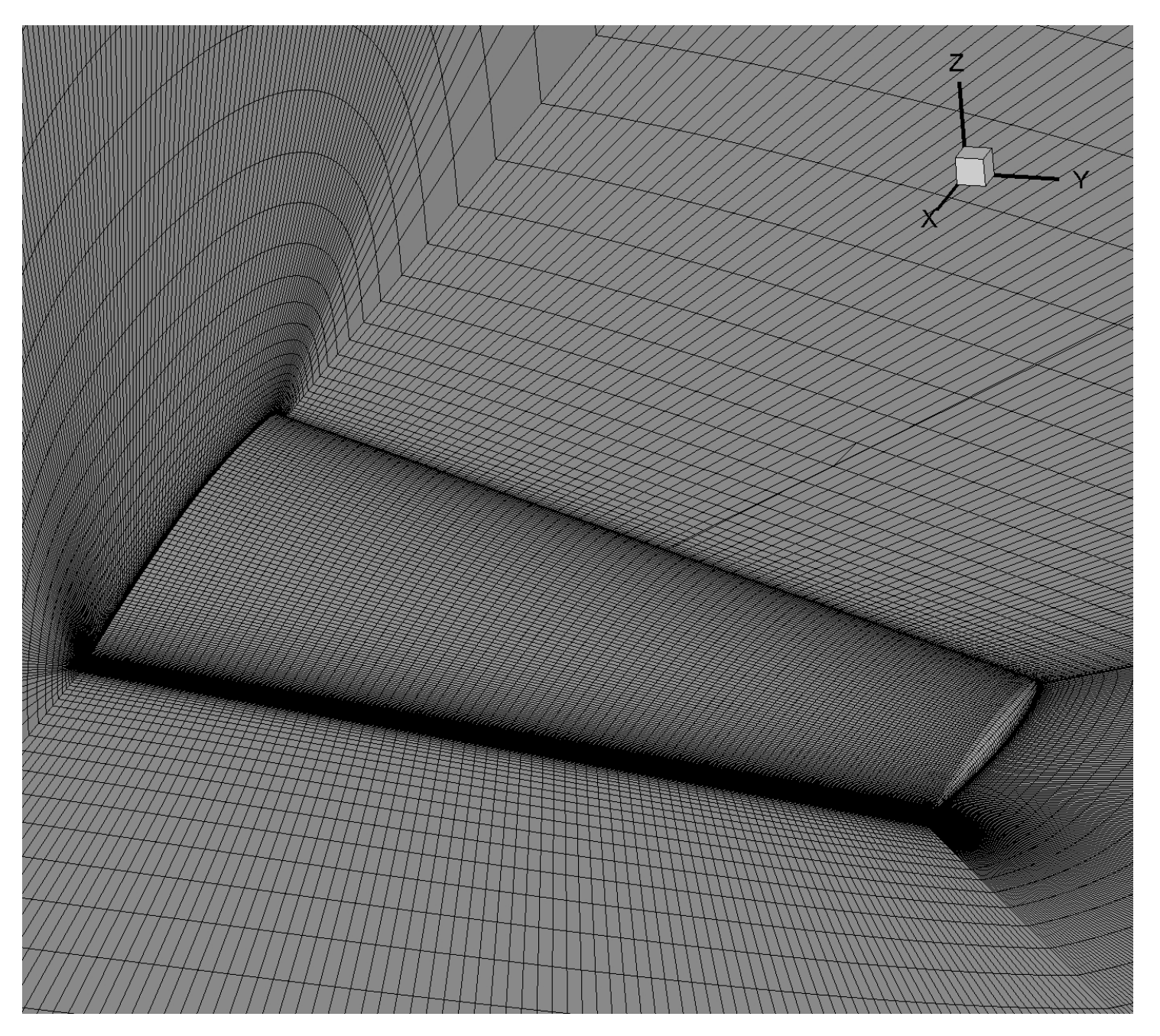}%
	}
	\caption{Computational mesh and boundary conditions for the ONERA M6 wing.}
	\label{fig:M6-geo}
\end{figure}

The ONERA M6 wing is considered to assess the performance of the proposed SAV-FV scheme for wall-distance computation on a practical three-dimensional configuration. The computational domain is a hemispherical region with a radius of 50 reference units. As shown in Figure \ref{fig:M6-geo}, the wing surface is prescribed as the wall boundary, the hemispherical outer boundary is treated as the far field, and a symmetry condition is imposed on the symmetry plane. The body-fitted mesh contains approximately 6.2 million hexahedral cells and can be directly employed in practical aerodynamic computations. The first-layer height is predominantly of order $10^{-7}L$, and the maximum cell aspect ratio is very close to $10^6$. Moreover, severe mesh distortion is present near the left edge of the wing. 
The global pseudo-time step is set to $\Delta\tau=0.1$ for the first 100 iterations and to $\Delta\tau=1$ thereafter until convergence. The upwind weighting parameter is fixed at $\alpha_{\phi}=1$ throughout. The normalized spanwise coordinate is defined as $\eta=y/b$, where $b$ denotes the wingspan.

The three-dimensional ONERA M6 wing poses a more demanding test than the preceding two-dimensional configurations. Its near-wall mesh contains highly elongated and distorted cells, and geometric singularities along the trailing edge give rise to a nonsmooth wall-distance gradient. Together, these features test whether the SAV-FV scheme can suppress spurious oscillations while preserving the accuracy of the computed wall-distance field.

\begin{figure}
	\centering
	\subfigure[$L_1(\nabla \phi)$  \label{fig:m6-L1}]{%
		\includegraphics[width=0.48\linewidth]{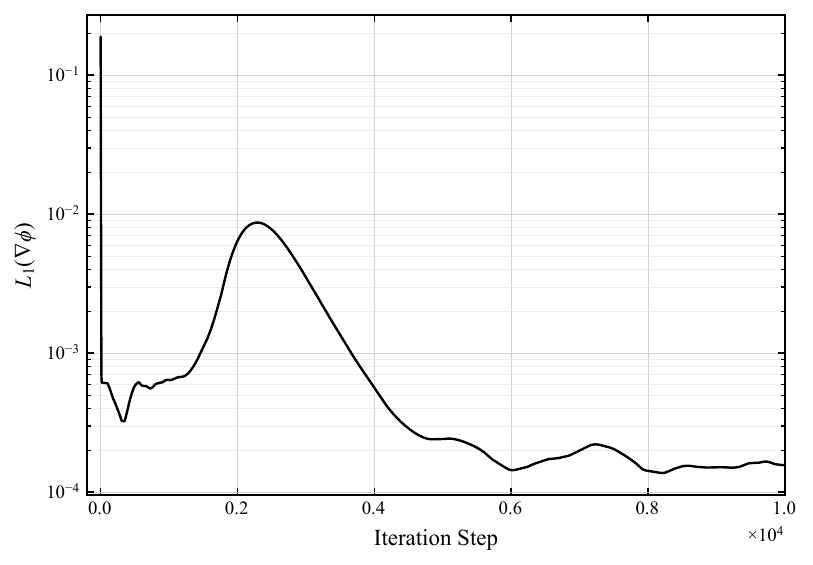}%
	}
	\hfill
	\subfigure[$L_{\infty}(\nabla \phi)$  \label{fig:m6-Linf}]{%
		\includegraphics[width=0.48\linewidth]{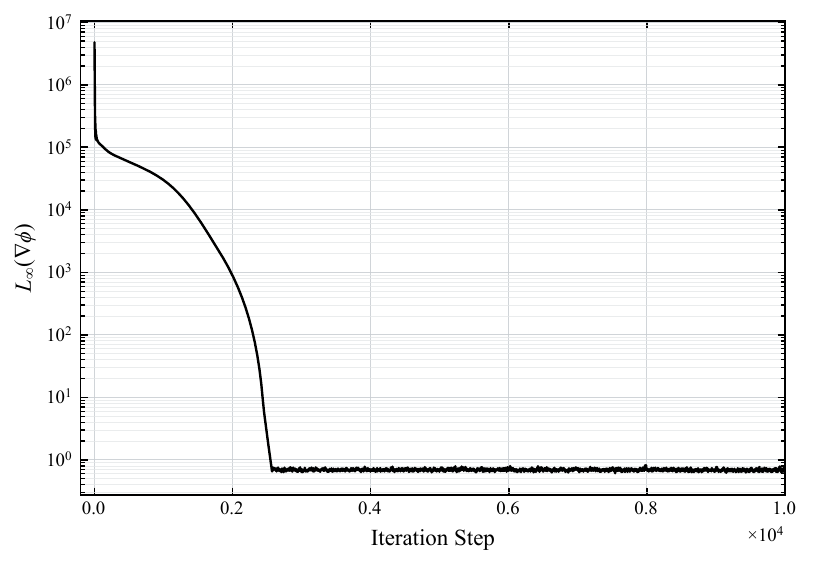}%
	}
	\caption{Convergence history of the ONERA M6 wing.}
	\label{fig:m6-Res}
\end{figure}

Figure \ref{fig:m6-Res} shows that the $L_1(\nabla \phi)$ residual decreases by approximately three orders of magnitude, whereas the $L_{\infty}(\nabla \phi)$ residual decreases by approximately seven orders of magnitude. The curves undergo a rapid initial reduction followed by a gradual approach to nearly flat levels, with small fluctuations in the later stage. This convergence behavior indicates that the method remains stable even for the highly distorted three-dimensional mesh and that the residuals associated with the most difficult local regions are effectively controlled.

\begin{figure}
	\centering
	\subfigure[Wall distance distribution at wall cell centers \label{fig:M6-wcell}]{%
		\includegraphics[width=0.53\linewidth]{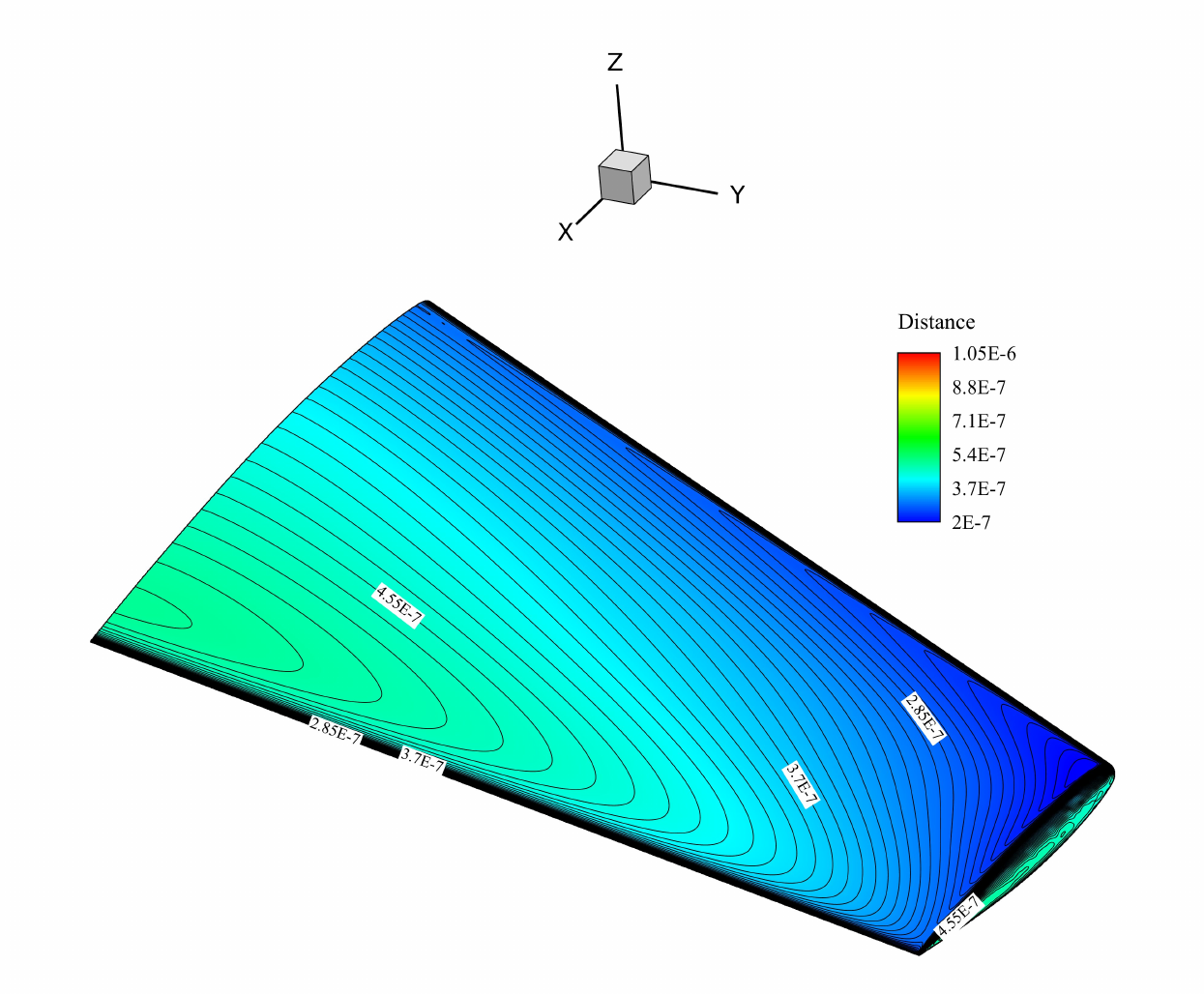}%
	}
	\hfill
	\subfigure[Slice at $\eta=0.5$ \label{fig:M6-dist-y0.6}]{%
		\includegraphics[width=0.42\linewidth]{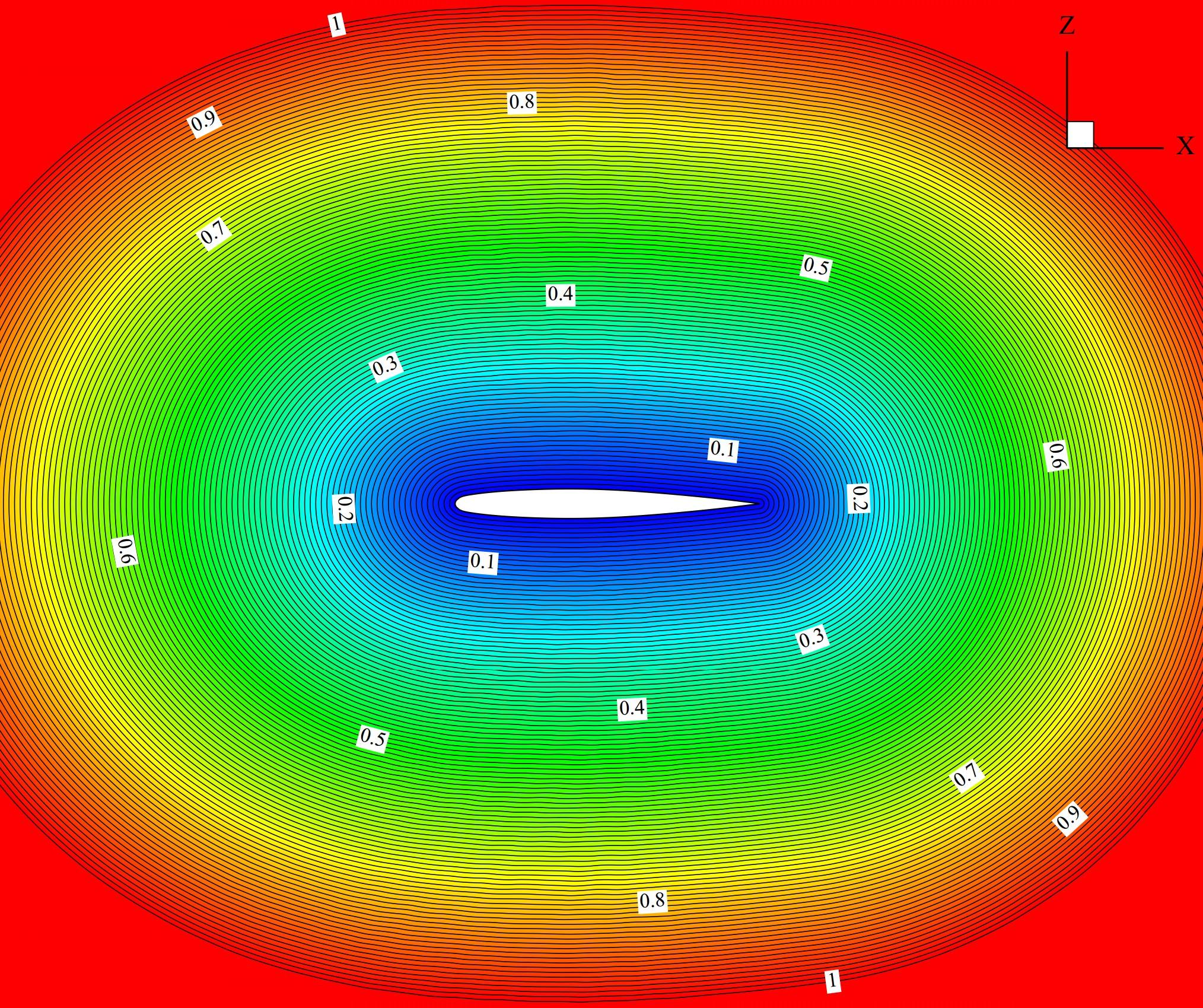}%
	}
	\caption{Computed wall distance of the ONERA M6 wing.}
	\label{fig:M6-dis}
\end{figure}

Figure~\ref{fig:M6-dis} shows the computed wall distance at the centers of the wall-adjacent cells and on the cross-section at $\eta=0.5$. The small distances in the first cell layer are consistent with the near-wall mesh spacing of order $10^{-7}$. Despite severe mesh distortion near the wing edge, the near-wall values vary smoothly along the wing, and the sectional contours follow the wing geometry even near the trailing-edge singularity. The absence of visible spurious oscillations in either view is consistent with the stabilizing effect of the vanishing artificial viscosity on this three-dimensional mesh.

\begin{figure}
	\centering
	\subfigure[Near head \label{fig:M6-Err-h}]{%
		\includegraphics[width=0.3\linewidth]{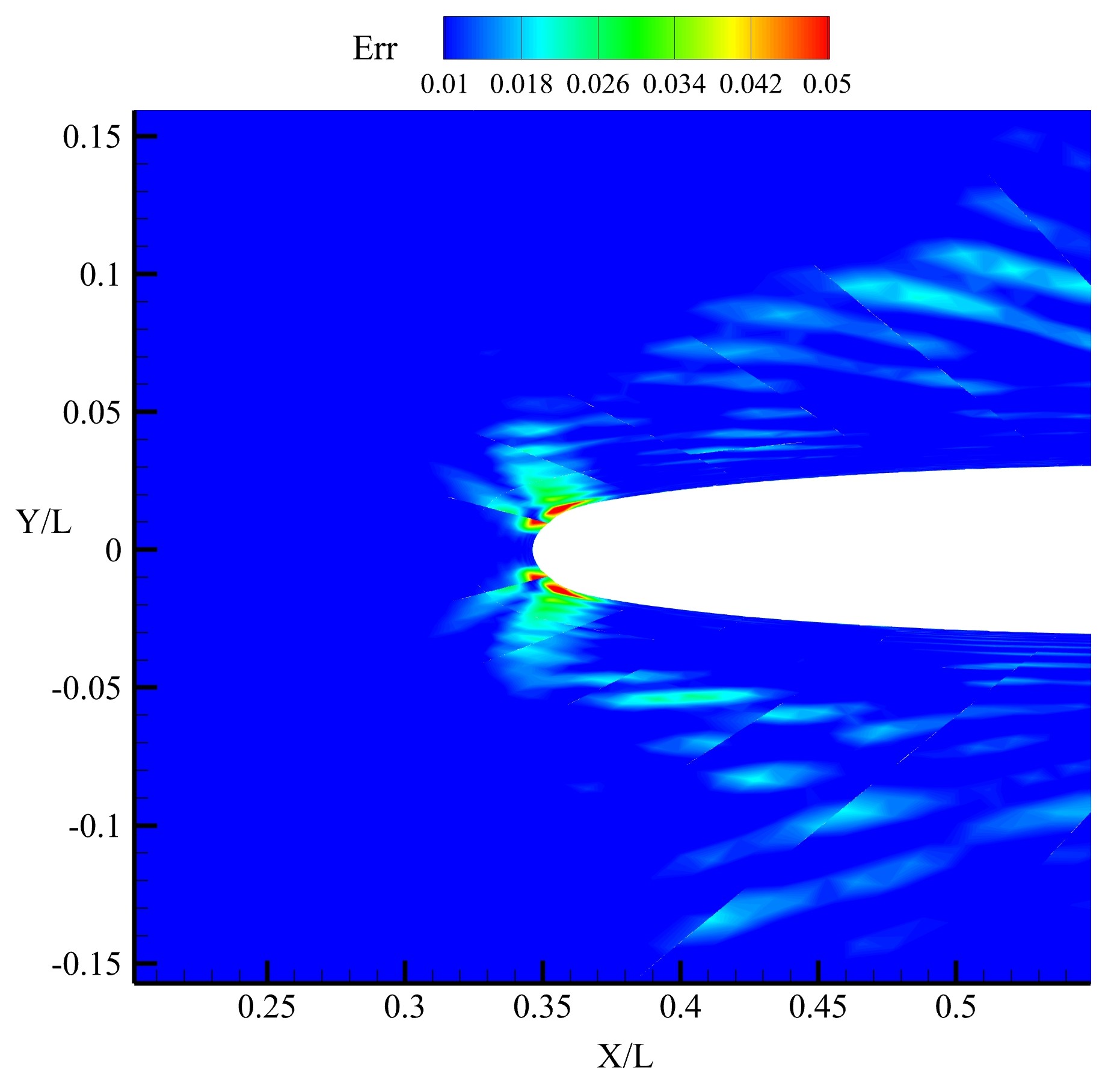}%
	}
	\hfill
	\subfigure[General view \label{fig:M6-Err-a}]{%
		\includegraphics[width=0.3\linewidth]{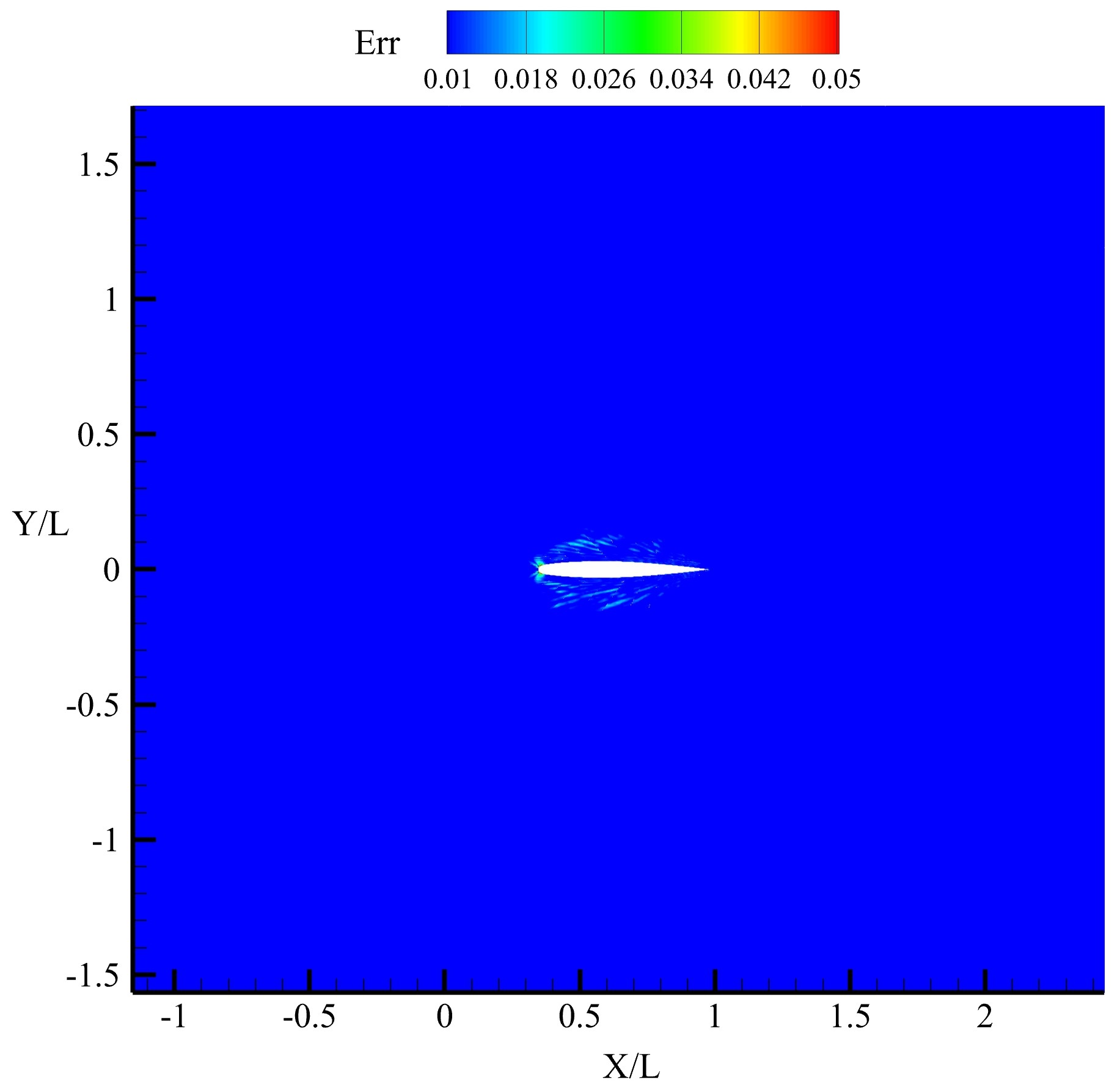}%
	}
	\hfill
	\subfigure[Near tail \label{fig:M6-Err-t}]{%
		\includegraphics[width=0.3\linewidth]{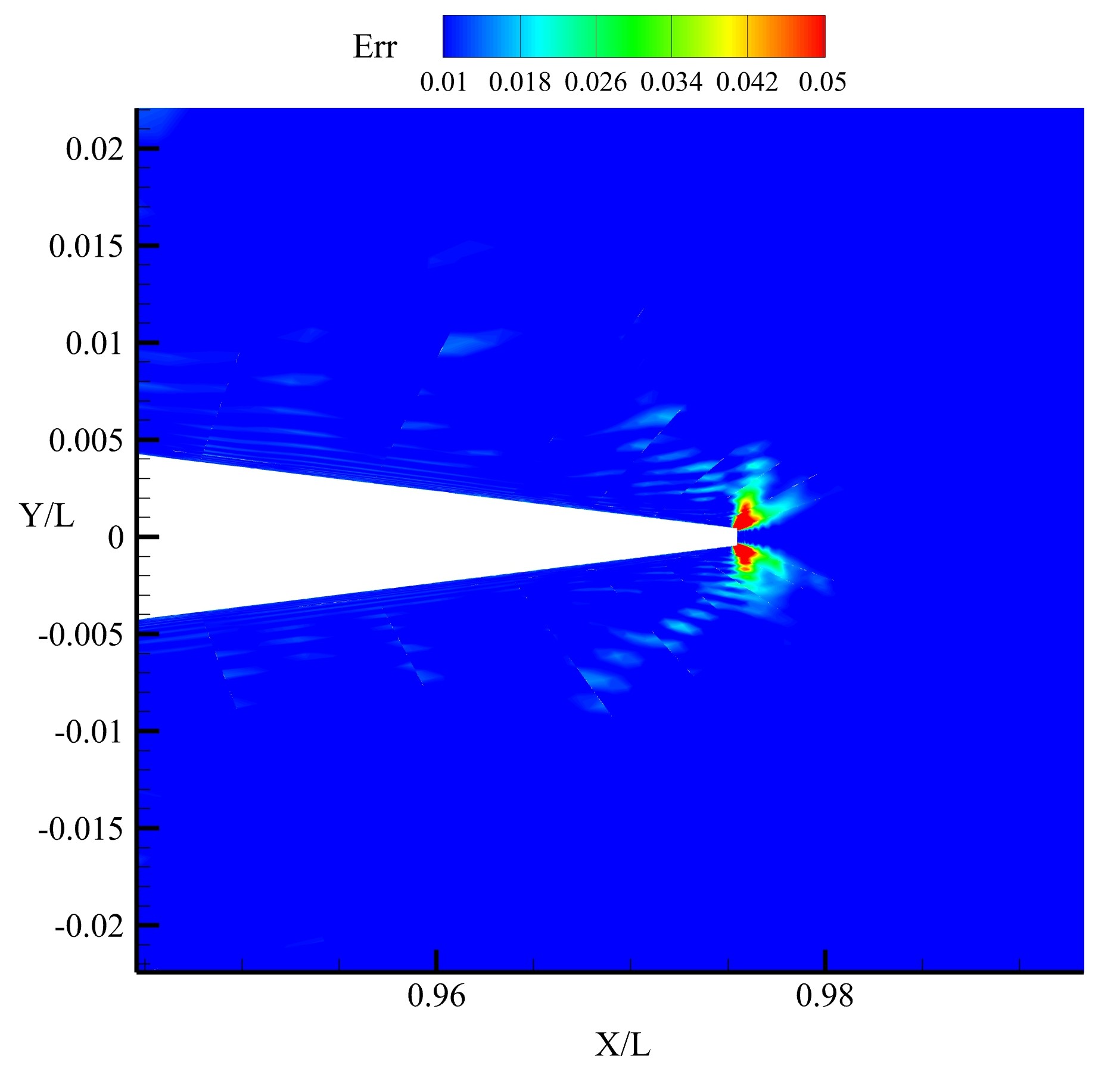}%
	}
	\caption{Relative error at $\eta=0.5$.}
	\label{fig:M6-Err}
\end{figure}

Figure~\ref{fig:M6-Err} shows that the relative wall-distance error is at most $3\%$ over nearly the entire cross-section at $\eta=0.5$. Errors exceeding $5\%$ are confined to small regions near the leading and trailing edges. Near the leading edge, high surface curvature causes the wall-distance gradient to change direction rapidly in a region where the mesh cells have high aspect ratios. Near the trailing edge, the larger errors are concentrated around the geometric singularity.

\begin{figure}
	\centering
	\subfigure[$\eta=0.2$ \label{fig:M6-0.2b}]{%
		\includegraphics[width=0.48\linewidth]{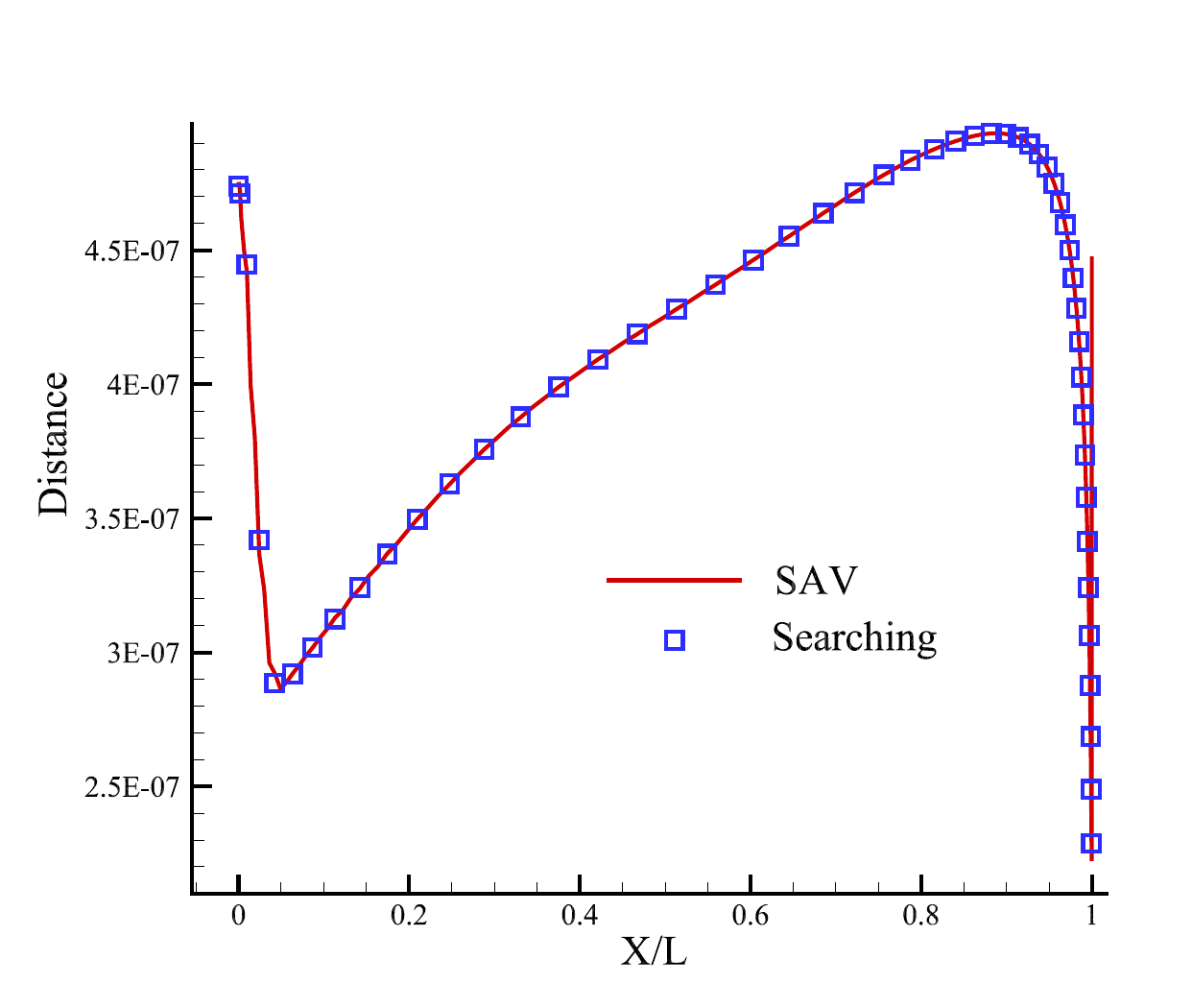}%
	}
	\hfill
	\subfigure[$\eta=0.4$ \label{fig:M6-0.4}]{%
		\includegraphics[width=0.48\linewidth]{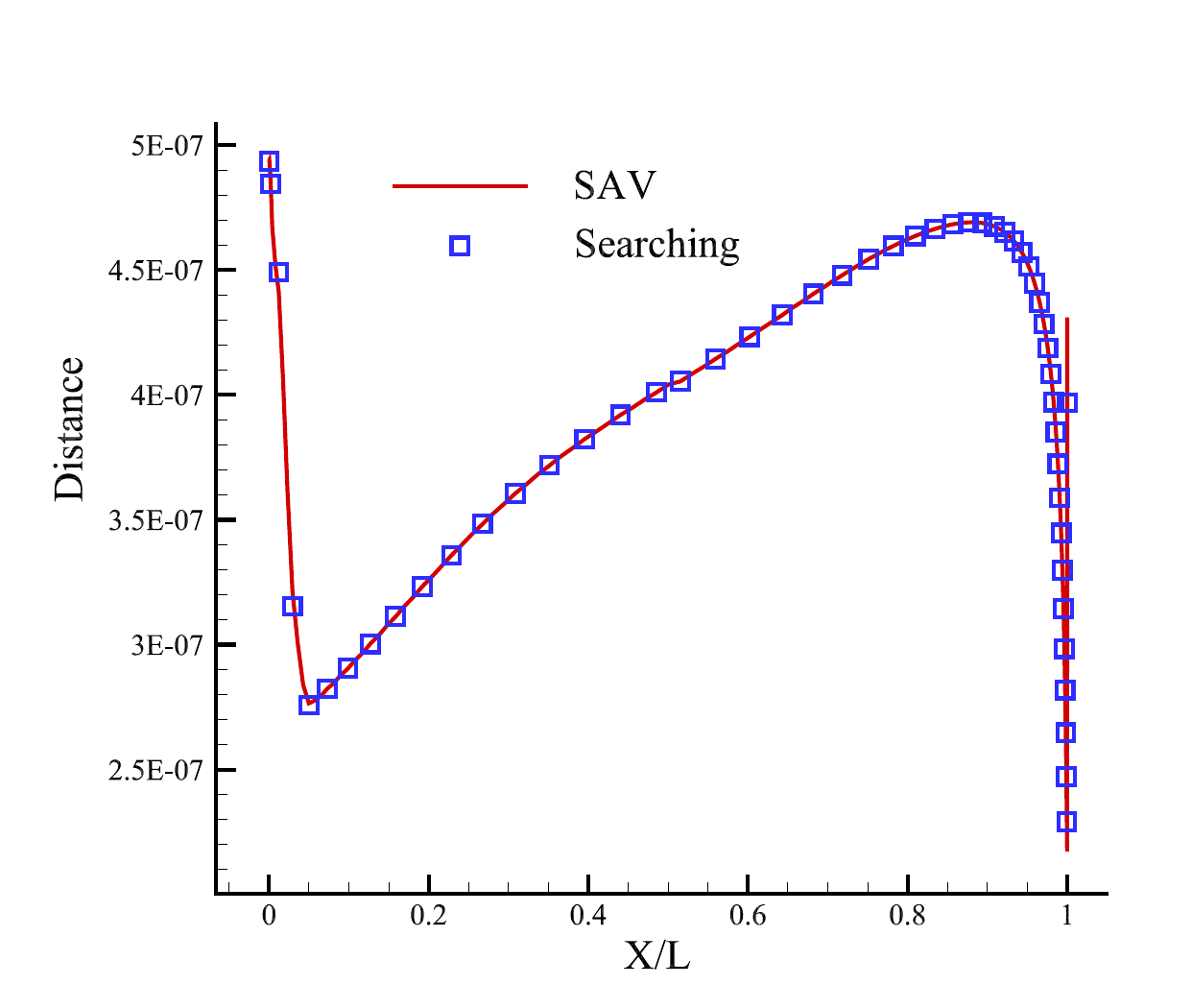}%
	}
	\hfill
	\subfigure[$\eta=0.6$ \label{fig:M6-0.6}]{%
		\includegraphics[width=0.48\linewidth]{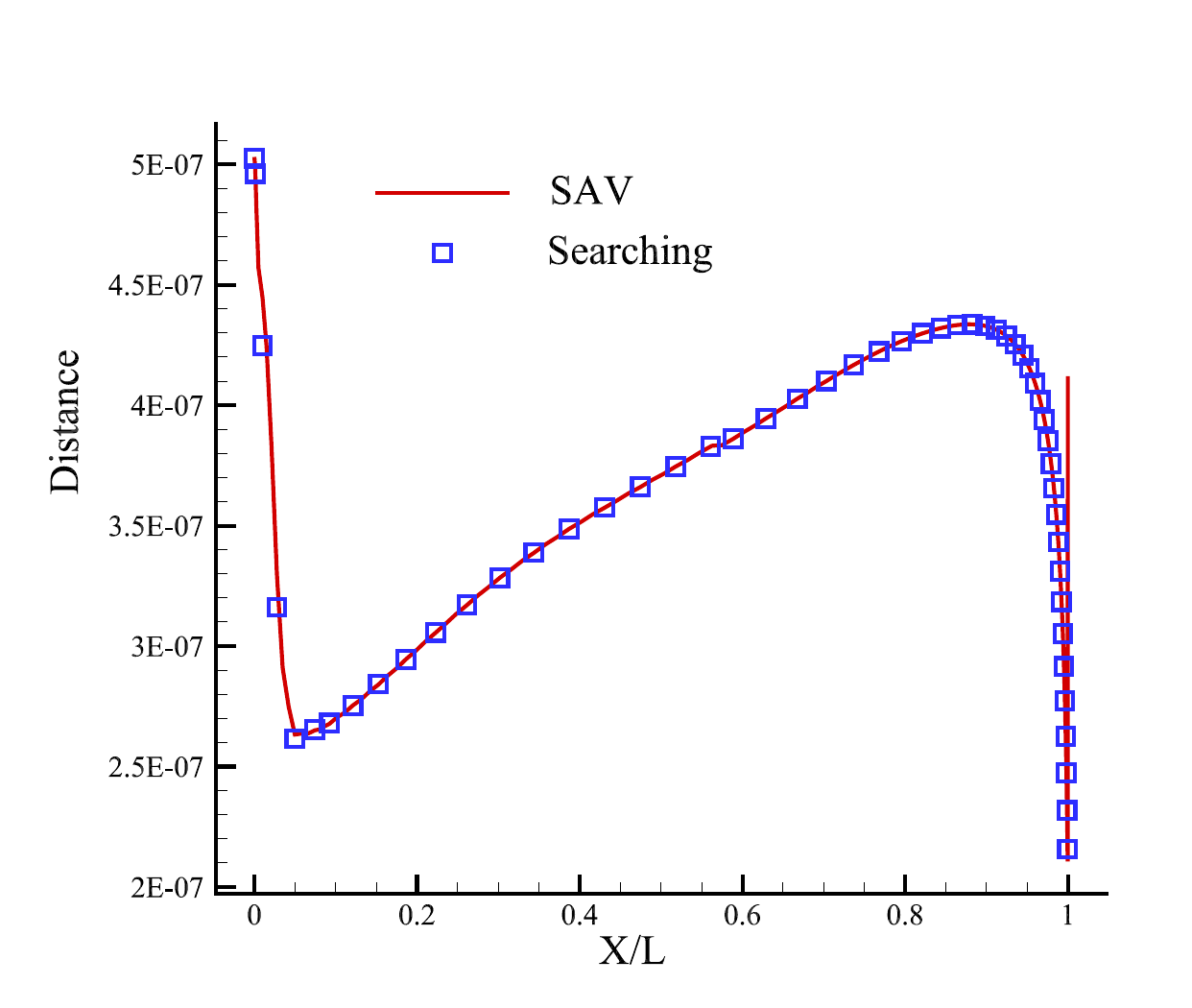}%
	}
	\hfill
	\subfigure[$\eta=0.8$ \label{fig:M6-0.8}]{%
		\includegraphics[width=0.48\linewidth]{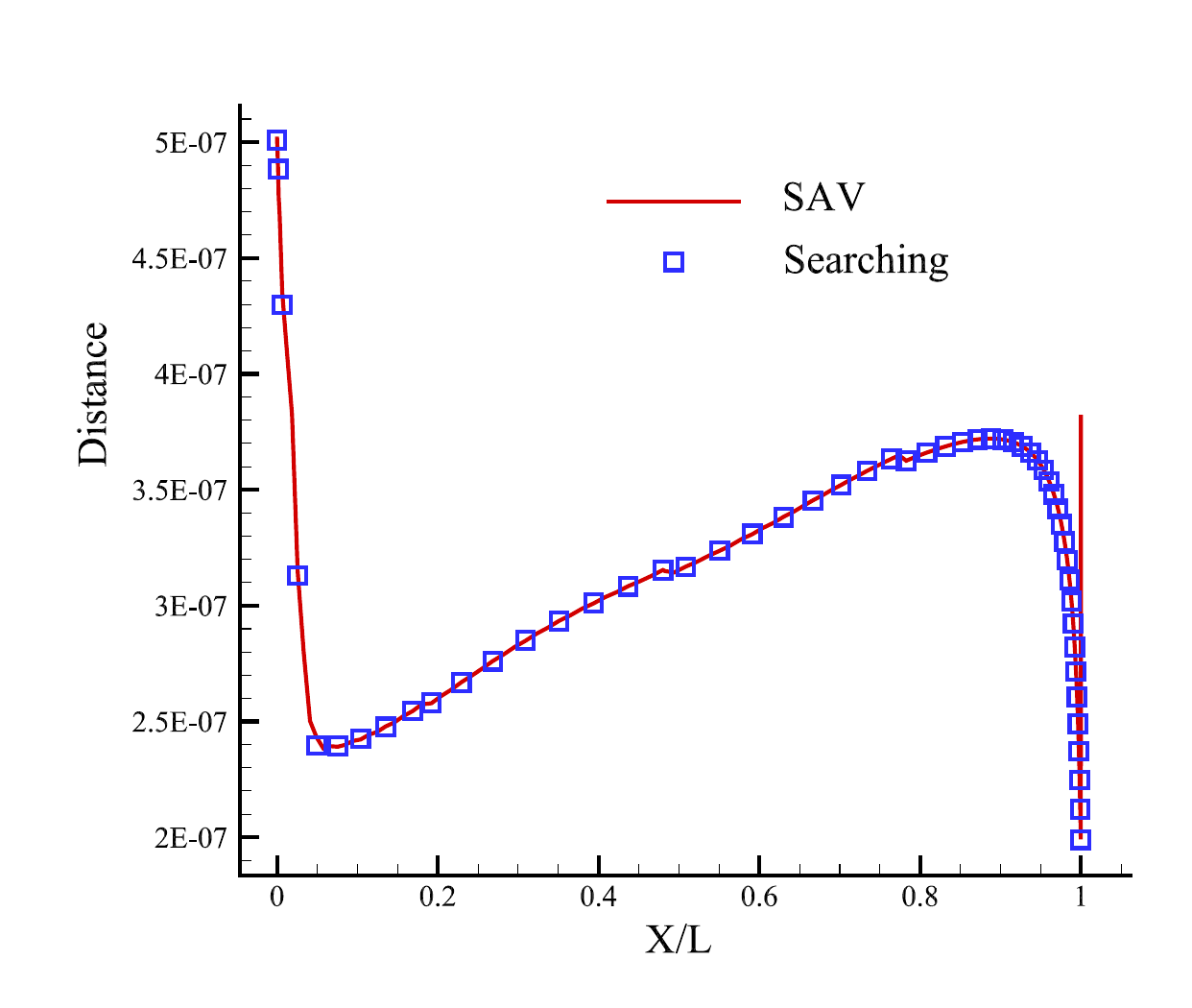}%
	}
	\caption{Wall distance profiles at various stations on the M6 wing upper surface.}
	\label{fig:M6-dis-span}
\end{figure}

Figure~\ref{fig:M6-dis-span} compares the wall distances computed by the SAV-FV scheme with direct-search reference values at the centers of cells adjacent to the upper wing surface. The chordwise profiles agree closely at all four spanwise stations ($\eta=0.2$, $0.4$, $0.6$, and $0.8$), including in highly stretched cells and near the trailing-edge singularity. This agreement suggests that the stabilization does not introduce appreciable over-dissipation into the converged solution. The ONERA M6 results demonstrate stable and accurate wall-distance computation on an engineering-scale three-dimensional mesh with complex geometry, extreme cell aspect ratios, and local mesh distortion.


\section{Conclusions} \label{section:conclusions}

This paper presents a finite-volume scheme that uses a scalar auxiliary variable (SAV) to advance the Eikonal equation in pseudo-time. Under homogeneous Dirichlet boundary conditions, the scheme is unconditionally energy-stable, permitting a global pseudo-time step independent of the grid spacing. 
The residual-based vanishing artificial viscosity suppresses spurious oscillations while preserving the formal order of spatial accuracy. The M-matrix analysis establishes a sufficient pseudo-time step condition for bound preservation. In practical computations with larger global pseudo-time steps, a simple truncation after each SAV update enforces a positive lower bound on the cell-averaged physical wall distance.
The quarter-cylinder test confirms the expected second-order spatial convergence. On complex geometries discretized with highly stretched meshes, the computed wall distances agree closely with direct-search reference values, with the largest discrepancies localized near geometric singularities.

Preserving spatial accuracy limits the artificial dissipation available to damp residuals during pseudo-time iteration. Future work will focus on accelerating residual convergence while retaining spatial accuracy, particularly on meshes with high cell aspect ratios. Since the SAV update is independent of the choice of reconstruction, the framework also provides a basis for extensions to other nonconservative scalar equations and higher-order finite-volume schemes. 

%

\section*{Acknowledgments}

This work was supported by the Science Challenge Project (Grant TZ2026004), and the National Natural Science Foundation of China (Grant 12372284).




\bibliographystyle{elsarticle-num}
\bibliography{ref}

\end{document}